\documentclass{pasj00}
\color{black}
\tenpoint

\SetRunningHead{K. Sato et al.}{
Temperature and Abundance Profiles of Abell~1060 with Suzaku}

\title{
X-Ray Study of Temperature and Abundance Profiles of
the Cluster of Galaxies Abell~1060 with Suzaku}
\author{
 Kosuke \textsc{Sato},\altaffilmark{1}
 Noriko \textsc{Y.~Yamasaki},\altaffilmark{2}
 Manabu \textsc{Ishida},\altaffilmark{2} 
 Yoshitaka \textsc{Ishisaki},\altaffilmark{1} \\ 
 Takaya \textsc{Ohashi},\altaffilmark{1} 
 Hajime \textsc{Kawahara},\altaffilmark{3}
 Takao \textsc{Kitaguchi},\altaffilmark{3} \\
 Madoka \textsc{Kawaharada},\altaffilmark{3}
 Motohide \textsc{Kokubun},\altaffilmark{3}
 Kazuo \textsc{Makishima},\altaffilmark{3} \\
 Naomi \textsc{Ota},\altaffilmark{4}
 Kazuhiro \textsc{Nakazawa},\altaffilmark{2} 
 Takayuki \textsc{Tamura},\altaffilmark{2}\\
 Kyoko \textsc{Matsushita},\altaffilmark{5} 
 Naomi \textsc{Kawano},\altaffilmark{6} 
 Yasushi \textsc{Fukazawa},\altaffilmark{6} \\
and  John P. \textsc{Hughes}\,\altaffilmark{7} 
}
\altaffiltext{1}{
Department of Physics, Tokyo Metropolitan University,
 1-1 Minami-Osawa, Hachioji, Tokyo 192-0397}
\email{ksato@phys.metro-u.ac.jp}
\altaffiltext{2}{
Institute of Space and Astronautical Science (ISAS),
Japan Aerospace Exploration Agency, \\
3-1-1 Yoshinodai, Sagamihara, Kanagawa 229-8510}
\altaffiltext{3}{
Department of Physics, University of Tokyo,
 7-3-1 Hongo, Bunkyo, Tokyo 113-0033} 
\altaffiltext{4}{
Cosmic Radiation Laboratory,
The Institute of Physical and Chemical Research (RIKEN),\\
 2-1 Hirosawa, Wako, Saitama 351-0198} 
\altaffiltext{5}{
Department of Physics, Tokyo University of Science,
1-3 Kagurazaka, Shinjuku-ku, Tokyo 162-8601}
\altaffiltext{6}{
Department of Physical Science, School of Science, Hiroshima University,\\
 1-3-1 Kagamiyama, Higashi-Hiroshima, Hiroshima 739-8526}
\altaffiltext{7}{
Department of Physics and Astronomy, Rutgers University,
 Piscataway, NJ 08854-8019, USA}
\Received{---}
\Accepted{---}
\Published{---}

\begin{document}
\maketitle

\begin{abstract}
We carried out observations of the central and $20'$ east offset 
regions of the cluster of galaxies Abell 1060 with Suzaku.
Spatially resolved X-ray spectral analysis has revealed
temperature and abundance profiles of Abell~1060
out to $27'\simeq 380\; h_{70}^{-1}$~kpc,
which corresponded to $\sim 0.25\; r_{180}$.
Temperature decrease of the intra cluster medium
from 3.4~keV at the center to 2.2~keV in the outskirt region
are clearly observed.
Abundances of Si, S and Fe also decrease by more than 50\%
from the center to the outer,
while Mg shows fairly constant abundance distribution at $\sim 0.7$ solar
within $r\lesssim 17'$.
O shows lower abundance of $\sim 0.3$ solar
in the central region ($r\lesssim 6'$),
and indicates a similar feature with Mg,
however it is sensitive to the estimated contribution of the Galactic
components of $kT_1\sim 0.15$~keV and $kT_2\sim 0.7$~keV
in the outer annuli ($r\gtrsim 13'$).
Systematic effects due to the point spread function tails,
contamination on the XIS filters, instrumental background,
cosmic and/or Galactic X-ray background,
and the assumed solar abundance tables are carefully examined.
Results on temperature and abundances of Si, S, and Fe are consistent
with those derived by XMM-Newton at $r\lesssim 13'$.
Formation and metal enrichment process of the cluster are discussed 
based on the present results.
\end{abstract}

\section{Introduction}

Clusters of galaxies, being the largest virialized system 
in the universe, are filled with  the intracluster medium (ICM), 
which consists of X-ray emitting hot plasma with typical temperatures
of a few times $10^7$~K\@. 
X-ray spectroscopy of the ICM can immediately
determine its temperature and metal abundances. 
The metal abundances of ICM have a lot of information to understand 
the chemical history and evolution of clusters.
A large amount of metals of the ICM are mainly produced by supernovae (SN) 
in early-type galaxies \citep{arnaud92,renzini93}, 
which are classified roughly as Type Ia (SN~Ia) and Type II (SN~II)\@.
Si, S and Fe are synthesized in both SN~Ia and SN~II, 
while $\alpha$ elements such as O, Ne, and Mg are mainly in SN~II, 
which are explosions of massive stars with initial mass above 
$\sim 10$~$M_\odot$. The metals produced in the galaxies are 
transfered into the ICM by galactic wind and/or ram pressure 
stripping.

ASCA firstly revealed the distribution of Si and Fe 
in the ICM \citep{fukazawa98,fukazawa00,finoguenov00,finoguenov01}.
The derived iron-mass-to-light ratios
(IMLR; \cite{Ciotti1991,renzini93,Renzini1997}) are nearly constant 
in rich clusters and decrease toward poorer systems \citep{makishima01}.
Recent observations with Chandra and XMM-Newton
allowed detailed studies of the metals in the ICM\@.
These observations, however, showed abundance profiles 
of O, Mg, Si and Fe 
only for the central regions of very bright clusters or groups of galaxies 
dominated by cD galaxies in a reliable manner
\citep{finoguenov02,fukazawa04,matsushita03,tamura03}. 
The abundance profiles of O and Mg, in particular 
for the cluster outer regions, are still poorly
determined, 
because these satellites are characterized by
relatively high intrinsic background levels.
\citet{tamura04} derived IMLR for five clusters within
250~$h_{100}^{-1}$~kpc to be $\sim0.01~M_{\odot}/L_{\odot}$,
and the oxygen mass within 50~$h_{100}^{-1}$~kpc for several clusters.
However, oxygen-mass-to-light ratios (OMLR) for rich clusters
are not reliable due to the lower emissivity of O\emissiontype{VII}
and O\emissiontype{VIII} lines in higher temperatures.
\citet{degrandi01,hayakawa06} found that clusters associated 
with cD galaxies and
central cool components showed abundance concentration 
in the cluster center, while clusters without cD galaxies showed 
flatter profiles. The central metallicity enhancement in the cool core
clusters were further studied and the excess metals were shown to be
supplied from the cD galaxies \citep{degrandi04}.

Abell~1060 (hereafter A~1060) is a nearby cluster of galaxies ($z=0.0114$)
characterized by a smooth and symmetric distribution
of intracluster medium (ICM), and has no cD-galaxy at the center.
Chandra observation detected very compact X-ray emissions from
the central two elliptical galaxies \citep{yamasaki02},
and the temperature and abundance distributions were shown to be somewhat
inhomogeneous at the cluster center \citep{hayakawa04}.
\citet{hayakawa06} detected a temperature drop by $\sim 30$\% from 
the central region to $r\sim13'$ with XMM-Newton,
while it had previously been considered as flat on the basis of 
the ASCA and ROSAT observations 
(\cite{tamura96}, \cite{furusho01}).

This paper reports results from Suzaku observations of A~1060. 
Owing to the low-background nature of the Suzaku XIS, 
we are able to measure the temperature and abundance profiles 
to a much outer region than the previous XMM-Newton study.
We use $H_0=70$ km~s$^{-1}$~Mpc$^{-1}$,
$\Omega_{\Lambda} = 1-\Omega_M = 0.73$ in this paper. 
At a redshift of $z=0.0114$, 1$'$ corresponds to 14~kpc, 
and the virial radius,
$r_{\rm 180} = 1.95\; h_{100}^{-1}\sqrt{k\langle T\rangle/10~{\rm keV}}$~Mpc
\citep{markevitch98}, is 1.53~Mpc for
the average temperature of $\langle T\rangle = 3$~keV \citep{hayakawa06}.
Throughout this paper we adopt the Galactic hydrogen column density 
of $N_{\rm H} = 4.9\times 10^{20}$ cm$^{-2}$ \citep{dickey90}
in the direction of A~1060.
Otherwise noted, the solar abundance table is given by \citet{anders89},
and errors are 90\% confidence
region for a single interesting parameter.

\section{Suzaku Observation and Data Reduction}\label{sec:obs}

\begin{figure*}
\begin{minipage}{\textwidth}
\vspace*{-1.5ex}
\FigureFile(0.5232936\textwidth,\textwidth){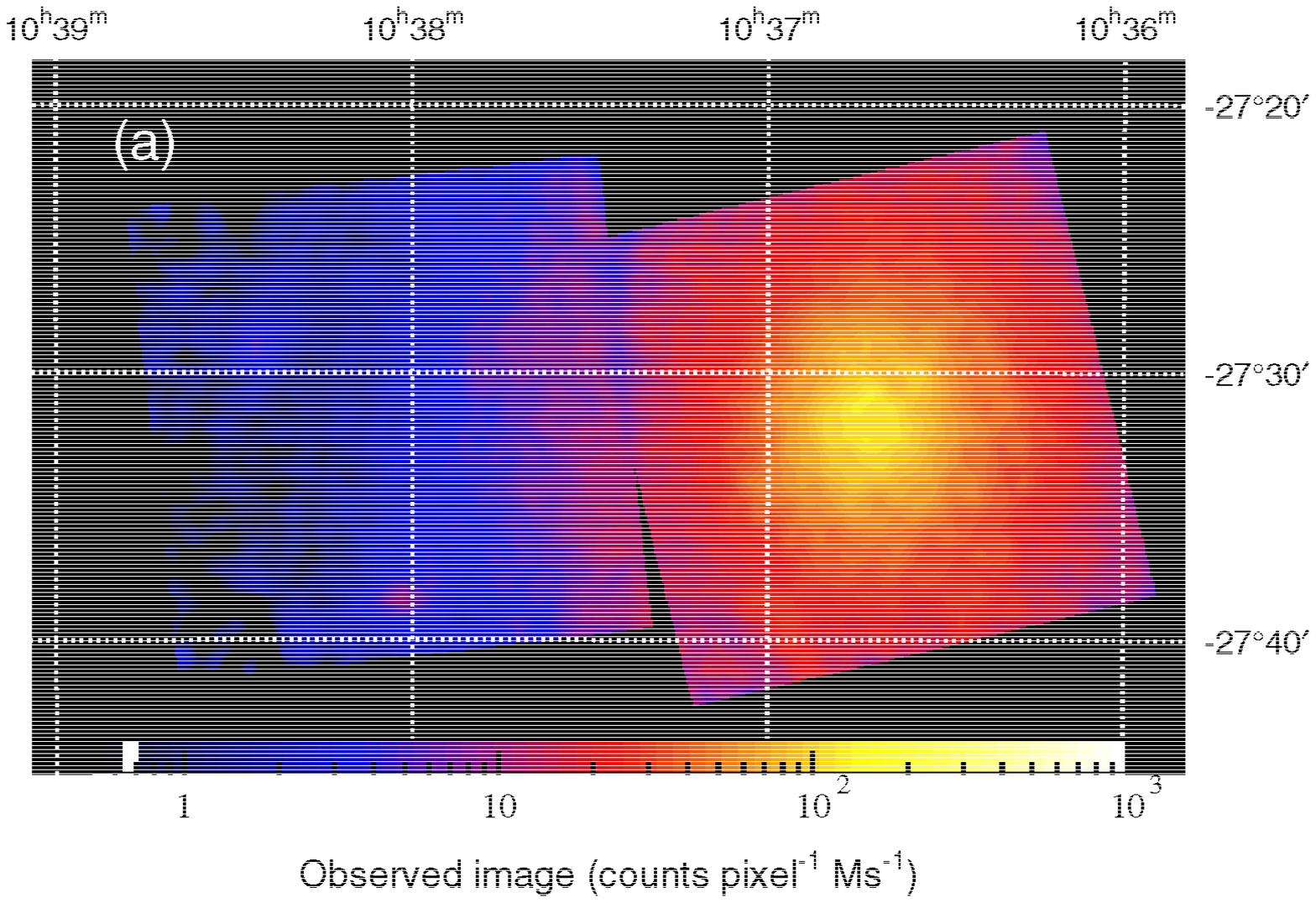}%
\FigureFile(0.4767064\textwidth,\textwidth){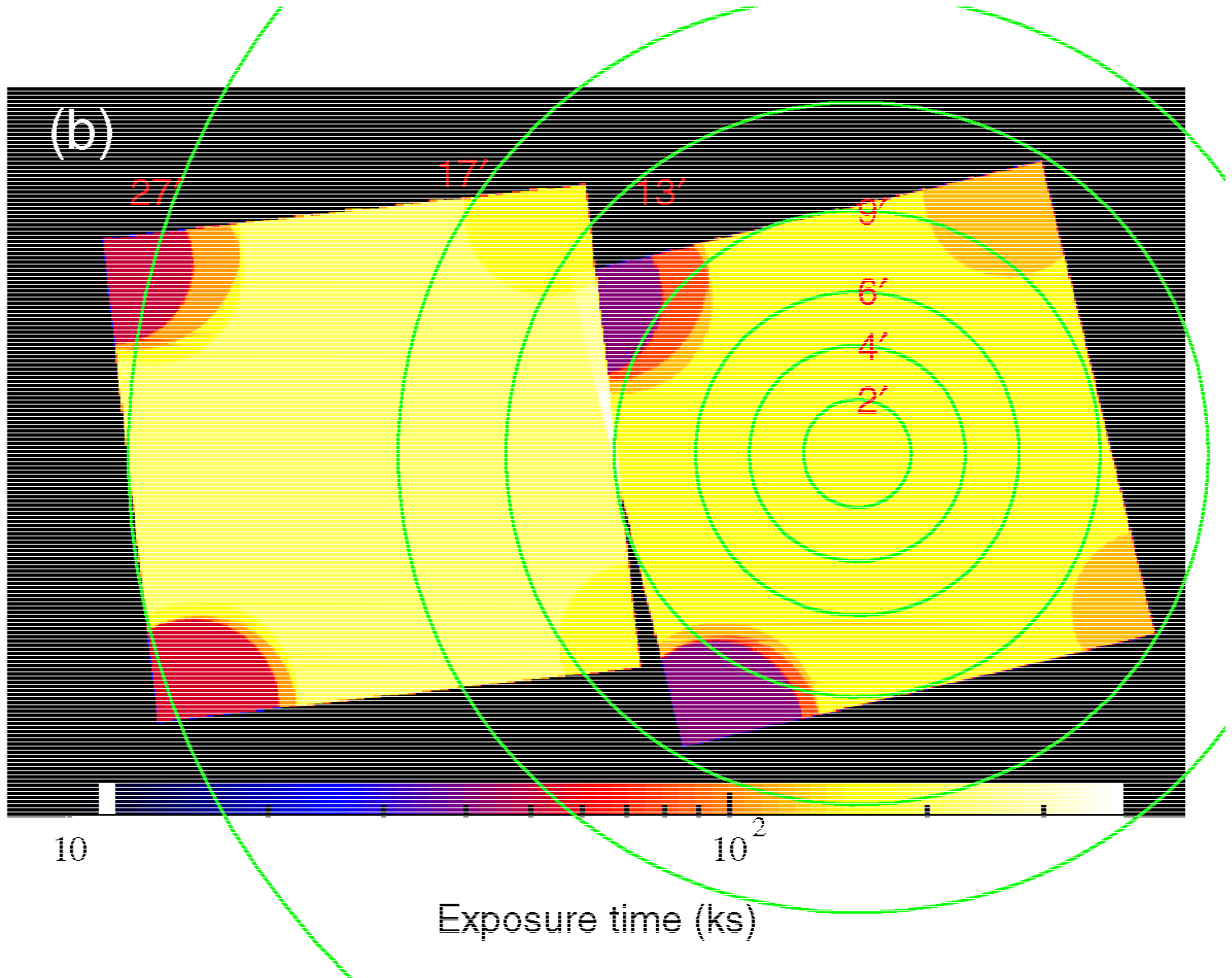}
\end{minipage}
\vspace*{-2ex}
\caption{
(a) Combined XIS image of the central and offset observations in the
0.5--7.0 keV energy range. The observed XIS0-3 images are added on the sky
coordinate after removing each calibration source region,
and smoothed with $\sigma=16$ pixel $\simeq 17''$ Gaussian.
Estimated components of extragalactic X-ray background (CXB)
and instrumental background (NXB) are subtracted,
and exposure and vignetting are corrected.
(b) Exposure map in ks, in which
coordinates of the image are the same with (a).
Four XIS sensors are treated separately.
Small exposure at the CCD corners correspond to
the $^{55}$Fe calibration source locations.
The annular regions utilized in the spectral analysis are
indicated by green circles.
}\label{fig:image}
\end{figure*}

\begin{table*}
\caption{Suzaku observation of A~1060 and NGC~2992}
\label{tab:ob}
\begin{tabular}{cccccc} 
\hline \hline
Target name & Sequence number & Date & Exposure time & (RA, Dec) in J2000 $^\ast$ & ($l$, $b$) \\ 
\hline
A~1060 center & 800003010 & 2005-Nov-22 & 40.5 ks & (\timeform{10h36m42.8s}, \timeform{-27D31'42''}) & (\timeform{269.D60}, \timeform{26.D49})\\
A~1060 offset & 800004010 & 2005-Nov-20 & 55.5 ks & (\timeform{10h38m03.8s}, \timeform{-27D31'42''}) & (\timeform{269.D88}, \timeform{26.D65}) \\
\hline
NGC~2992 $^\dagger$ & 700005010 & 2005-Nov-06 & 38.8 ks & (\timeform{09h45m45.2s}, \timeform{-14D16'10''}) & (\timeform{249.D69}, \timeform{28.D83})  \\
NGC~2992 $^\dagger$ & 700005020 & 2005-Nov-19 & 38.7 ks & (\timeform{09h45m41.7s}, \timeform{-14D16'05''}) & (\timeform{249.D66}, \timeform{28.D82})  \\
NGC~2992 $^\dagger$ & 700005030 & 2005-Dec-13 & 48.3 ks & (\timeform{09h45m51.0s}, \timeform{-14D16'51''}) & (\timeform{249.D70}, \timeform{28.D84})  \\
\hline\\[-1ex]
\multicolumn{6}{l}{\parbox{0.95\textwidth}{\footnotesize 
\footnotemark[$\ast$]
Average pointing direction of the XIS, written in the 
RA\_NOM and DEC\_NOM keywords of the event FITS files.}}\\
\multicolumn{6}{l}{\parbox{0.95\textwidth}{\footnotesize
\footnotemark[$\dagger$]
We used NGC~2992 data to estimate the Galactic component
in subsection \ref{subsec:direct}.}}\\
\end{tabular}
\end{table*}

\subsection{Observation}
\label{subsec:obs}

Suzaku carried out two pointing observations
for A~1060 in November 2005, 
the central region and 20$'$ east offset region, 
with exposure of 40.5 and 55.5~ks, respectively. 
The observation log is summarized in table~\ref{tab:ob},
and the combined X-ray Imaging Spectrometers (XIS; \cite{koyama06}) image
in the 0.5--7~keV range is shown in figure~\ref{fig:image}.
We utilize only the XIS data in this paper.
The XIS is an X-ray CCD camera, which consists of
one back-illuminated sensor (BI = XIS1) and
three front-illuminated sensors (FI = XIS0, XIS2, XIS3).
The BI and FI sensors have different advantages.
The former has higher quantum efficiency in the soft energy band
($E \lesssim 1$~keV),
while the latter shows lower instrumental non X-ray background (NXB)\@.
The XIS was operated in the Normal clocking mode
(no window nor burst option, so 8~s exposure per frame),
with the standard $5\times 5$ or $3\times 3$ editing mode \citep{koyama06}.

\begin{figure*}
\begin{minipage}{0.48\textwidth}
\centerline{\FigureFile(\textwidth,\textwidth){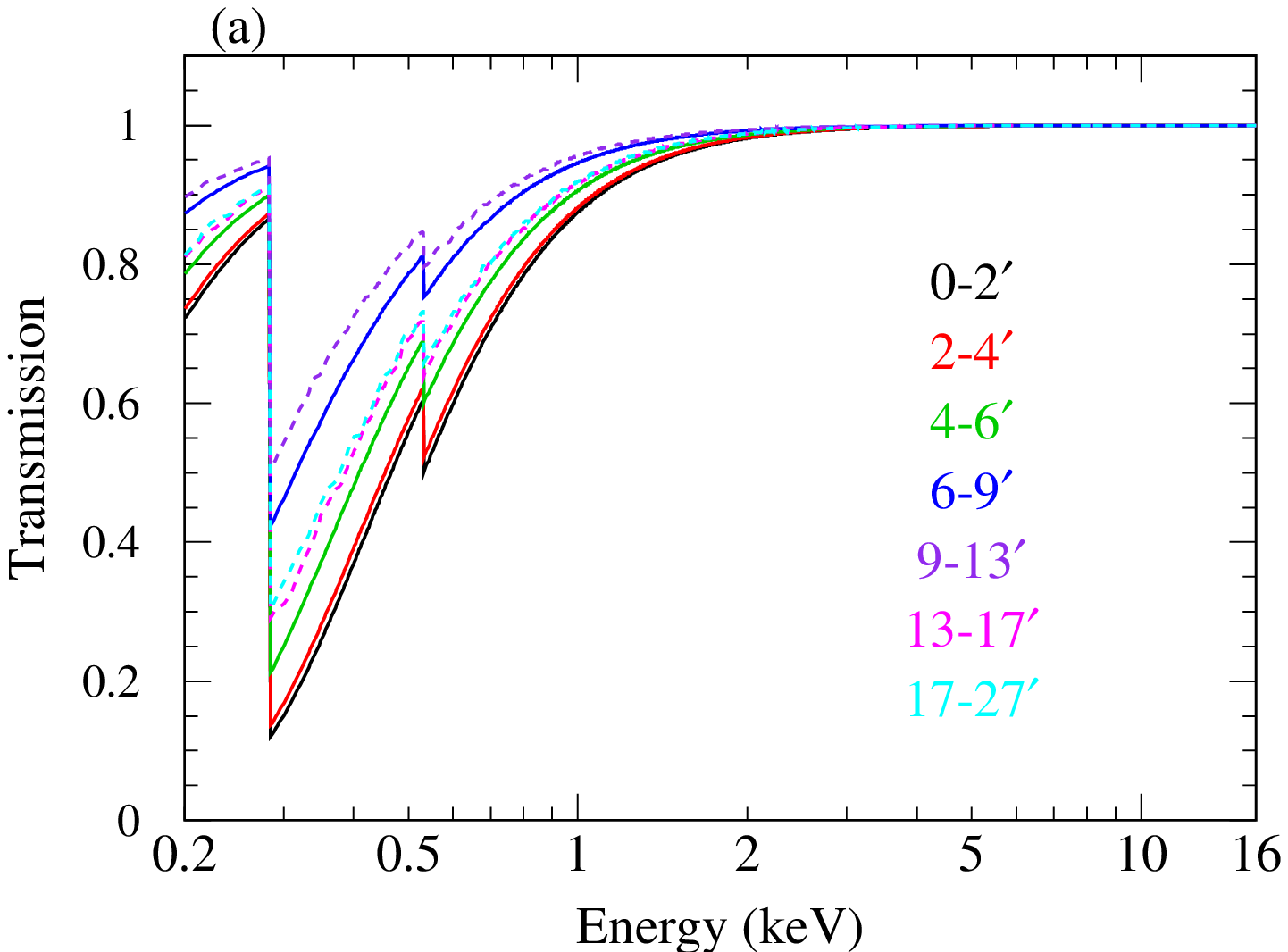}}
\end{minipage}\hfill
\begin{minipage}{0.48\textwidth}
\centerline{\FigureFile(\textwidth,\textwidth){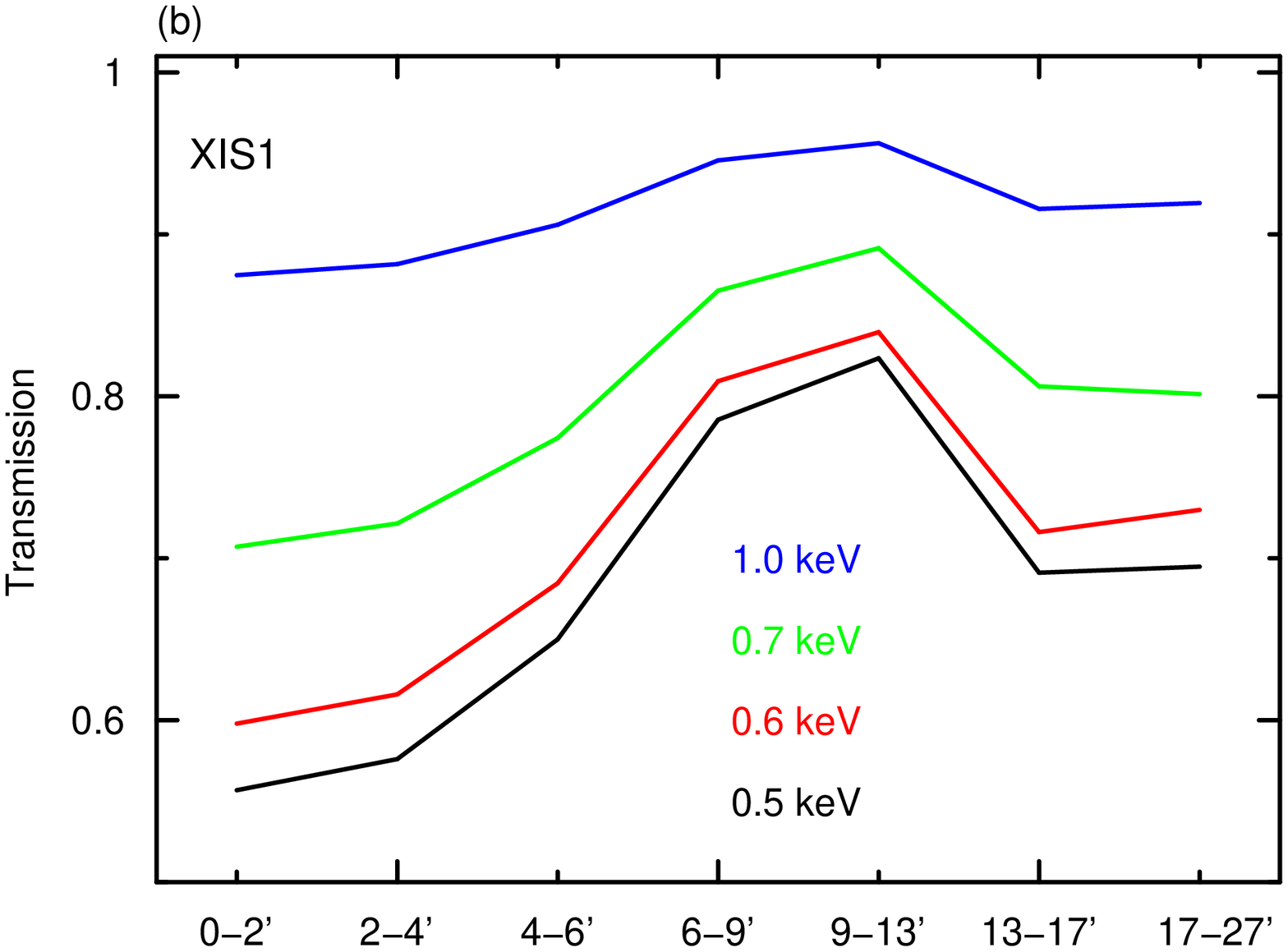}}
\end{minipage}\hfill
\caption{
(a) Estimated transmission of the contaminant on the BI (XIS1) sensor
for each annular region used in the spectral analysis
plotted against the X-ray energy.
Transmissions in the central observations are drawn by solid lines,
and the offset observations are by dashed lines.
These transmissions are calculated by ``xissimarfgen''
with a calibration file of {\tt ae\_xi1\_contami\_20060525.fits},
and written in the {\sc contami\_transmis} column of the ARF response.
(b) The calculated transmission plotted against each annular region
in the energies of 0.5, 0.6, 0.7, and 1.0~keV\@.
}\label{fig:trans}
\end{figure*}

The optical blocking filters (OBF) of the XIS have been gradually
contaminated in time by out-gas from the satellite,
and the degradation of the low energy transmission was
already significant in November 2005.
The thickness of the contaminant is different among sensors,
and is also dependent on the location on the CCD\@.
The estimated column density (C/O=6 in number ratio is assumed)
at the center of the CCD is listed in table~\ref{tab:contami},
and the calculated X-ray transmission for BI (XIS1)
at each annular region is plotted in figure~\ref{fig:trans}.\footnote{
The calibration database file of {\tt ae\_xi{\it N}\_contami\_20060525.fits}
was used for the estimation of the XIS contamination
($N=0,1,2,3$ corresponding to the XIS sensor).}
Though the thickness of the OBF contaminant
is different on each annulus,
this effect is considered in the calculation of
the Ancillary Response File (ARF) by the ``xissimarfgen'' Ftools
task \citep{ishisaki06}.
The energy resolution was also degraded slightly
(FWHM $\sim 150$~eV at 5.9~keV) after the launch,
due to the radiation damage of the CCD\@.

\begin{table}
\caption{Estimated column density of the contaminant for each sensor at 
the center of CCD in unit of 10$^{18}$~cm$^{-2}$.
}\label{tab:contami}
\begin{center}
\begin{tabular}{cllll} 
\hline \hline
\hspace*{8em} &XIS0 & XIS1 & XIS2 &XIS3\\
\hline
Carbon $\dotfill$ & 1.37   & 1.83  & 2.52 & 4.04 \\
Oxygen $\dotfill$ & 0.228  & 0.305 & 0.420 & 0.673 \\
\hline 
\end{tabular}
\end{center}
\end{table}

\subsection{Data Reduction}

We used the version 0.7 processing data \citep{mitsuda06},
and the analysis was performed with HEAsoft version 6.0.6
and XSPEC 11.3.2t. We started the event screening from the
cleaned event file, in which selection of the event grade
and bad CCD column, disposal of non-observational intervals
(during maneuver, data-rate low, South Atlantic Anomaly,
Earth occultation, bright Earth rim to avoid scattered solar X-ray),
and removal of hot and flickering pixels by the ``cleansis'' Ftools,
were already conducted.
The exposure time given in table~\ref{tab:ob} is for the cleaned event file.
We further applied the Good-Time Intervals (GTI) given for excluding 
the telemetry saturation by the XIS team.
The light curve of each sensor in the 0.3--10~keV range
with 16~s time bin was also examined to reject periods of
anomalous event rate greater or less than $\pm 3\sigma$ around the mean.
After the above screenings, remaining exposure of 
the central observation was 40.2~ks and 
that of the offset observation was 52.9 or 54.6 ks (FI or BI)\@.
These exposures are not so different from those in table~\ref{tab:ob},
which represent that the NXB was almost stable during the both observations.
The event screening with the cut-off rigidity (COR)
was not performed in our data.

\subsection{NXB \& CXB Subtraction}\label{sec:background}

In order to subtract the NXB and the extra-galactic
cosmic X-ray background (CXB; see \cite{Brandt2005} for review), 
we used the night earth database of 770~ks exposure
provided by the XIS team for the NXB,
and estimated the CXB component using the ASCA results.
 
The night-earth spectra were extracted from
the same detector region as the A~1060 observation
in order to cancel positional variation of the NXB\@.
Furthermore, we divided the night-earth data referring to COR in the 
ranges of $< 4$~GV, 4--13~GV in 1~GV step, and $> 13$~GV,
because it is known that the intensity and energy spectrum of the NXB
are primarily correlated with COR at the orbital location of the satellite.
The NXB spectra to be subtracted in the spectral fitting were
estimated by adding these COR-sorted night-earth spectra
weighted with exposure times of the A~1060 observation
in the corresponding COR range.

The CXB component was estimated
by the ``fake'' command of XSPEC using uniform-sky ARFs,
which is generated by ``xissimarfgen'' assuming
that the whole sky (practically, $r<20'$) has
the constant intensity and the same X-ray spectrum.
We assumed a power-law spectrum for the CXB
with the values by \citet{kushino02},
$\Gamma=1.4$ and $S_{\rm X} = 5.97\times 10^{-8}$
erg~cm$^{-2}$~s$^{-1}$~sr$^{-1}$ (2--10~keV),
absorbed with the neutral hydrogen column of
$N_{\rm H}= 4.9\times 10^{20}$ cm$^{-2}$ \citep{dickey90}.
The above CXB intensity is taken from table 3 of \citet{kushino02},
for the integrated spectrum with source elimination
brighter than $S_0=2\times 10^{-13}$ erg~cm$^{-2}$~s$^{-1}$ (2--10~keV)
in the GIS filed of view with $\Gamma=1.4$ (fix)
and the nominal NXB level (0\%).
It is confirmed that this gives a reasonable estimate of the CXB
contribution for the XIS in subsection 6.2 of \citet{ishisaki06}
and \citet{fujimoto06}

\begin{figure*}
\begin{minipage}{0.33\textwidth}
\FigureFile(\textwidth,\textwidth){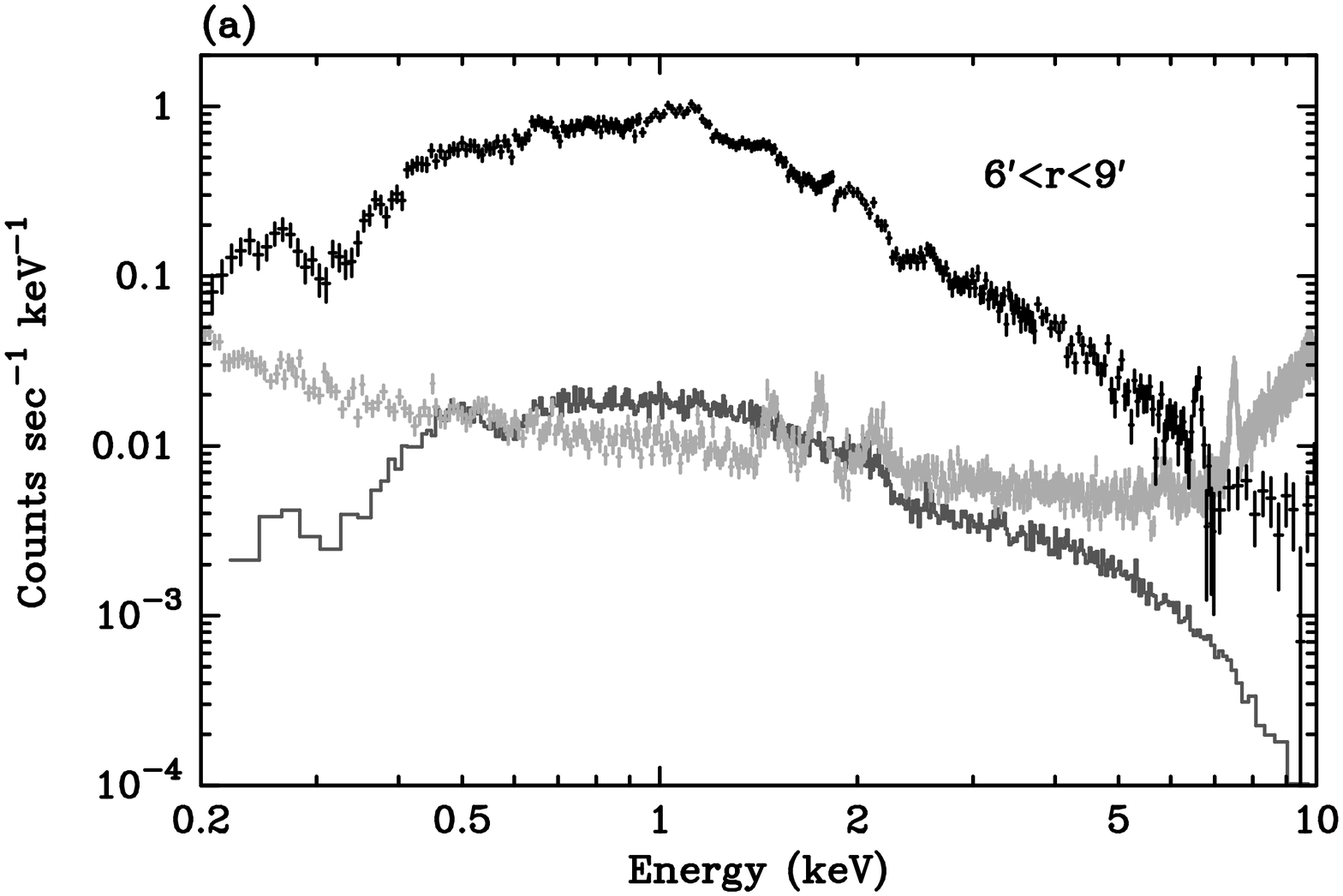}
\end{minipage}\hfill
\begin{minipage}{0.33\textwidth}
\FigureFile(\textwidth,\textwidth){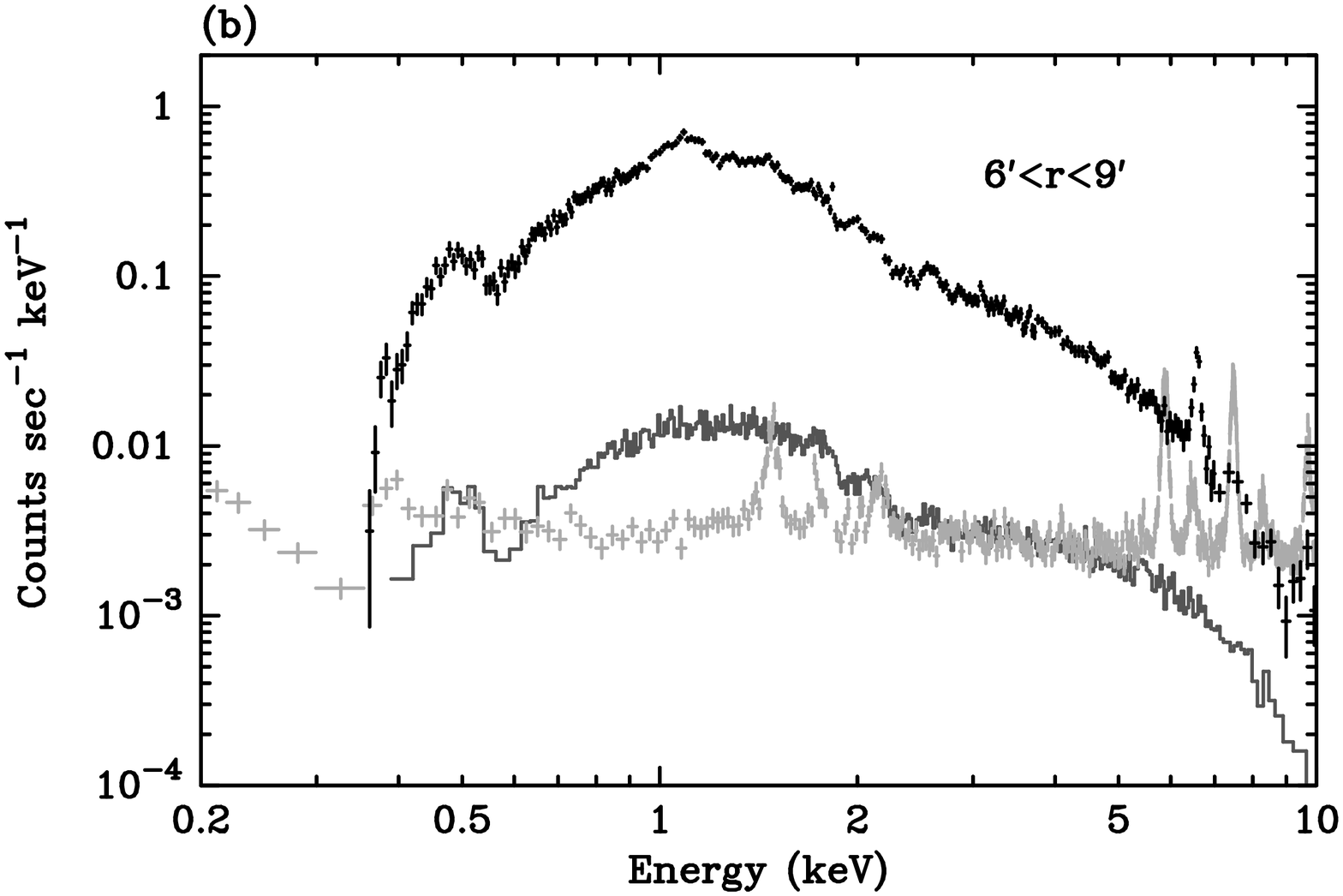}
\end{minipage}\hfill
\begin{minipage}{0.33\textwidth}
\FigureFile(\textwidth,\textwidth){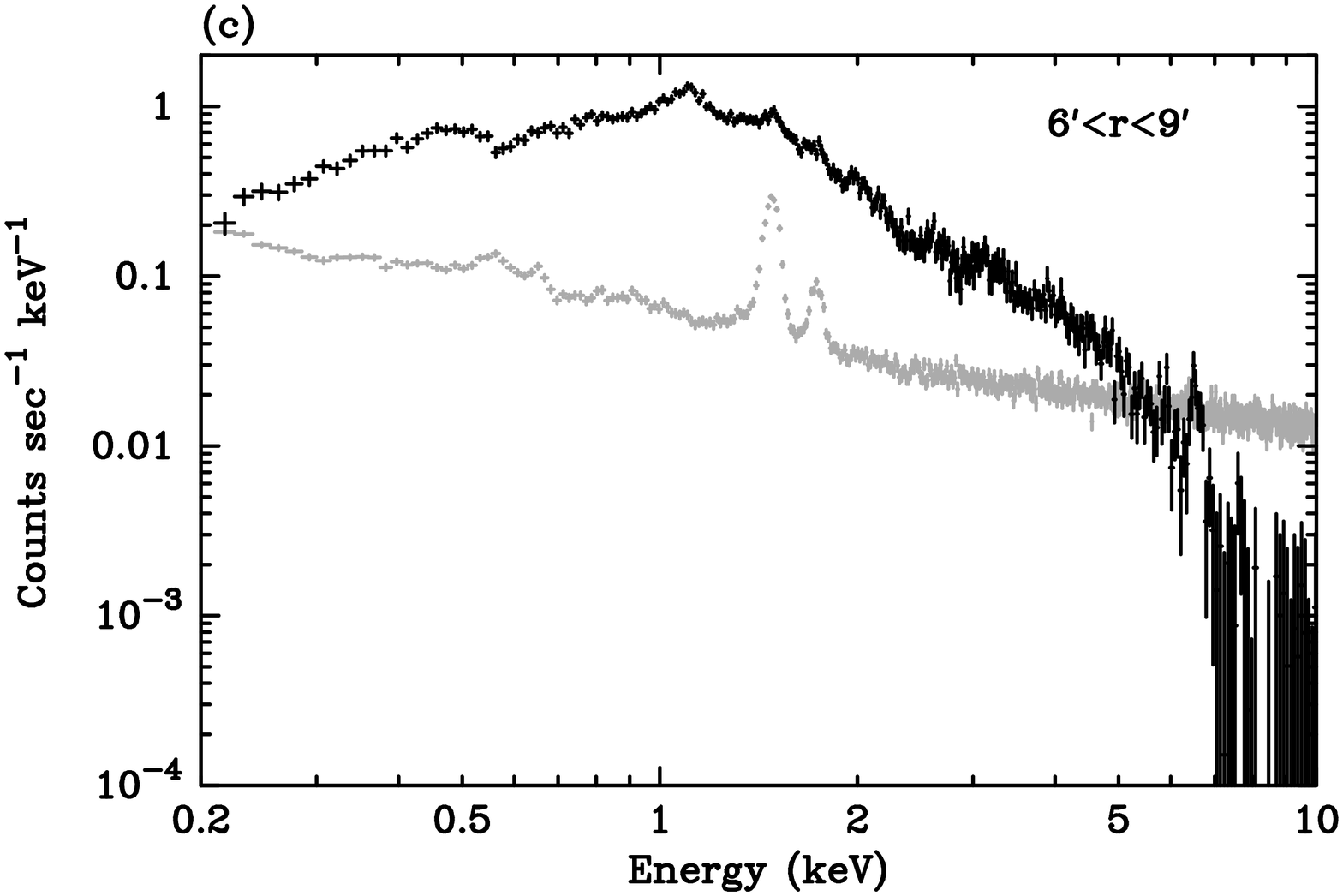}
\end{minipage}
\caption{
(a) The observed spectrum at the annular region of 6--9$'$
with the Suzaku BI (XIS1) sensor
is plotted in black crosses after subtracting the estimated
CXB and NXB components, which are plotted by gray crosses and
a black histogram, respectively.
(b) Same as (a) but for the FI (XIS0+XIS2+XIS3) sensors.
The count rate drop below $\lesssim 0.4$~keV is due to
the event threshold of the FI CCDs \citep{koyama06}.
Although 0.2--10~keV energy range is shown here, only 0.4--7.1~keV
band was utilized for the spectral fitting for both BI and FI\@.
(c) Same as (a) but for the XMM-Newton MOS1 sensor,
and the estimated CXB+NXB component using blank-sky observations
are plotted in gray. The 0.5--8.0~keV energy range was used for
the spectral fitting with XMM-Newton.
}\label{fig:cxbnxb}
\end{figure*}

Figures~\ref{fig:cxbnxb} (a)--(c) show the background level for 
Suzaku BI, FI sensors, and XMM-Newton MOS1,
at the same 6--9$'$ annulus of the cluster.
The observed spectrum after the CXB and NXB subtraction
are compared with the estimated CXB and NXB spectra separately
for Suzaku, while sum of the CXB and NXB are indicated for
the XMM-Newton because the blank-sky data was used as the background.
These figures show the background level of Suzaku XIS is 
$\sim 50$\% lower than
that of XMM-Newton. It is also notable that there is a strong
Al-K$_\alpha$ peak at 1.49~keV for the XMM-Newton background,
which makes the determination of the Mg abundance quite difficult.
The S/N ratio of the Suzaku XIS BI/FI sensors are
by about 1.8/2.0 times higher than that of the XMM-Newton MOS
at 1~keV, and 1.3/1.8 times higher at 4~keV\@.
The particle background is more stable for Suzaku
than XMM-Newton due to its low-earth orbit.

\section{XMM-Newton Observation \& Analysis}

\begin{figure}[t]
\centerline{\FigureFile(0.45\textwidth,8cm){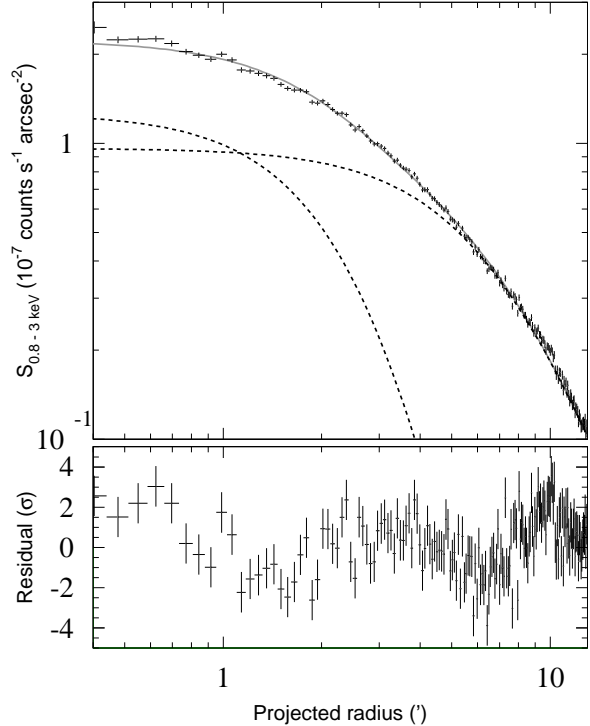}}
\vspace*{-1ex}
\caption{
In the upper panel, a radial profile of the surface brightness 
of A~1060 in the 0.8--3~keV band are plotted for XMM-Newton MOS1+2 ($r<13'$). 
The best-fit double-$\beta$ model is shown by the solid gray line, and 
the two $\beta$-components are indicated by dashed lines.
In the bottom panel, fit residuals are shown in unit of $\sigma$.
}\label{fig:2beta}
\end{figure}

\begin{table}[t]
\caption{The best-fit parameters of figure~\ref{fig:2beta}.
}\label{tab:2beta}
\begin{center}
\begin{tabular}{cllc}
\hline\hline
\hspace*{6em} & \multicolumn{1}{c}{$\beta$} & \multicolumn{1}{c}{$r_{\rm c}$} &$S_{\makebox{\tiny 0.8--3~keV}}\,^*$ \\
\hline
narrower $\dotfill$ & $1.3\pm0.1$& $3.7'\pm 0.3'$ & $1.26\pm 0.14$ \\
wider $^{\dagger}$ $\dotfill$ & 0.69 (fix)&7.3$'$ (fix) & $0.96\pm 0.05$ \\
\hline\\[-1ex]
\multicolumn{4}{l}{\parbox{0.45\textwidth}{\footnotesize
\footnotemark[*]
$S_{\makebox{\tiny 0.8--3~keV}}$ at the center in unit of $10^{-7}$ counts~s$^{-1}$~arcsec$^{-2}$.}}\\
\multicolumn{4}{l}{\parbox{0.45\textwidth}{\footnotesize
\footnotemark[$\dagger$]
$\beta$ and $r_{\rm c}$ is fixed to the values by \citet{hayakawa06}.}}
\end{tabular}
\end{center}
\end{table}

The XMM-Newton observation of A~1060 
was carried out on 2004 June 29 with total exposure of 64~ks.
The same dataset with \citet{hayakawa06} was utilized.
We used only the MOS data, and data reduction and analysis 
were performed with SAS version 6.0 and HEAsoft version 6.0.6.
As for the XMM-Newton data, data reduction and spectral analysis
were based on \citet{sato05} and \citet{hayakawa06}.
\citet{hayakawa06} have reported the temperature and abundance 
profiles with XMM-Newton observation,
however, we reanalyzed the data in order to 
compare with the Suzaku results on the same criterion.
The differences form \citet{hayakawa06} are: 
{\it (1)}\/ the event extraction region,
{\it (2)}\/ using the {\it vapec} thin thermal plasma model
\citep{smith01} instead of {\it vmekal}\/ \citep{mewe85,mewe86}, and
{\it (3)}\/ deriving the element abundances separately.

Utilizing the XMM-Newton image,
we also derived the surface brightness profile of A~1060
needed to generate the Suzaku ARFs,
because the spatial resolution of XMM-Newton is superior to that of Suzaku.
Figure~\ref{fig:2beta} shows a radial profile of A~1060
in the 0.8--3~keV energy range, fitted with a double-$\beta$ model.
The origin of the radial profile is placed at
(RA, Dec) = (\timeform{10h36m43.1s}, \timeform{-27D31'46''}) in J2000.
The best-fit parameters are summarized in table~\ref{tab:2beta},
in which $\beta$ and $r_{\rm c}$ of the wider $\beta$-component
are fixed to the values by \citet{hayakawa06}.

\section{Spectral Analysis and Results}\label{sec:spec}

\subsection{Suzaku XIS Spectra}
\label{sec:icm}

We extracted spectra from seven annular regions
of 0--2$'$, 2--4$'$, 4--6$'$, 6--9$'$, 9--13$'$, 13--17$'$ and 17--27$'$,
centered on (RA, Dec) = (\timeform{10h36m42.8s}, \timeform{-27D31'42''}).
The first four annuli were taken from the central observation,
and the rest of thee annuli were from the offset observation.
Table~\ref{tab:region} lists areas of the extraction regions (arcmin$^2$),
coverage of the whole annulus (\%),
the {\sc source\_ratio\_reg} values (\%; see caption for its definition)
and the observed counts in 0.4--7.1~keV including NXB and CXB
for the BI and FI sensors.

\begin{figure*}
\begin{minipage}{0.48\textwidth}
\FigureFile(\textwidth,\textwidth){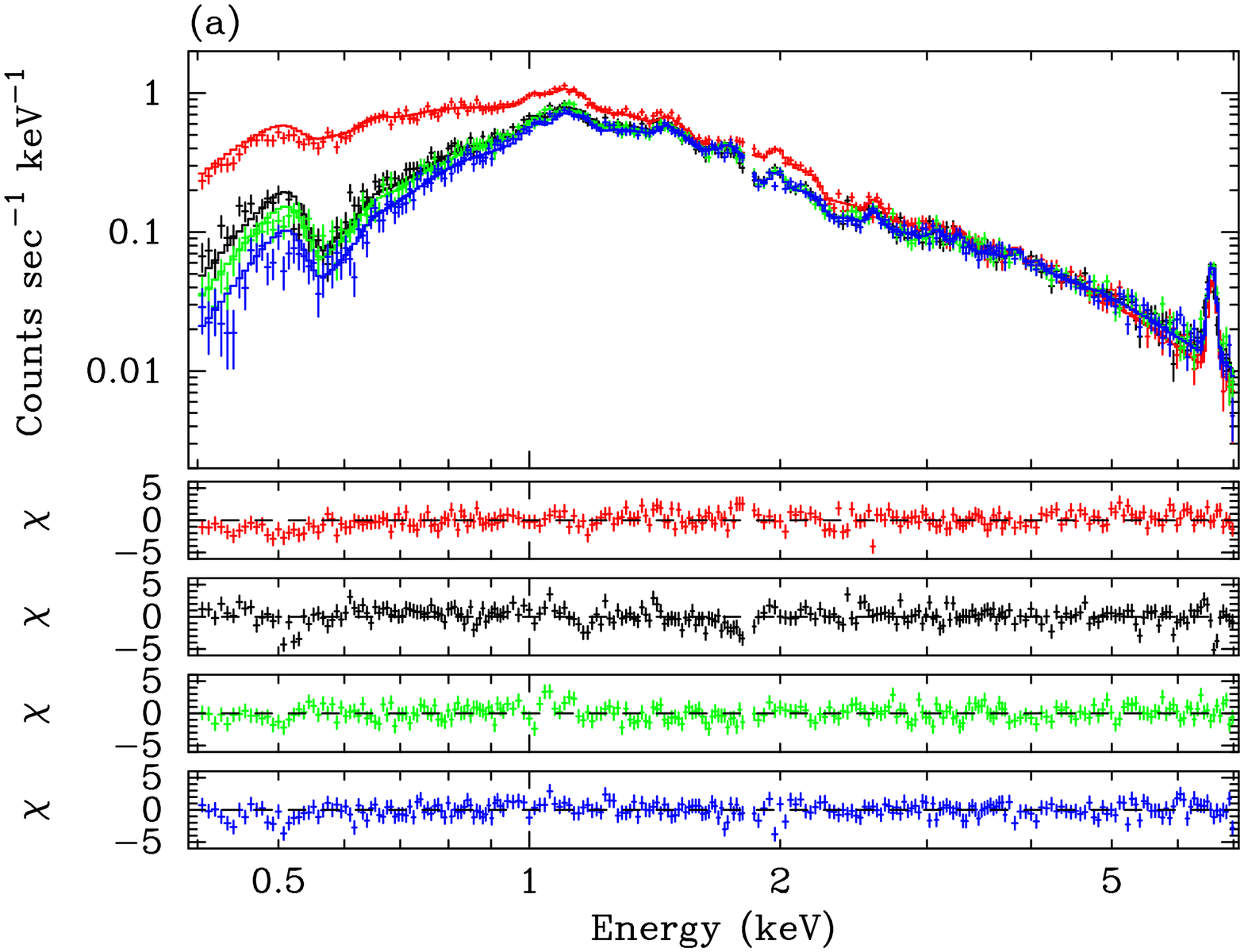}
\end{minipage}\hfill
\begin{minipage}{0.48\textwidth}
\FigureFile(\textwidth,\textwidth){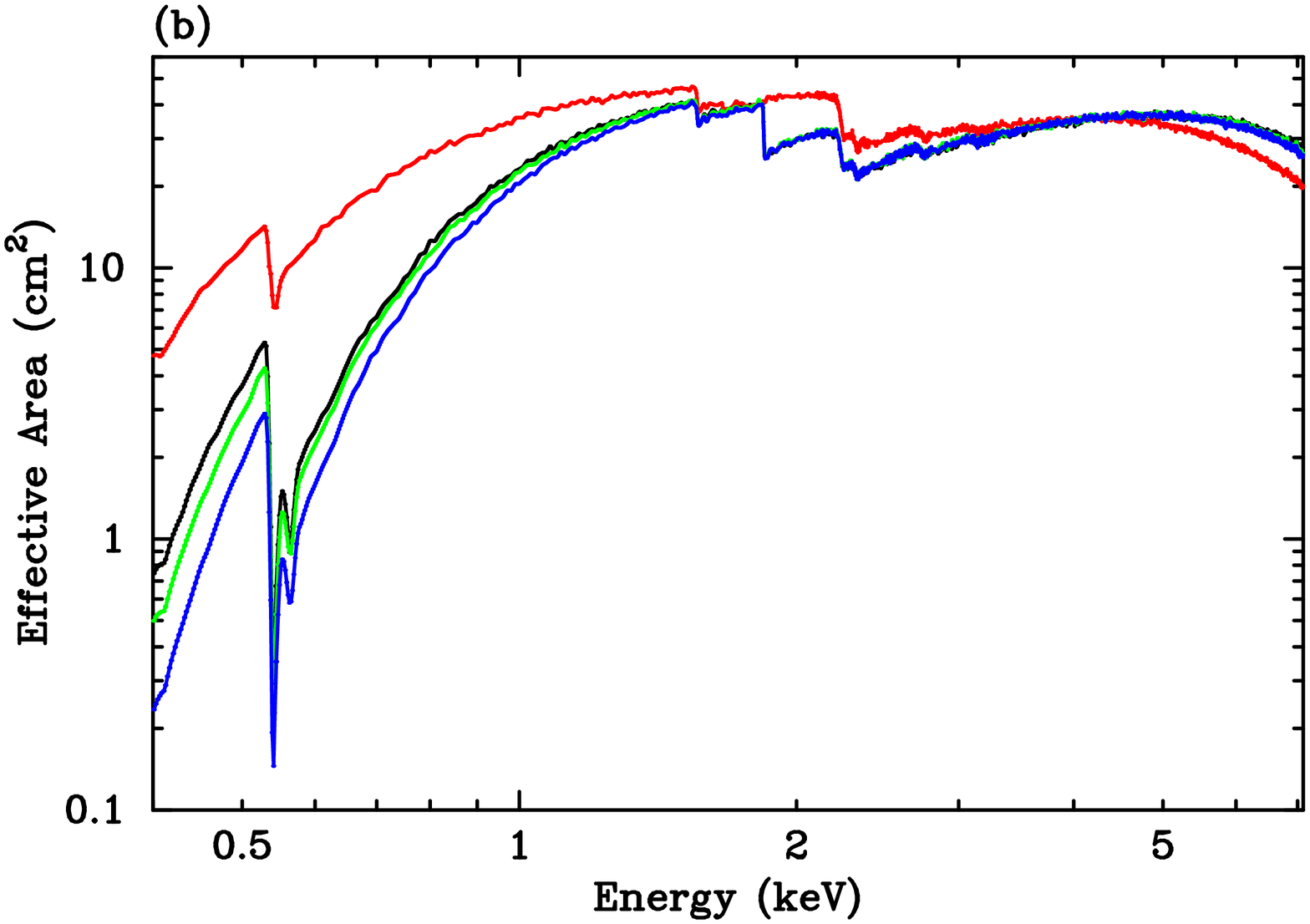}
\end{minipage}
\caption{
(a) The uppermost panel show the observed spectra
after subtracting the estimated CXB and NXB components
at the annular region of 2--4$'$ plotted separately
for the XIS0 (black), XIS1 (red), XIS2 (green),
and XIS3 (blue) sensors. The cross markers denote the observed spectra
and the solid lines show the best-fit model with
${\it apec}_1 + {\it apec}_2 + {\it phabs} \times {\it vapec}$.
The energy range around the Si K-edge (1.825--1.840 keV) is ignored
for the spectral fit.
Lower three panels show the residuals of the fit in unit of $\sigma$
for XIS1, XIS0, XIS2, and XIS3 from upper to lower, respectively.
(b) Plots of the calculated XIS effective area (ARF + RMF)
including the XIS quantum efficiency.
These responses are used in the spectral fit in (a) with the same colors.
The quantum efficiency is much higher for BI (XIS1; red) than FI sensors,
while the OBF contamination is thicker in the order of
XIS3, XIS2, XIS1, and XIS0\@.
\ifnum1=0
(b) Plots of the calculated ARF responses which are used
in the spectral fit in (a) with the same colors.
Because quantum efficiency of the CCD is
considered in the RMF response, difference in the effective area 
below $\sim 1$~keV is due entirely to the OBF contamination.
The quantum efficiency is much higher for BI (XIS1) than FI sensors.
\fi
}\label{fig:xis}
\end{figure*}

Although the thickness of the OBF contamination is different
among sensors as shown in table~\ref{tab:contami} and
figure~\ref{fig:xis}(b), we have confirmed that the four sensors
give quite consistent fit results after incorporating the contamination
effect into the ARFs, as demonstrated in figure~\ref{fig:xis}(a).
We therefore add the three FI spectra (XIS0, XIS2, XIS3), hereafter.

\begin{figure*}
\begin{minipage}{0.33\textwidth}
\FigureFile(\textwidth,\textwidth){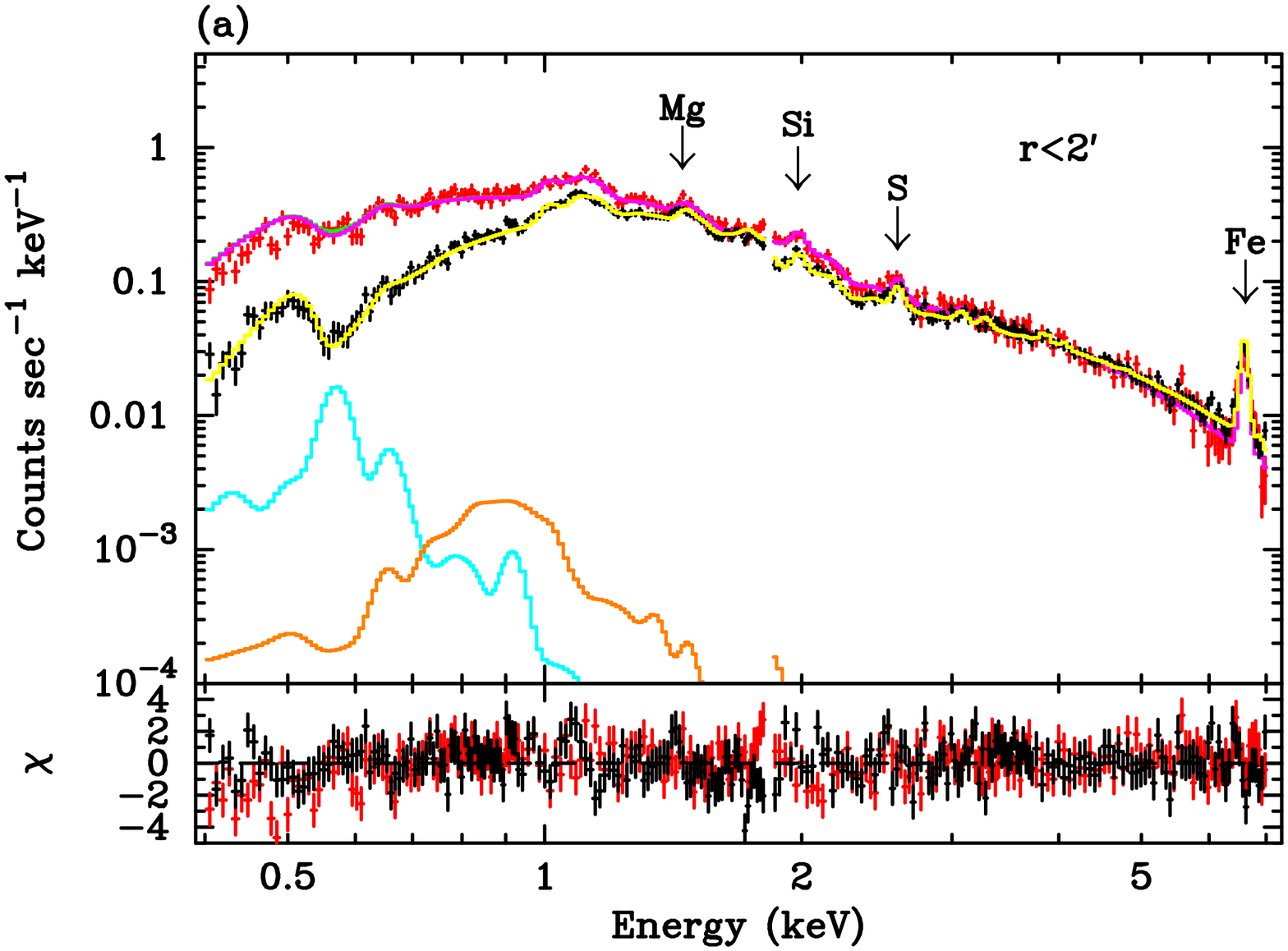}
\end{minipage}\hfill
\begin{minipage}{0.33\textwidth}
\FigureFile(\textwidth,\textwidth){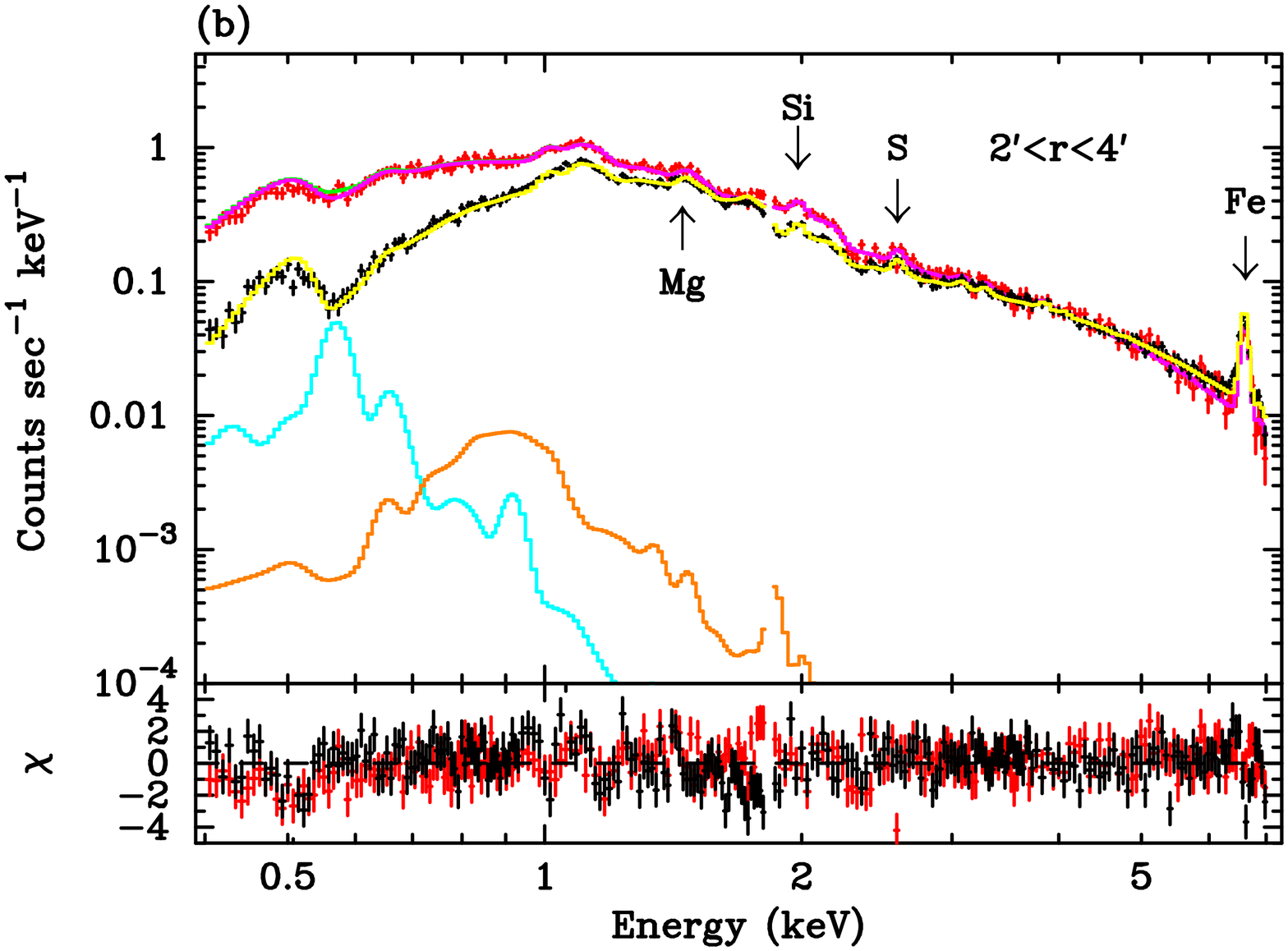}
\end{minipage}\hfill
\begin{minipage}{0.33\textwidth}
\FigureFile(\textwidth,\textwidth){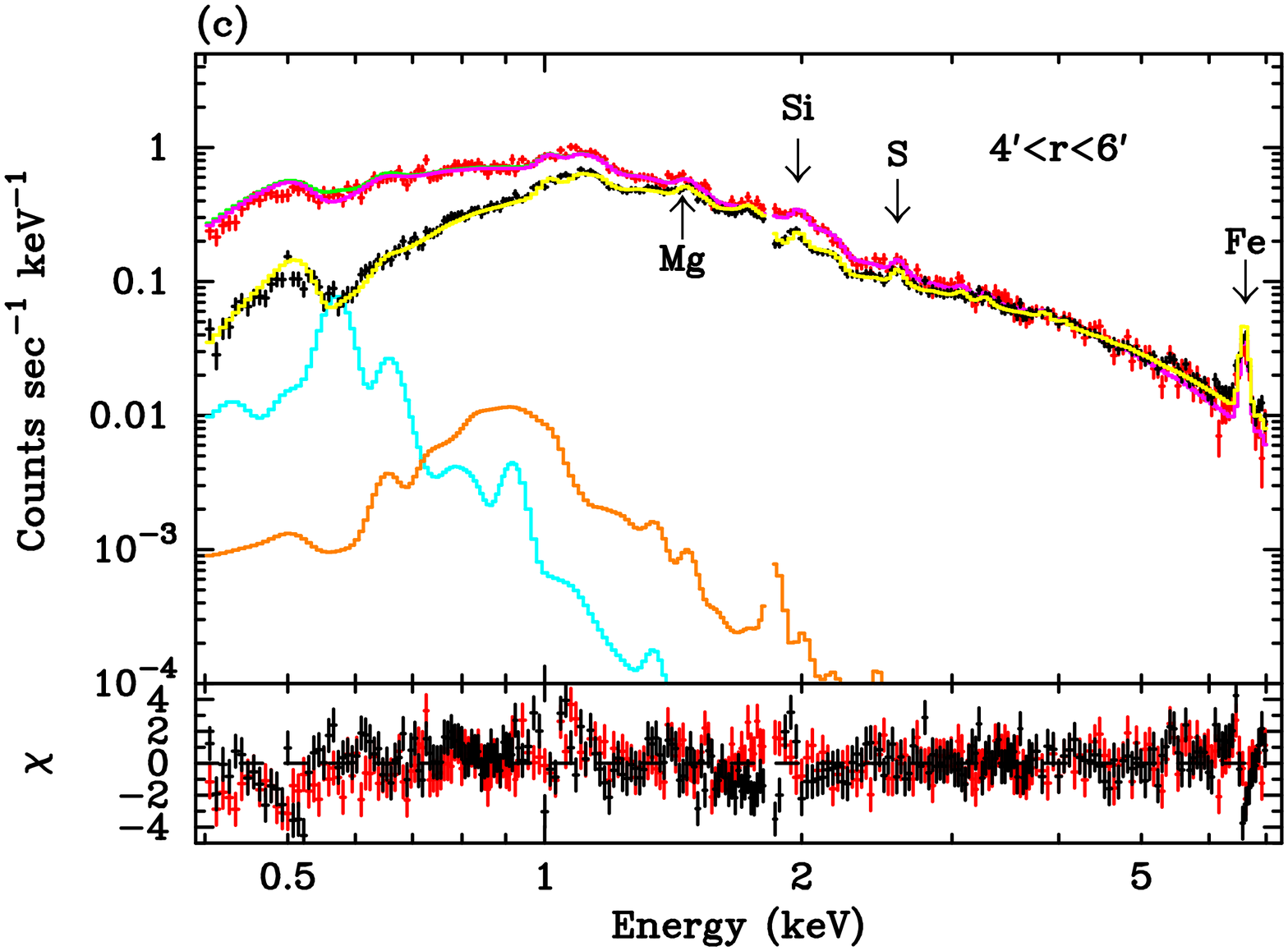}
\end{minipage}

\begin{minipage}{0.33\textwidth}
\FigureFile(\textwidth,\textwidth){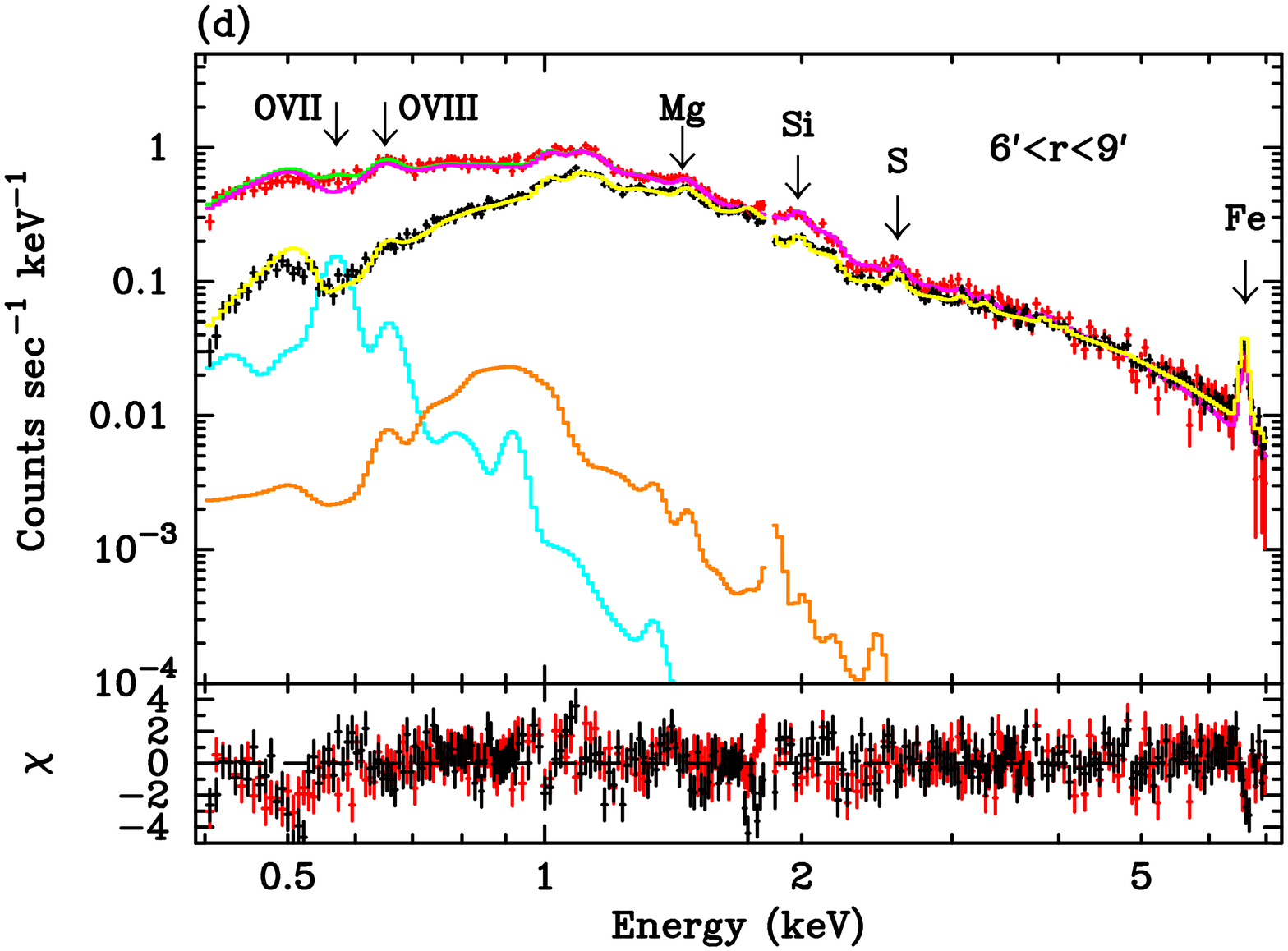}
\end{minipage}\hfill
\begin{minipage}{0.33\textwidth}
\FigureFile(\textwidth,\textwidth){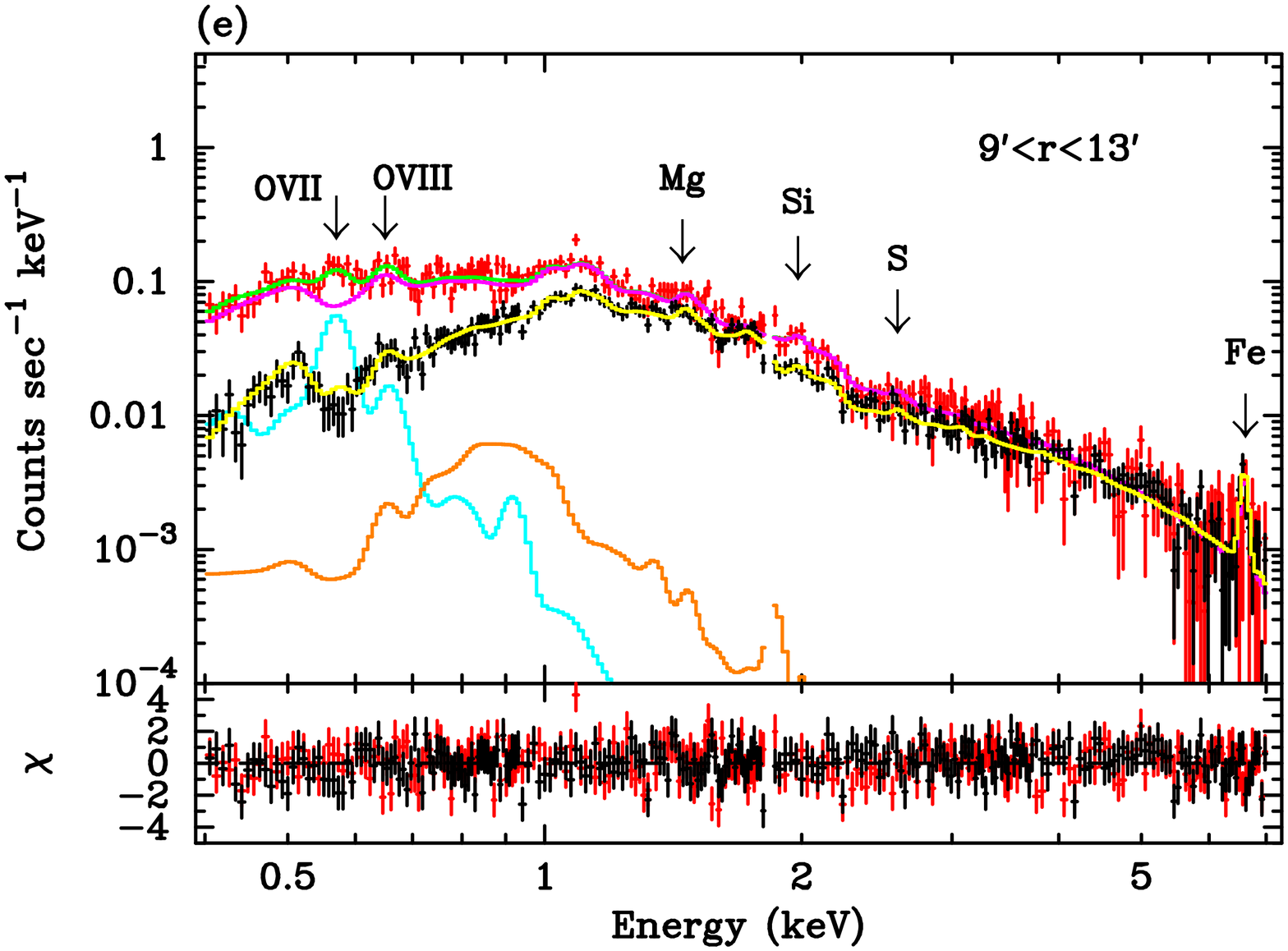}
\end{minipage}\hfill
\begin{minipage}{0.33\textwidth}
\FigureFile(\textwidth,\textwidth){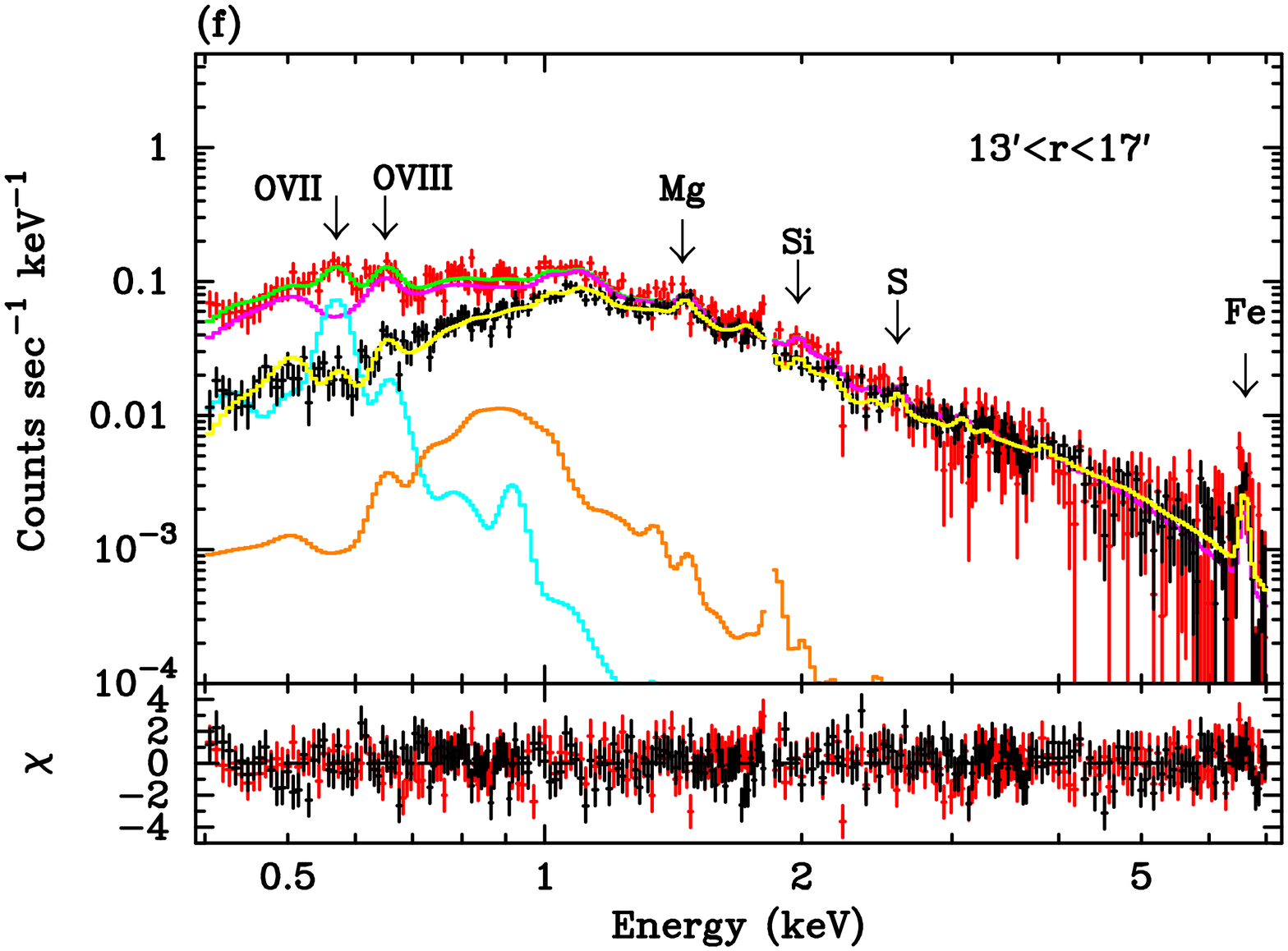}
\end{minipage}

\begin{minipage}{0.33\textwidth}
\FigureFile(\textwidth,\textwidth){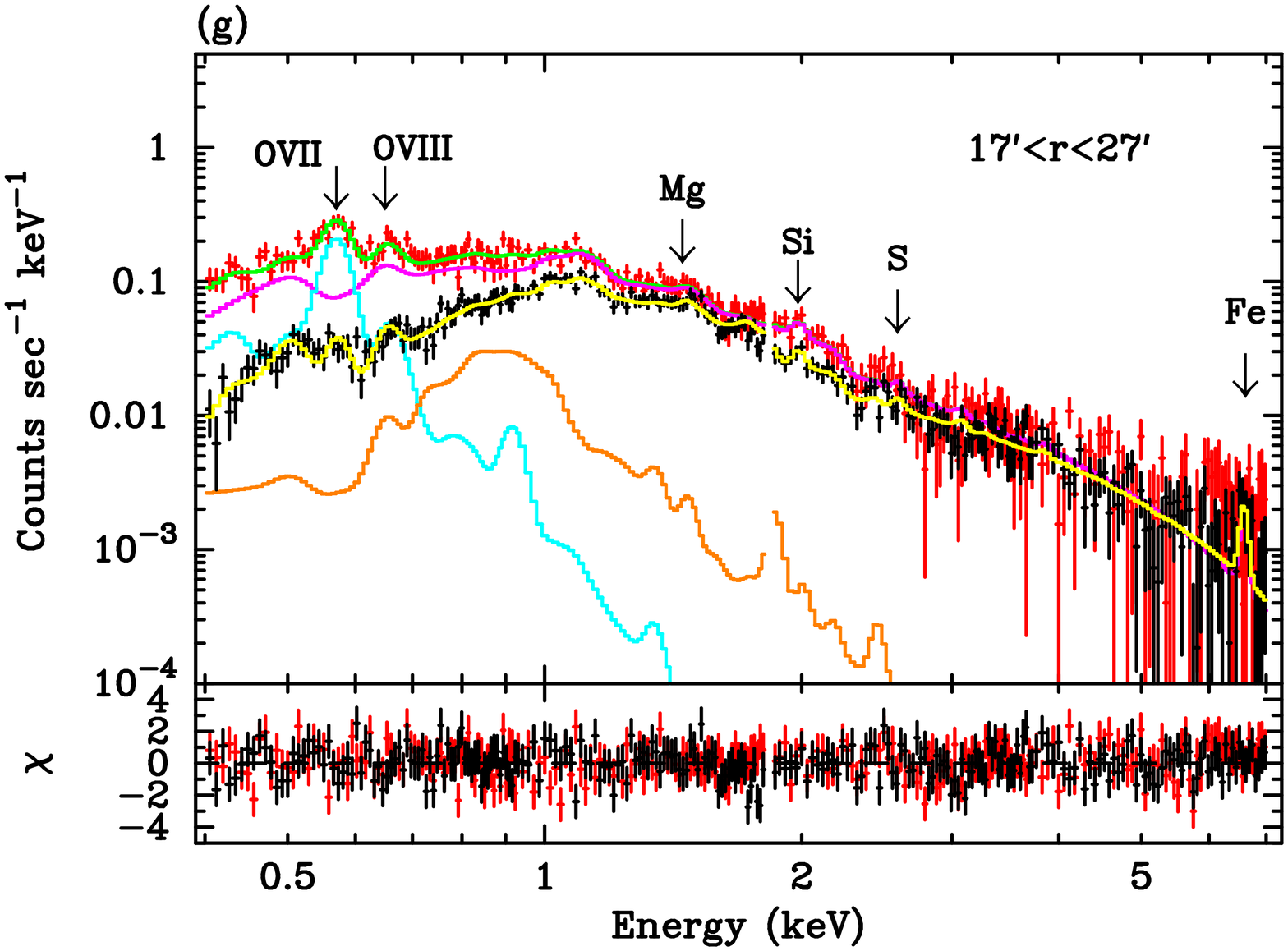}
\end{minipage}\hfill
\begin{minipage}{0.66\textwidth}
\caption{
The upper panels of (a)--(g) show
the background (NXB+CXB) subtracted spectra
at the annular regions which is denoted in the panels,
and they are plotted by red and black crosses for BI and FI, respectively.
The estimated CXB and NXB components are subtracted,
and the plotted data are fitted with the
${\it apec}_1 + {\it apec}_2 + {\it phabs} \times {\it vapec}$ model
drawn by green and yellow lines for the BI and FI spectra.
The ${\it apec}_1$ and ${\it apec}_2$ components
representing the local Galactic emission for the BI spectra
are indicated by cyan and orange lines.
The energy range around the Si K-edge (1.825--1.840 keV) is ignored
for the spectral fit.
The energies of the several prominent lines are
also indicated in the panels.
The lower panels show the fit residuals in unit of $\sigma$.
}\label{fig:spec}
\end{minipage}
\end{figure*}

Each annular spectrum is shown in figure~\ref{fig:spec}.
The ionized Mg, Si, S, Fe lines are clearly seen in each ring.
The O\emissiontype{VII} and O\emissiontype{VIII} lines are prominent
in the outer rings, however, most of the O\emissiontype{VII} emission
is supposed to come from the local Galactic emission,
which will be examined in detail in
subsections \ref{subsec:galactic}--\ref{subsec:direct}.

\begin{table}
\caption{
Area, coverage of whole annulus, {\sc source\_ratio\_ reg}
and observed counts for each annular region.
{\sc source\_ ratio\_reg} represents the
flux ratio in the assumed spatial distribution on the sky
(double-$\beta$ model) inside the accumulation region
to the entire model, and written in the header keyword of
the calculated ARF response by ``xissimarfgen''.
}\label{tab:region}
\begin{center}
\begin{tabular}{lrrrcc} 
\hline \hline
\makebox[3em][l]{Region\,$^\ast$} & \multicolumn{1}{c}{Area\makebox[0in][l]{\,$^\dagger$}} & Coverage\makebox[0in][l]{\,$^\dagger$}\hspace*{-0.5em} & \makebox[4.2em][r]{\sc source\_\makebox[0in][l]{\,$^\ddagger$}}\hspace*{-1em} & \multicolumn{2}{c}{Counts\makebox[0in][l]{\,$^\S$}} \\
& \makebox[2em][c]{(arcmin$^2$)} &      & \makebox[4.2em][r]{\sc ratio\_reg}\hspace*{-1em} & BI     & FI \\
\hline
0--2$'$   & 12.6 &100.0\%       &  7.7\% & 30,549 & 64,606 \\
2--4$'$   & 37.7 &100.0\%       & 13.7\% & 55,186 & \makebox[0in][r]{1}14,220 \\
4--6$'$   & 62.8 &100.0\%       & 13.8\% & 49,259 & \makebox[0in][r]{1}00,080 \\
6--9$'$   &137.0 & 96.9\%       & 17.2\% & 52,719 & \makebox[0in][r]{1}00,750 \\[-1.5ex]
\multicolumn{6}{l}{\hspace*{-0.6em}$\dotfill$\hspace*{-0.6em}} \\
9--13$'$  & 49.1 & 17.8\%       &  2.8\% & 10,696 & 17,329 \\
13--17$'$ & 76.2 & 20.2\%       &  2.2\% & 11,032 & 21,010\\
17--27$'$ &167.3 & 12.1\%       &  2.0\% & 18,055 & 28,834 \\
\hline\\[-1ex]
\multicolumn{6}{l}{\parbox{0.45\textwidth}{\footnotesize
\footnotemark[$\ast$]
The first four annuli are extracted from the central observation,
and others are from the offset observation.}} \\[1.2ex]
\multicolumn{6}{l}{\parbox{0.45\textwidth}{\footnotesize
\footnotemark[$\dagger$]
The largest values among four sensors are presented.}} \\
\multicolumn{6}{l}{\parbox{0.45\textwidth}{\footnotesize
\footnotemark[$\ddagger$]
$\makebox{\sc source\_ratio\_reg}\equiv
\makebox{Coverage}\; \int_{r_{\rm in}}^{r_{\rm out}} S(r)\,r\,dr / 
\int_{0}^{r_{\rm max}} S(r)\,r\,dr$,
where $S(r)$ represents the radial profile
shown in figure~\ref{fig:2beta} and table~\ref{tab:2beta},
and the $r_{\rm max}$ is set to 30$'$.}}\\
\multicolumn{6}{l}{\parbox{0.45\textwidth}{\footnotesize
\footnotemark[$\S$]
Observed counts including NXB and CXB in 0.4--7.1 keV.}}
\end{tabular}
\end{center}
\end{table}

\subsection{Strategy of Spectral Fit}
\label{subsec:strategy}

The basic strategy of the spectral fit is
described in subsection 6.3 of \citet{ishisaki06}.
The observed spectrum is assumed to contain
{\sc(a)} thin thermal plasma emission from the ICM,
{\sc(b)} local Galactic emission, {\sc(c)} CXB, and {\sc(d)} NXB\@.
The estimation of {\sc(c)} and {\sc(d)} is described
in subsection \ref{sec:background}, and both are subtracted
from the observed spectrum.
The spectrum of {\sc(b)} can be represented
by one or two thin-thermal plasma model(s), {\it apec}, with 1~solar abundance,
however, it may vary from field to field by more than an order of magnitude
\citep{kushino02}.
The ICM spectrum {\sc(a)} can be represented by
a variable abundance thin-thermal plasma model, {\it vapec},
whose best-fit parameters are what we want.
An important point is that the spatial distributions
are different between {\sc(a)} and {\sc(b)}.
The former follows the surface brightness of the cluster,
while the latter is supposed to have almost a uniform distribution
in the XIS field of view.

We therefore generated two different ARFs for the spectrum of each annulus,
$A^{\makebox{\small\sc u}}$ and
$A^{\makebox{\small\sc b}}$,
which respectively assume the uniform-sky emission and
$\sim 1^{\circ} \times 1^{\circ}$ size of 
the double-$\beta$ surface brightness profile
obtained with the XMM-Newton data (table~\ref{tab:2beta}).
$A^{\makebox{\small\sc u}}$ was used to evaluate the CXB,
and the surface brightness of the Galactic component {\sc(b)}
at each annulus in combination with the XSPEC ``fakeit'' command,
and $A^{\makebox{\small\sc b}}$ was used for the actual fitting.
We have confirmed that the assumed double-$\beta$ surface brightness profile
is consistent with the observed Suzaku image by about $\pm 15\%$
in figure~5 of \citet{ishisaki06}.
We further constrained that the surface brightness of
the Galactic component {\sc(b)} is nearly constant
among all the extraction annuli,
whereas its spectral shape is determined from the spectral fit
of our Suzaku data. Details will be described in the next subsection.

We adopted the nominal Redistribution Matrix Files (RMF) of
{\tt ae\_xi$N$\_20060213.rmf} for spectral fitting,
although slight degradation in energy resolution is
expected (subsection \ref{subsec:obs}).
The ARFs were generated by ``xissimarfgen'',
and were convolved with the RMFs and added for three FI sensors,
using the ``marfrmf'' and ``addrmf'' tasks in Ftools.
The spectra from BI and FI are fitted
simultaneously in the 0.4--7.1 keV band except for
energy range of anomalous response around the Si K-edge (1.825--1.840~keV)\@.
We ignored below 0.4~keV because the C edge (0.284~keV) seen
in the BI spectra could not be reproduced perfectly in our data.
Energy range above 7.1~keV was also ignored because background
Ni line ($\sim 7.5$~keV) left artificial structures after the NXB
subtraction at large radii. It is also known that the XIS response
in $E\gtrsim 8$~keV are not fully understood at the present stage.
In the simultaneous fit of BI and FI,
only the normalization are allowed to be different between them,
although we found that the derived normalizations are
quite consistent between the two.

\subsection{Estimation of Galactic Component}
\label{subsec:galactic}

\begin{table*}
\caption{
The best-fit parameters of the {\it apec} component(s)
for the simultaneous fit of the spectra in 13--17$'$ and 17--27$'$ annuli.
}\label{tab:fit_gal}
\begin{center}
\begin{tabular}{lccccc}
\hline\hline
\makebox[19em][l]{Fit model} & ${\it Norm}_1\,^\ast$ & $kT_1$ & ${\it Norm}_2\,^\ast$ & $kT_2$ & $\chi^2$/dof\\
 & & (keV) & & (keV) \\
\hline
(a) ${\it apec}_1 + {\it phabs}\times {\it vapec}$ $\dotfill$ &
        $1.20\pm 0.07$ & $0.179^{+0.006}_{-0.012}$ & --- & --- & 1128/994\\
(b) ${\it apec}_1 + {\it apec}_2 + {\it phabs}\times {\it vapec}$ $\dotfill$ &
        $1.66\pm 0.35$ & $0.143^{+0.014}_{-0.015}$ & $0.21\pm 0.05$ & $0.737^{+0.063}_{-0.088}$ &  1082/992\\
\hline\\[-1ex]

\multicolumn{6}{l}{\parbox{0.94\textwidth}{\footnotesize 
\footnotemark[$*$] Normalization of the {\it apec} component
divided by the solid angle,
$\Omega^{\makebox{\tiny\sc u}} = \pi\times (20')^2$,
assumed in the uniform-sky ARF calculation
(20$'$ radius from the optical axis of each XIS sensor),
${\it Norm} = \int n_{\rm e} n_{\rm H} dV \,/\,
(4\pi\, (1+z)^2 D_{\rm A}^{\,2}) \,/\, \Omega^{\makebox{\tiny\sc u}}$
$\times 10^{-20}$ cm$^{-5}$~arcmin$^{-2}$,
where $D_{\rm A}$ is the angular distance to the source.}}
\end{tabular}
\end{center}
\end{table*}

\begin{figure}[t]
\centerline{
\FigureFile(0.48\textwidth,0.48\textwidth){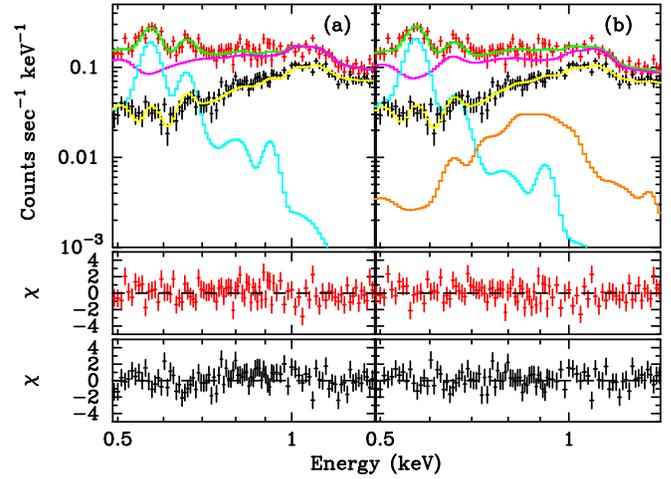}}
\caption{
(a) A magnification of 0.49--1.4~keV range of the BI (red) and
FI (black) spectra in the 17--27$'$ annulus. They are fitted with the
${\it apec}_1 + {\it phabs} \times {\it vapec}$ model,
and the best-fit model are drawn by green and yellow lines.
The estimated CXB and NXB components were subtracted,
and a simultaneously fit with the 13--17$'$ annulus
was conducted in the entire energy range of 0.4--7.1~keV\@.
The ${\it apec}_1$ or {\it vapec} component for BI is
indicated by a cyan or magenta line, respectively.
Lower two panels show the fit residuals for BI and FI in unit of $\sigma$.
The orange dotted lines in figure~\ref{fig:result} correspond to
this modeling of the Galactic component.
(b) Same as (a) but fitted with the
${\it apec}_1 + {\it apec}_2 + {\it phabs} \times {\it vapec}$ model,
and the ${\it apec}_2$ component for BI is indicated by an orange line.
We adopt this model in the spectral fit.
See subsection \ref{subsec:galactic} for details.
}\label{fig:fe-l}
\end{figure}

\begin{figure}[t]
\centerline{
\FigureFile(0.45\textwidth,0.45\textwidth){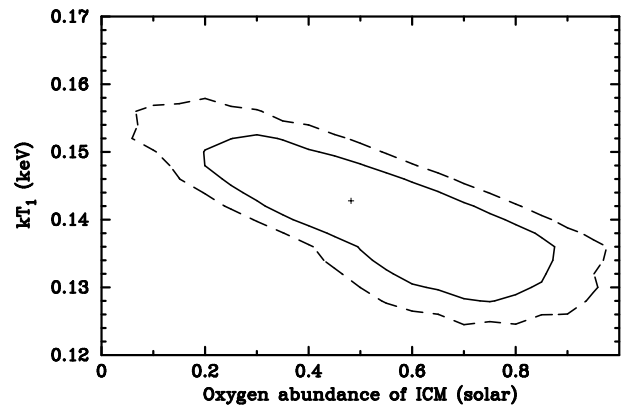}}
\caption{
A plot of confidence contour between $kT_1$
(temperature of the cooler part of the two {\it apec} components)
and the O abundance of {\it vapec} for the 17--27$'$ annulus,
in the simultaneous fitting of 13--17$'$ and 17--27$'$ annuli with the
${\it apec}_1 + {\it apec}_2 + {\it phabs} \times {\it vapec}$ model.
The cross denotes the best-fit location,
and the two contours represent 1$\sigma$ and 90\% confidence
ranges, from inner to outer, respectively.
}\label{fig:cont}
\end{figure}

We found that the estimation of the Galactic component
significantly affect the determination of the oxygen abundance,
because both the ICM and the Galactic component contribute to
the O\emissiontype{VIII} emission line (0.653~keV),
and the XIS cannot resolve them by redshift
due to the limited energy resolution.
On the other hand, most of the O\emissiontype{VII} emission is supposed
to originate in the Galactic component, because the ICM temperature
$kT\gtrsim 2$~keV $\simeq 2.3\times 10^7$~K is too high to emit
the O\emissiontype{VII} lines (0.561, 0.568, 0.574~keV)\@.
See, e.g., figure~3 of \citet{yoshikawa03}
for the oxygen line emissivity.

It is therefore important to estimate the Galactic component
precisely, which is possible using the offset observation of
A~1060 with Suzaku. In order to determine the surface
brightness and the spectral shape of the Galactic component,
we performed the simultaneous fit of the 13--17$'$ and 17--27$'$ annuli.
The Galactic component is prominent in these annuli as
shown in figure~\ref{fig:spec}(f) and (g),
however the ICM component is still dominant almost all the energy range
except for the O\emissiontype{VII} line.
We made the simultaneous fit in the whole 0.4--7.1~keV range
(except 1.825--1.840~keV),
assuming one or two {\it apec} models for the Galactic component,
and the fit results are presented in table~\ref{tab:fit_gal}
and figure~\ref{fig:fe-l}. The resultant normalization of
the {\it apec} model in table~\ref{tab:fit_gal} is scaled
so that it gives the surface brightness in the unit solid angle of
arcmin$^2$.

It may appear that the difference between these two models are not
large in figure~\ref{fig:fe-l}, however, it is notable that
the derived O\emissiontype{VIII} line intensity for the Galactic component is
by about twice larger for (a).
We examined the improvement of the $\chi^2$ ($\Delta\chi^2 = 46$)
with the $F$-test, and adding the ${\it apec}_2$ component was
justified with a large significance (false probability $\sim 10^{-9}$).
We further allowed the normalization of the ${\it apec}_2$
to become free between the two annuli.
The derived surface brightness ratio of
the 13--17$'$ annulus to 17--27$'$ was $0.9\pm 0.2$,
hence it was consistent to be constant.
We also confirmed that the 9--13$'$ annulus gave the same result.
We therefore concluded that the two {\it apec} models are required
to account for the Galactic component. Hereafter, the
${\it apec}_1 + {\it apec}_2 + {\it phabs}\times {\it vapec}$ model
are utilized for the spectral fitting, otherwise stated.

To demonstrate how sensitive the O abundance of the ICM is
to the assumed Galactic component model,
we present a confidence contour between $kT_1$ (keV) of the ${\it apec}_1$
component and the O abundance (solar) of ${\it vapec}$
for the outermost annulus (17--27$'$) in figure~\ref{fig:cont}.
There appears to exist a negative correlation between the two parameters,
because higher temperature of the Galactic component produces
more O\emissiontype{VIII} emission line relative to O\emissiontype{VII},
which contribute to reduce the O\emissiontype{VIII} line from
the ICM ({\it vapec} component).
Influences on the derived temperature and abundance
by the modeling of the Galactic component
will be tested in subsection \ref{subsec:radial}, too.

In order to take into account both existence of
the Galactic component itself and propagation of its statistical error,
we simultaneously fitted each annulus with the outermost annulus of 17--27$'$.
As mentioned in the previous subsection,
the normalization of the ${\it apec}_1 + {\it apec}_2$ component
are constrained to give the same surface brightness between the two annuli.
The temperatures of the two {\it apec} models were also common
between the two, however their values (two normalizations and two temperatures)
themselves are left free. It is confirmed that the derived normalizations,
${\it Norm}_1$ and ${\it Norm}_2$, and temperatures, $kT_1$ and $kT_2$,
are consistent with the values in table~\ref{tab:fit_gal}(b)
within the quoted errors.

\subsection{Radial Temperature \& Abundance Profiles}
\label{subsec:radial}

\begin{table*}
\caption{
Result of the spectral fit in the central region of 0--6$'$, with the 
${\it apec}_1 + {\it apec}_2 + {\it phabs} \times {\it vapec}$ model.
Only the parameters for the ${\it vapec}$ component are presented.
The abundance of each element (except for He, C, and N, which are
fixed to 1~solar with the assumed abundance ratio of {\it angr})
is allowed to be free in the spectral fit.
Errors are 90\% confidence range of statistical errors,
and do not include systematic errors.
}\label{tab:fit_center}
\begin{center}
\begin{tabular}{cccccccc}
\hline\hline\\[-2ex]
\makebox[11em][l]{Region}&$kT$&O&Ne&Mg&Al&Si & $\chi^2$/dof \\
&(keV)&(solar)&(solar)&(solar)&(solar)&(solar)& \\[0.5ex]
\hline\\[-2.0ex]
0--6$'$ $\dotfill$ &3.54$^{+0.03}_{-0.03}$&0.31$^{+0.08}_{-0.08}$&1.21$^{+0.10}_{-0.10}$&0.70$^{+0.10}_{-0.09}$&0.00$^{+0.13}_{-0.00}$ & 0.57$^{+0.05}_{-0.05}$&1871/986 \\[0.5ex]
\cline{2-8}\\[-2.0ex]
& {\it Norm} $^{*}$ &S&Ar&Ca&Fe&Ni& \\
&&(solar)&(solar)&(solar)&(solar)&(solar)&\\[0.5ex]
\cline{2-8}\\[-2.0ex]
&$302\pm 3$&0.57$^{+0.07}_{-0.07}$&0.53$^{+0.16}_{-0.16}$&0.44$^{+0.17}_{-0.17}$&0.43$^{+0.01}_{-0.01}$&0.83$^{+0.23}_{-0.23}$&\\[0.5ex]
\hline\\[-1ex]
\multicolumn{8}{l}{\parbox{0.94\textwidth}{\footnotesize
\footnotemark[$*$] Normalization of the {\it vapec} component
scaled with a factor of {\sc source\_ratio\_reg} / {\sc area}
in table~\ref{tab:region},\\
${\it Norm}=\frac{\makebox{\sc source\_ratio\_reg}}{\makebox{\sc area}}
\int n_{\rm e} n_{\rm H} dV \,/\, (4\pi\, (1+z)^2 D_{\rm A}^{\,2})$
$\times 10^{-20}$~cm$^{-5}$~arcmin$^{-2}$,
where $D_{\rm A}$ is the angular distance to the source.}}
\end{tabular}
\end{center}
\end{table*}

\begin{table*}
\caption{
Results of the spectral fits in the central region of 0--6$'$
by changing the amount of the OBF contaminant.
}\label{tab:contami_cen}
\begin{center}
\begin{tabular}{lcccccccccc}
\hline\hline\\[-2ex]
\makebox[7em][l]{Contaminant}     & O    & Ne   & Mg   & Al   & Si   & S    & Ar   &Ca    & Fe   & Ni \\
&(solar)&(solar)&(solar)&(solar)&(solar)&(solar)&(solar)&(solar)&(solar)&(solar)\\
\hline
nominal $\dotfill$\hspace*{-1em}  & 0.31 & 1.21 & 0.70 & 0.00 & 0.57 & 0.57 & 0.53 & 0.44 & 0.43 & 0.83 \\
$+20$\% $\dotfill$\hspace*{-1em}  & 0.43 & 1.05 & 0.48 & 0.00 & 0.47 & 0.50 & 0.45 & 0.46 & 0.41 & 0.41 \\
free $^{*}\dotfill$\hspace*{-1em} & 0.38 & 1.12 & 0.57 & 0.00 & 0.51 & 0.53 & 0.47 & 0.43 & 0.42 & 0.54 \\
\hline\hline\\[-2ex]
Contaminant  &  $kT$ & $Norm$\makebox[0in][l]{$\,^{\dagger}$} & $kT_1$  & $Norm_1$\makebox[0in][l]{\,$^{\ddagger}$} & $kT_2$ & $Norm_2$\makebox[0in][l]{\,$^{\ddagger}$} &  \multicolumn{2}{c}{$N_{\rm \,C\,O_{1/6}}$\makebox[0in][l]{\,$^\S$}} & \multicolumn{2}{l}{\ \ $\chi^2$/dof} \\
 & (keV) &   & (keV) &  & (keV) &  & BI & FI & \\
\hline
nominal $\dotfill$\hspace*{-1em}  & 3.54 & 302 & 0.159  & 1.00 & 0.752  & 1.00  & 1.74 & 2.32 & \multicolumn{2}{l}{1878/986}\\
$+20$\% $\dotfill$\hspace*{-1em}  & 3.36 & 298 & 0.148  & 1.40 & 0.721  & 1.09  & 2.11 & 2.72 & \multicolumn{2}{l}{1746/986}\\
free $^{*}\dotfill$\hspace*{-1em} & 3.50 & 301 & 0.159  & 0.96 & 0.765  & 0.95  & 2.18 & 2.38 & \multicolumn{2}{l}{1681/982}\\
\hline\\[-1ex]
\multicolumn{11}{l}{\parbox{0.94\textwidth}{\footnotesize
\footnotemark[$*$] We fitted the amount of the OBF contaminant as a free 
parameter with the C/O number ratio fixed to 6.}}\\
\multicolumn{11}{l}{\parbox{0.94\textwidth}{\footnotesize
\footnotemark[$\dagger$] Normalization of the {\it vapec} model, 
calculated in the same way with table~\ref{tab:fit_center}}.}\\
\multicolumn{11}{l}{\parbox{0.94\textwidth}{\footnotesize
\footnotemark[$\ddagger$] Ratio to the normalization of
the ``nominal'' ARF\@.}}\\
\multicolumn{11}{l}{\parbox{0.94\textwidth}{\footnotesize
\footnotemark[$\S$] Column density of the OBF contaminant 
with chemical composition of C\,O$_{1/6}$ for BI and FI
in unit of $10^{18}$ cm$^{-2}$.}}\\
\end{tabular}
\end{center}
\end{table*}

\begin{table*}
\caption{
Summary of the best-fit parameters of the {\it vapec} component
for each annular region with the
${\it apec}_1 + {\it apec}_2 + {\it phabs} \times {\it vapec}$ model.
Each annulus is simultaneous fitted with the the outermost 17--27$'$ annulus.
Errors are 90\% confidence range of statistical errors,
and do not include systematic errors.
The solar abundance ratio of {\it angr} is assumed.
These results are plotted in figure~\ref{fig:result}.
The O abundance in the outer two regions becomes lower when
the Galactic component is expressed by a single {\it apec} model,
which is drawn by orange dotted lines in figure~\ref{fig:result}.
The $kT$, O, and Mg columns at $r<6'$ are slightly different
when the OBF contaminant is increased by ``+20\%'',
which is drawn by black dotted lines in figure~\ref{fig:result}.
The Ne abundance is probably not reliable because the Suzaku XIS cannot
resolve the ionized Ne lines from the Fe-L line complex.
}\label{tab:fit_icm}
\centerline{
\begin{tabular}{lrccccccccc}
\hline\hline\\[-2ex]
\makebox[2em][l]{Region} & {\it Norm}\makebox[0in][l]{\,$^\ast$} & $kT$&O&Ne&Mg&Si&\makebox[0in][c]{S, Ar, Ca}&Fe&Ni & $\chi^2$/dof \\
& &(keV)&(solar)&(solar)&(solar)&(solar)&(solar)&(solar)&(solar) & \\
\hline\\[-2ex]
0--2$'$    &  $605\pm1\makebox[0in][l]{2}$  & 3.34$^{+0.04}_{-0.04}$ & 0.38$^{+0.16}_{-0.15}$ & 1.20$^{+0.20}_{-0.20}$ & 0.76$^{+0.19}_{-0.19}$ & 0.69$^{+0.11}_{-0.11}$ & 0.82$^{+0.12}_{-0.12}$ & 0.49$^{+0.03}_{-0.03}$ & 1.13$^{+0.46}_{-0.45}$  &  1240/992\\\\[-2ex]
2--4$'$    &  $352\pm5$  & 3.39$^{+0.03}_{-0.03}$ & 0.28$^{+0.11}_{-0.12}$ & 1.11$^{+0.16}_{-0.15}$ & 0.65$^{+0.14}_{-0.14}$ & 0.52$^{+0.08}_{-0.08}$ & 0.52$^{+0.08}_{-0.08}$ & 0.42$^{+0.02}_{-0.02}$ & 0.90$^{+0.34}_{-0.34}$  & 1290/992\\\\[-2ex]
4--6$'$    &  $198\pm3$  & 3.42$^{+0.04}_{-0.04}$ & 0.29$^{+0.13}_{-0.13}$ & 1.24$^{+0.18}_{-0.16}$ & 0.67$^{+0.15}_{-0.15}$ & 0.58$^{+0.09}_{-0.09}$ & 0.53$^{+0.09}_{-0.10}$ & 0.41$^{+0.02}_{-0.02}$ & 0.87$^{+0.38}_{-0.37}$  & 1345/992\\\\[-2ex]
6--9$'$    &  $109\pm2$  & 3.25$^{+0.04}_{-0.04}$ & 0.58$^{+0.14}_{-0.14}$ & 1.17$^{+0.16}_{-0.15}$ & 0.59$^{+0.15}_{-0.15}$ & 0.50$^{+0.08}_{-0.08}$ & 0.55$^{+0.09}_{-0.09}$ & 0.40$^{+0.02}_{-0.02}$ & 1.17$^{+0.36}_{-0.35}$  & 1329/992\\\\[-2ex]
\makebox[0in][l]{9--13$'$}   &  $ 44\pm2$  & 2.87$^{+0.10}_{-0.10}$ & 0.65$^{+0.34}_{-0.29}$ & 0.89$^{+0.35}_{-0.33}$ & 0.75$^{+0.33}_{-0.32}$ & 0.27$^{+0.17}_{-0.17}$ & 0.19$^{+0.20}_{-0.19}$ & 0.40$^{+0.05}_{-0.05}$ & 1.13$^{+0.77}_{-0.72}$  & 1057/992\\\\[-2ex]
\makebox[0in][l]{13--17$'$}  &  $ 26\pm1$  & 2.50$^{+0.09}_{-0.09}$ & 0.72$^{+0.29}_{-0.38}$ & 0.44$^{+0.29}_{-0.28}$ & 0.78$^{+0.29}_{-0.27}$ & 0.30$^{+0.15}_{-0.14}$ & 0.44$^{+0.18}_{-0.17}$ & 0.25$^{+0.05}_{-0.05}$ & 0.16$^{+0.60}_{-0.16}$  & 1082/992\\\\[-2ex]
\makebox[0in][l]{17--27$'$}  &  $ 13\pm1$  & 2.22$^{+0.11}_{-0.12}$ & 0.48$^{+0.41}_{-0.37}$ & 0.30$^{+0.29}_{-0.26}$ & 0.21$^{+0.22}_{-0.21}$ & 0.28$^{+0.12}_{-0.12}$ & 0.21$^{+0.14}_{-0.14}$ & 0.21$^{+0.04}_{-0.04}$ & 0.00$^{+0.41}_{-0.00}$  &  ---\makebox[0in][l]{\,$^\dagger$} \\\\[-2ex]
\hline\\[-1ex]
\multicolumn{11}{l}{\parbox{0.96\textwidth}{\footnotesize
\footnotemark[$\ast$] Normalization of the {\it vapec} model,
calculated in the same way with table~\ref{tab:fit_center}.}}\\
\multicolumn{11}{l}{\parbox{0.96\textwidth}{\footnotesize
\footnotemark[$\dagger$] The 17--27$'$ annulus was fitted simultaneously
with other annuli, and the best-fit values with the 13--17$'$ annulus
are presented here.
}}
\end{tabular}}
\end{table*}

\begin{figure*}
\begin{minipage}{0.33\textwidth}
\FigureFile(\textwidth,\textwidth){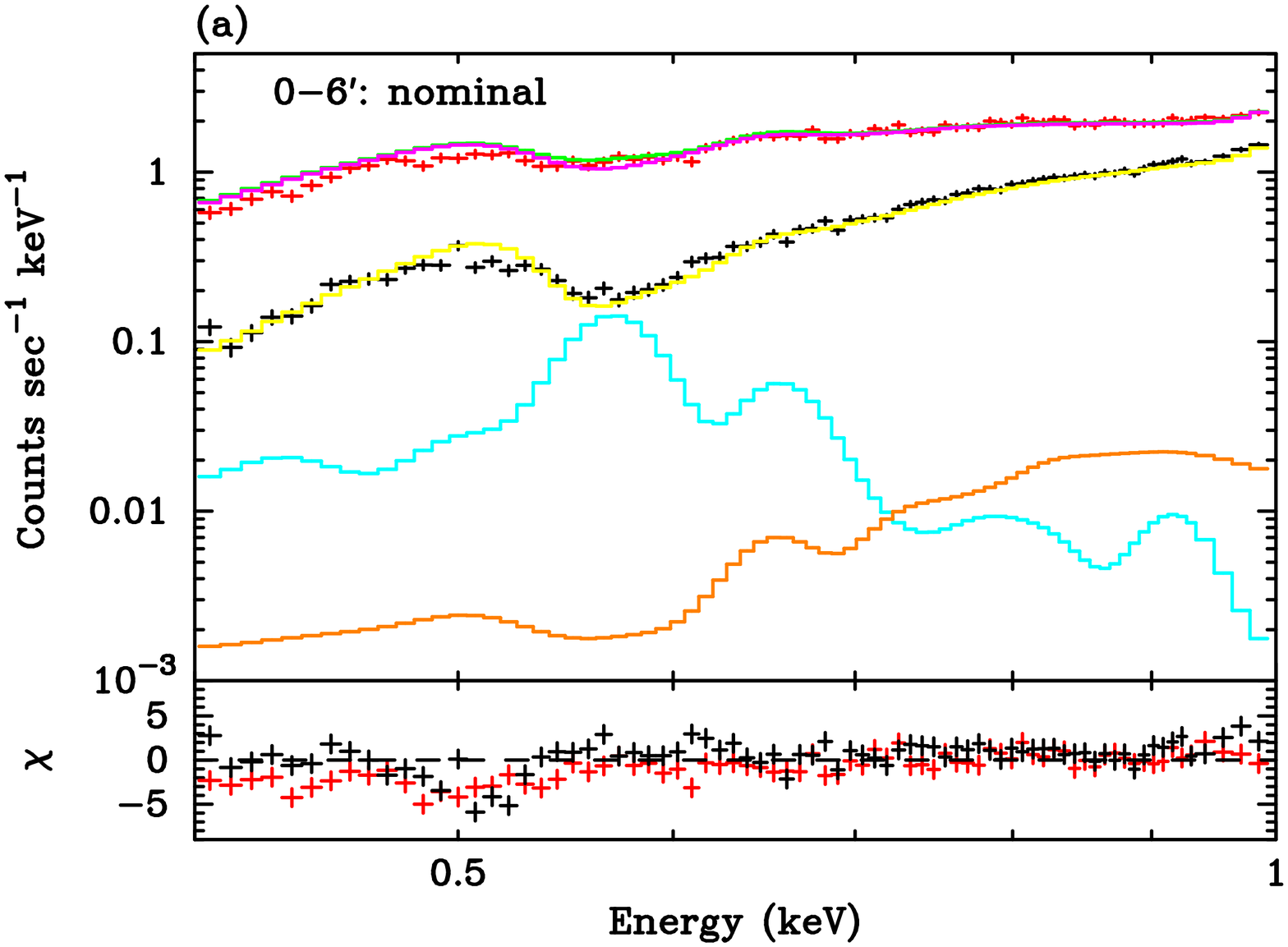}
\end{minipage}\hfill
\begin{minipage}{0.33\textwidth}
\FigureFile(\textwidth,\textwidth){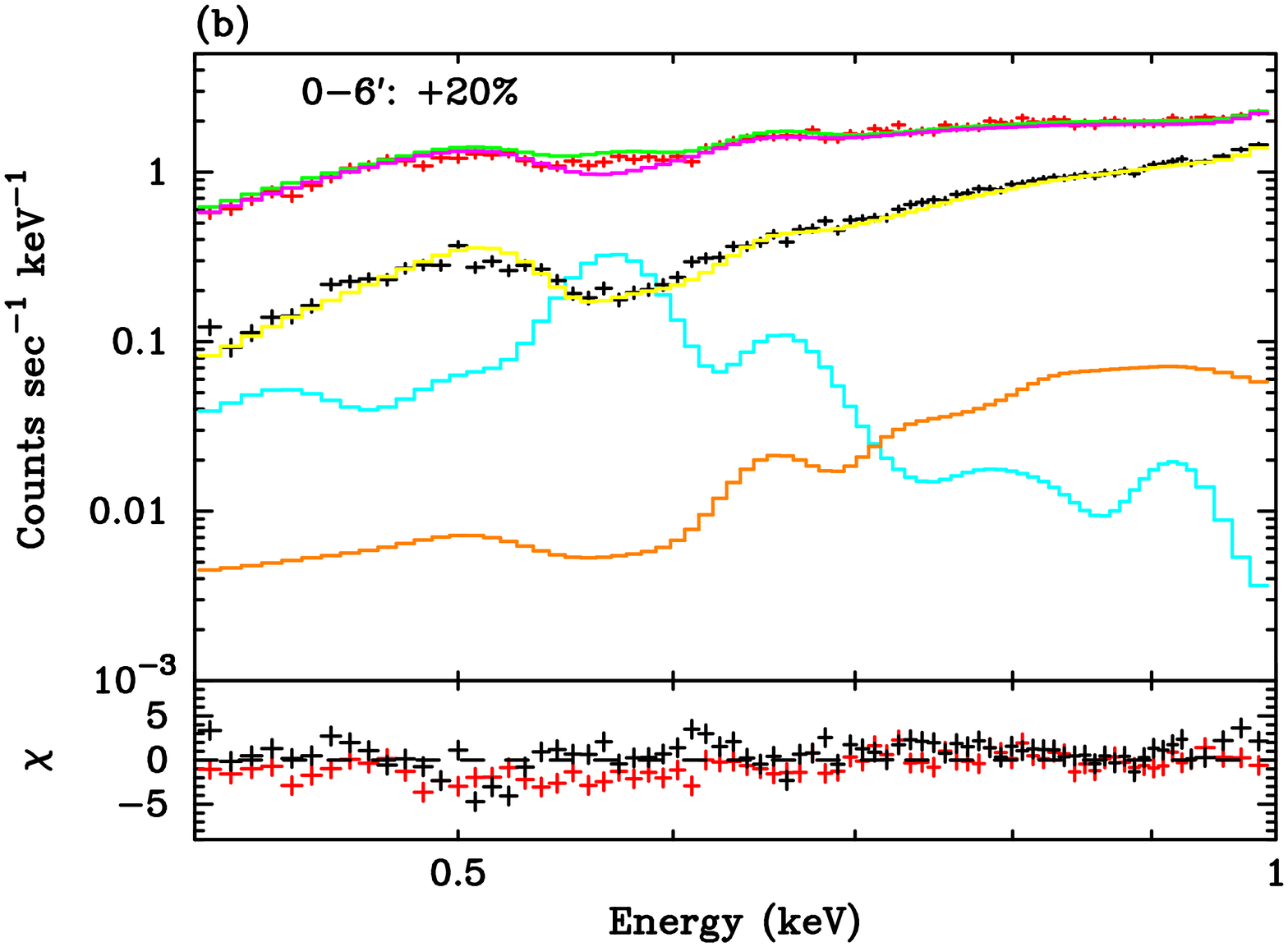}
\end{minipage}\hfill
\begin{minipage}{0.33\textwidth}
\FigureFile(\textwidth,\textwidth){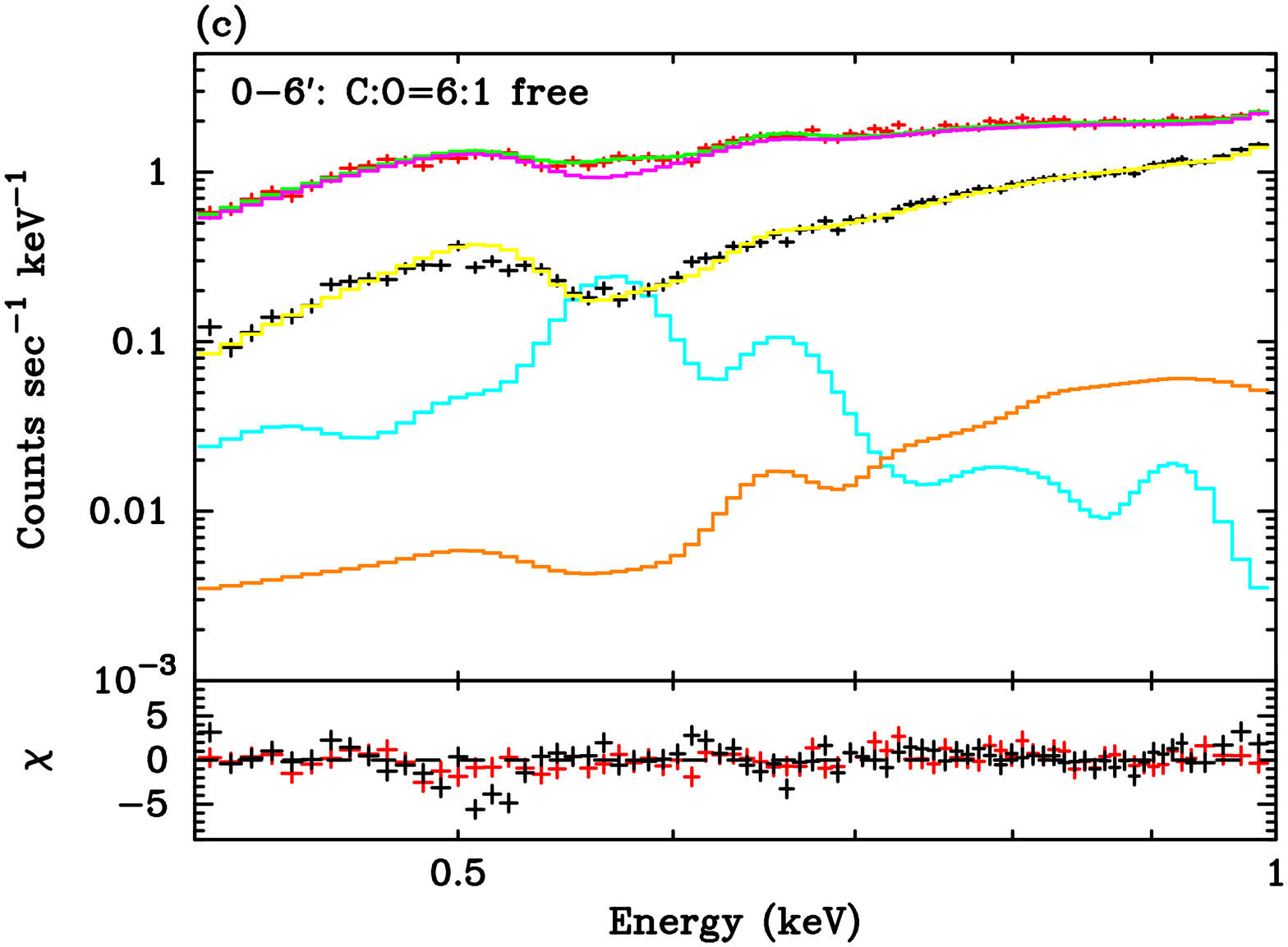}
\end{minipage}

\begin{minipage}{0.33\textwidth}
\FigureFile(\textwidth,\textwidth){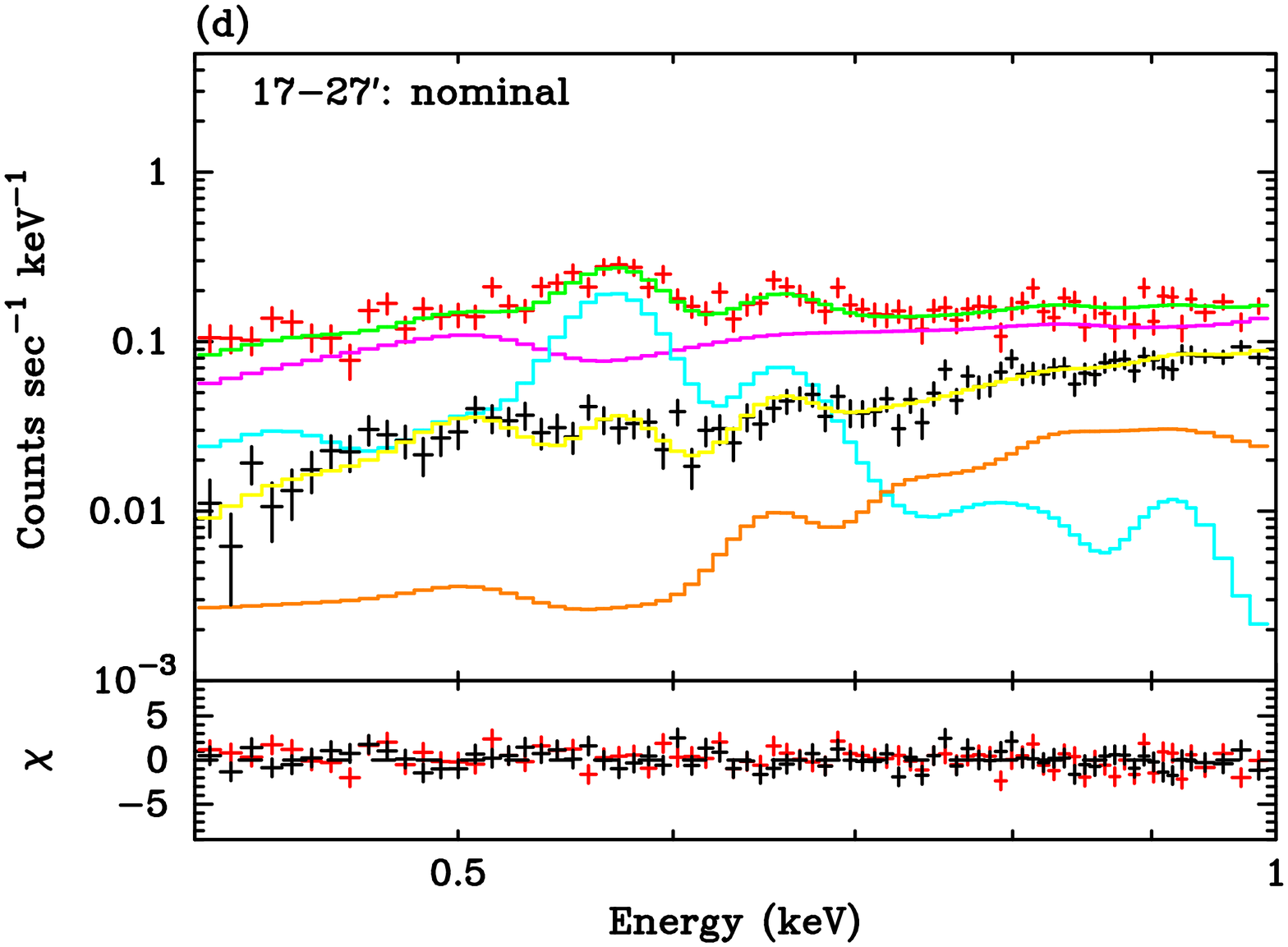}
\end{minipage}\hfill
\begin{minipage}{0.33\textwidth}
\FigureFile(\textwidth,\textwidth){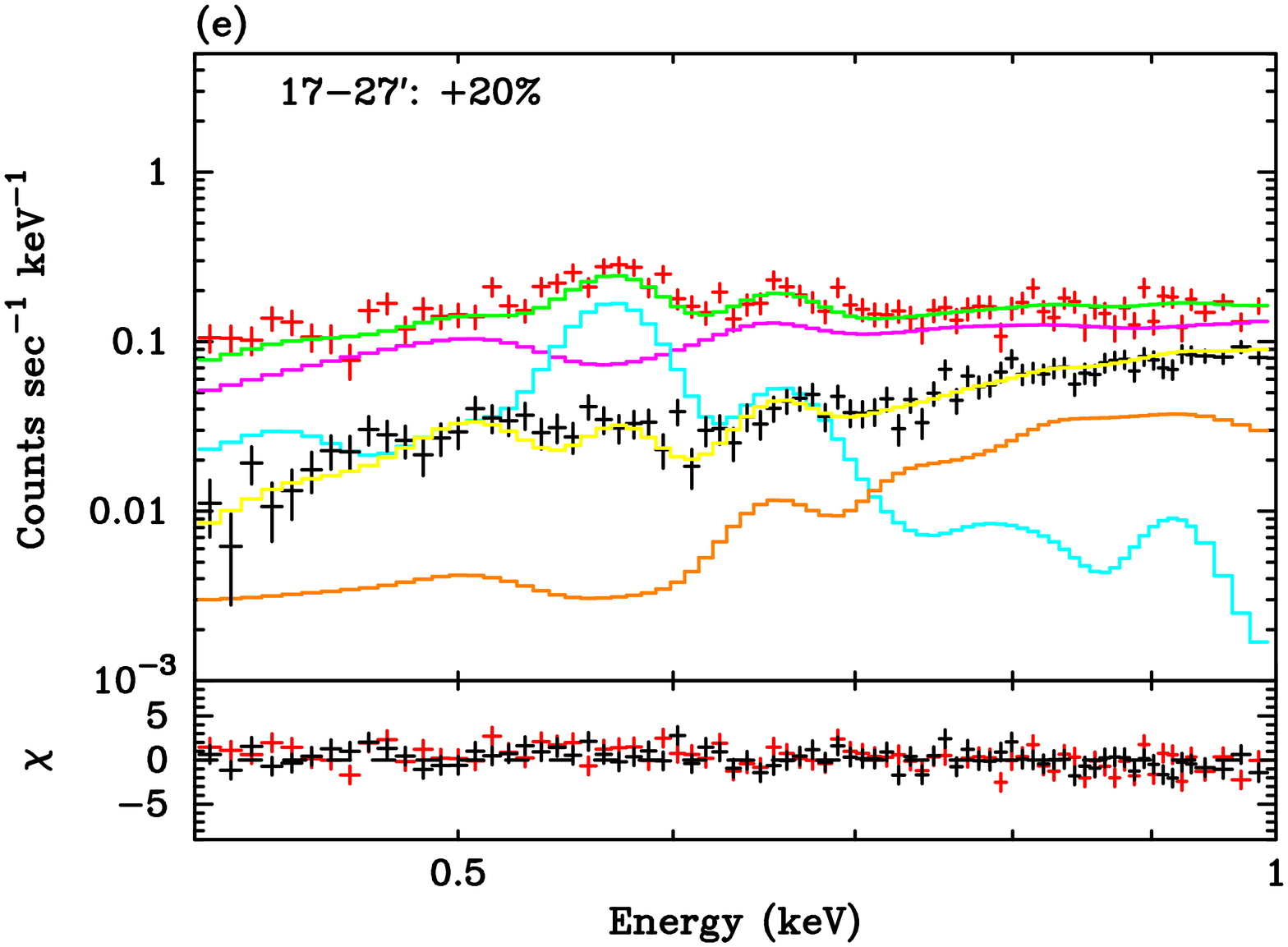}
\end{minipage}\hfill
\begin{minipage}{0.33\textwidth}
\FigureFile(\textwidth,\textwidth){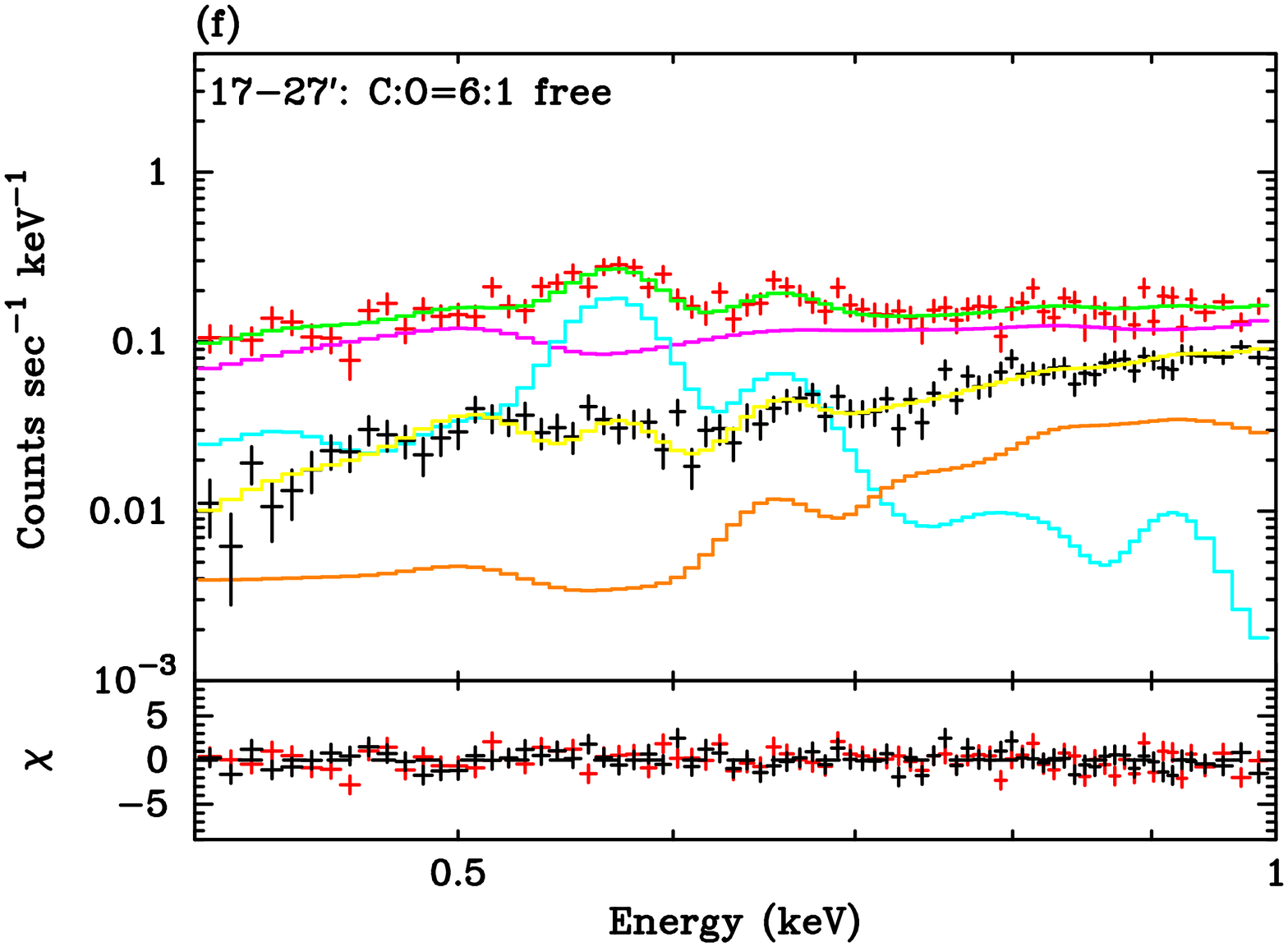}
\end{minipage}
\caption{
(a) A magnification of the best-fit spectrum at 0--6$'$ annulus
in the energy range of 0.4--1.0~keV with the ``nominal'' ARF response,
in which transmission of the OBF contaminant was estimated
according to the calibration files
{\tt ae\_xi$N$\_contami\_20060525.fits} ($N=0,1,2,3$).
Table~\ref{tab:fit_center} and
the ``nominal'' row of table~\ref{tab:contami_cen}
show the best-fit parameters.
(b) The best-fit spectrum with the ``+20\%'' ARF,
in which amount of the OBF contaminant was increased by +20\%
than the ``nominal'' ARF\@.
The ``+20\%'' row of table~\ref{tab:contami_cen} shows the best-fit parameters.
(c) The best-fit spectrum with the no-contaminant ARF but fitted with
${\it varabs}\times
({\it apec}_1 + {\it apec}_2 + {\it phabs} \times {\it vapec})$ model
to consider the transmission of the OBF contaminant
in the XSPEC $varabs$ model with chemical composition of C\,O$_{1/6}$.
The ``free'' row of table~\ref{tab:contami_cen} shows the best-fit parameters.
(d), (e), (f) Same as (a), (b), (c) but for the 17--27$'$ annulus,
respectively. Each spectrum was simultaneously fitted with (a), (b), (c).
}\label{fig:contami_cen}
\end{figure*}

\begin{figure*}[t]
\begin{minipage}{0.33\textwidth}
\FigureFile(\textwidth,\textwidth){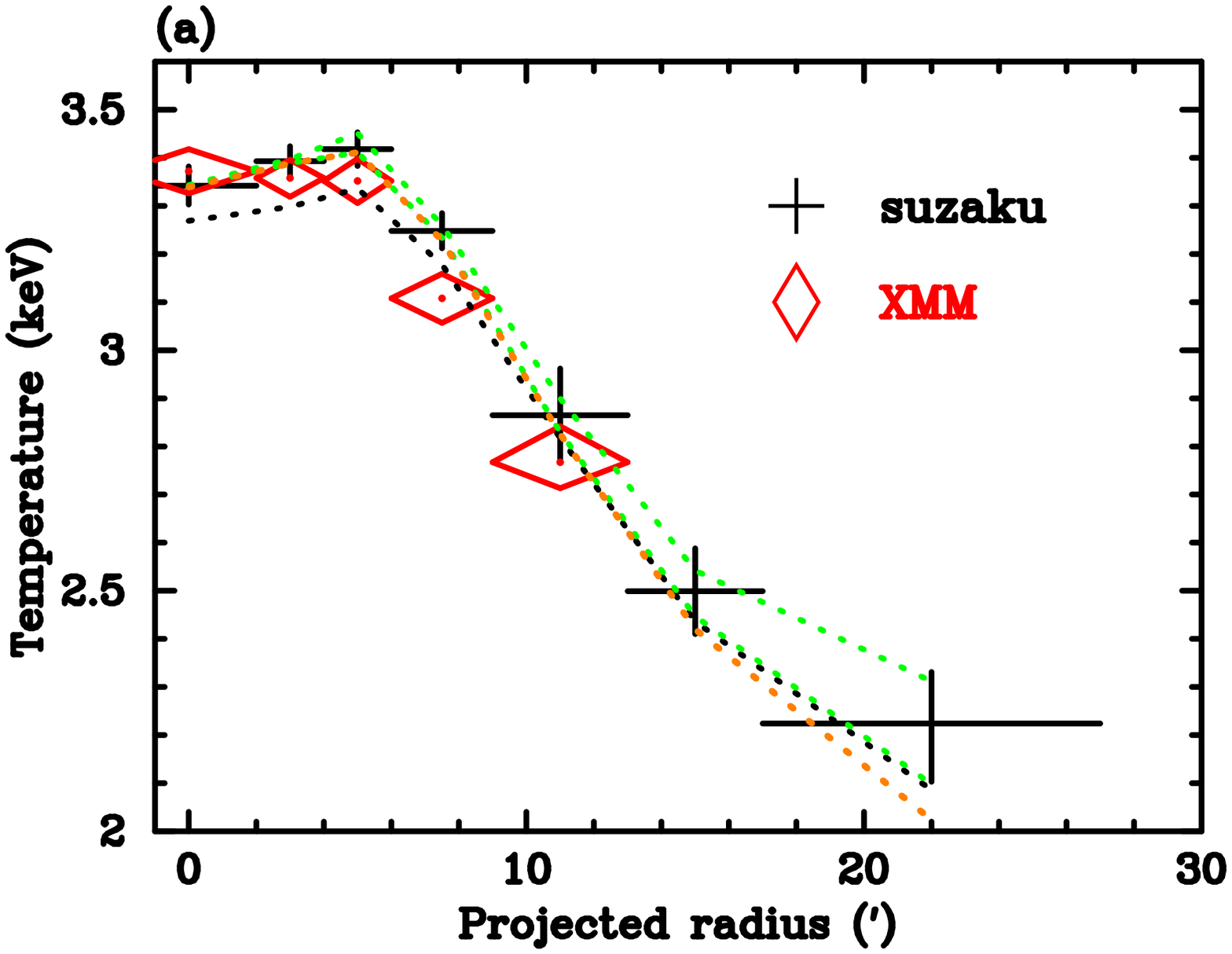}
\end{minipage}%
\begin{minipage}{0.33\textwidth}
\FigureFile(\textwidth,\textwidth){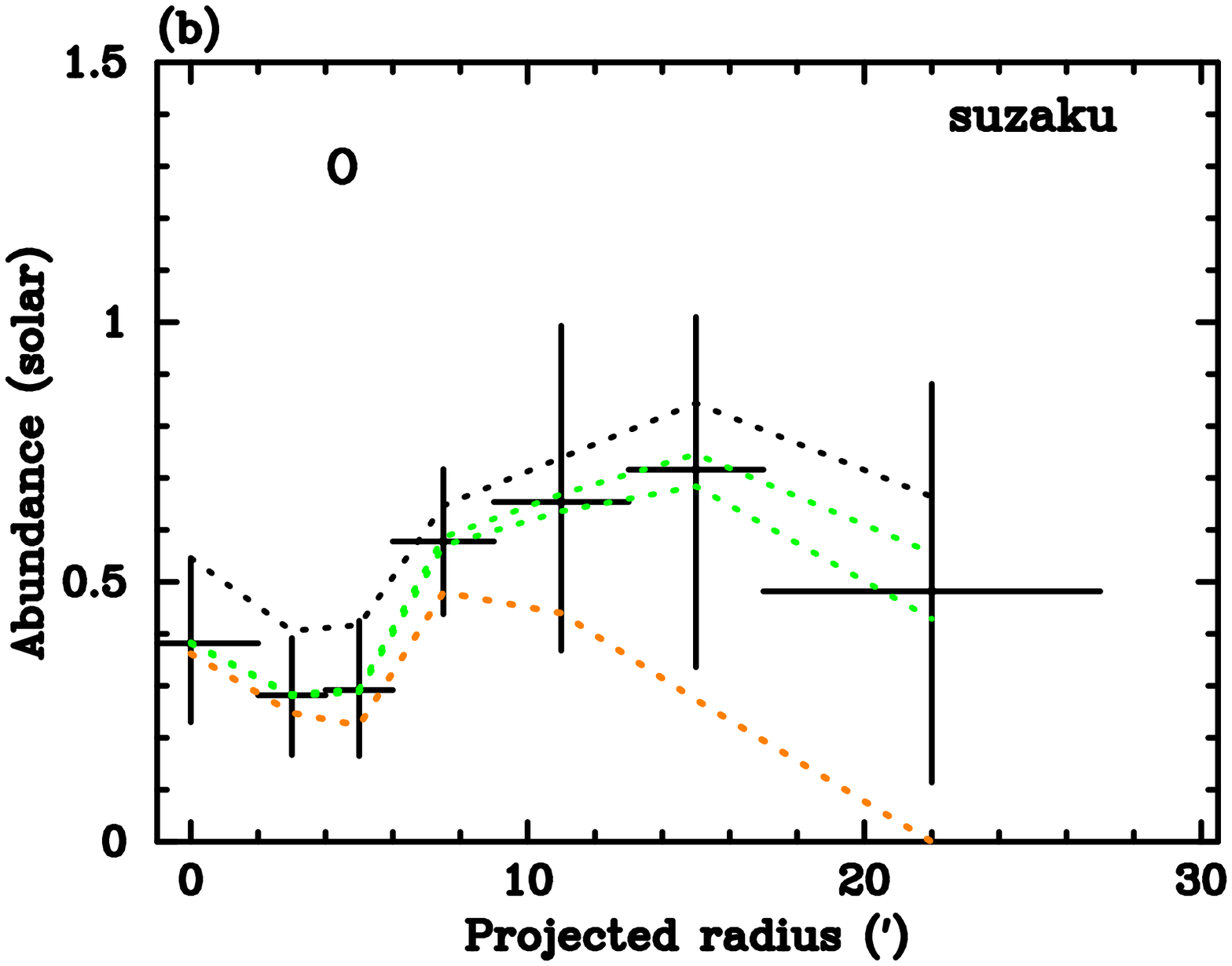}
\end{minipage}%
\begin{minipage}{0.33\textwidth}
\FigureFile(\textwidth,\textwidth){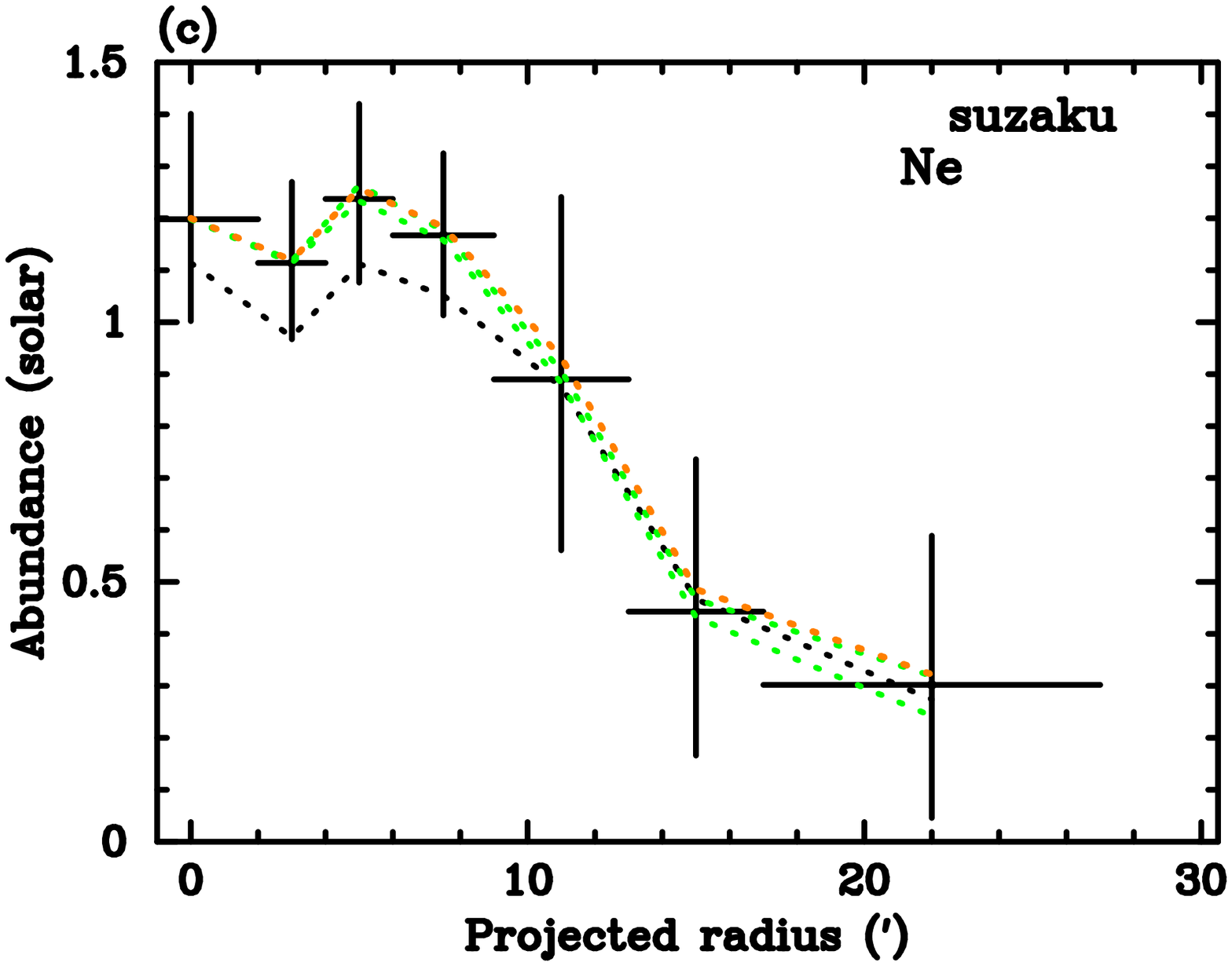}
\end{minipage}%

\begin{minipage}{0.33\textwidth}
\FigureFile(\textwidth,\textwidth){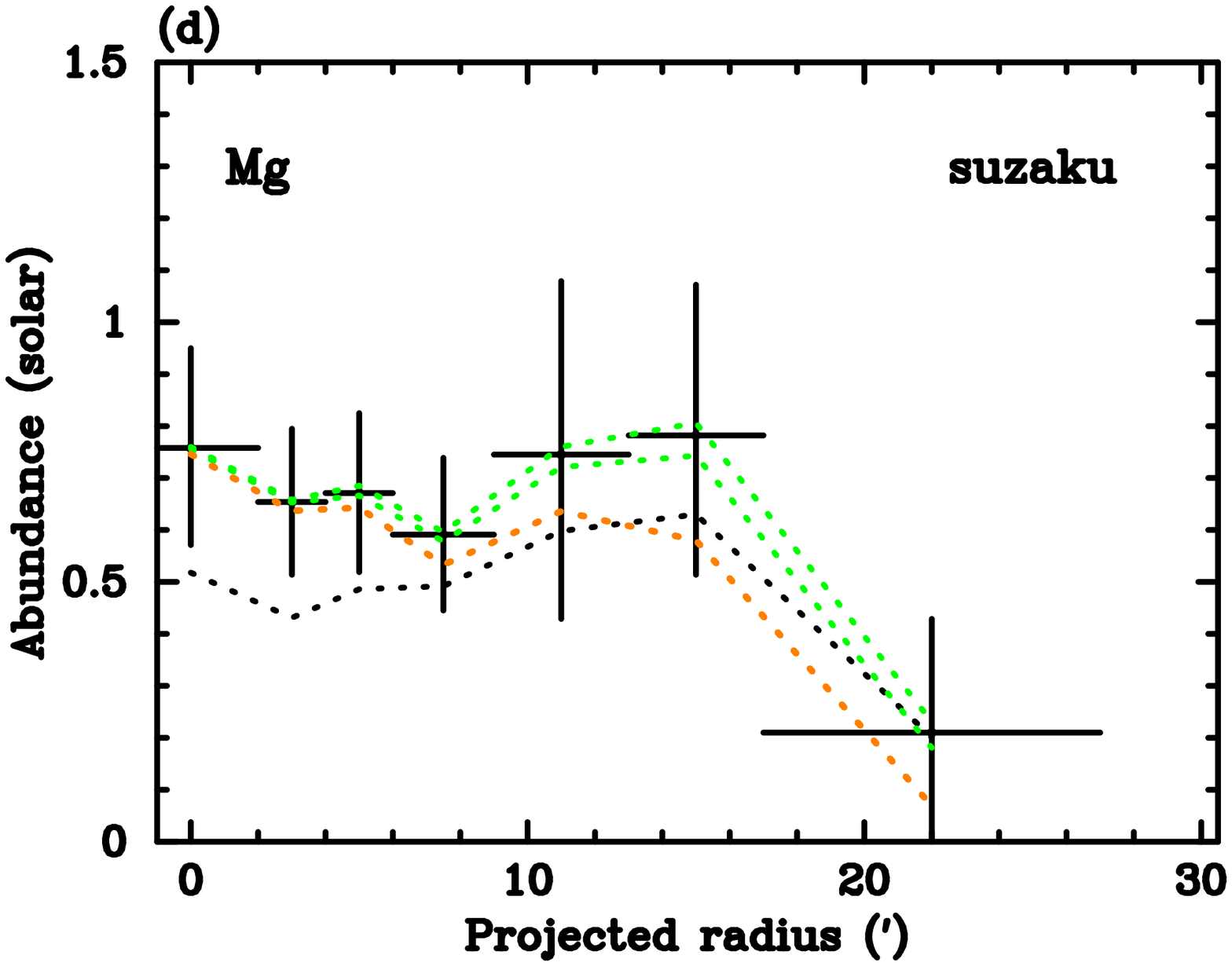}
\end{minipage}%
\begin{minipage}{0.33\textwidth}
\FigureFile(\textwidth,\textwidth){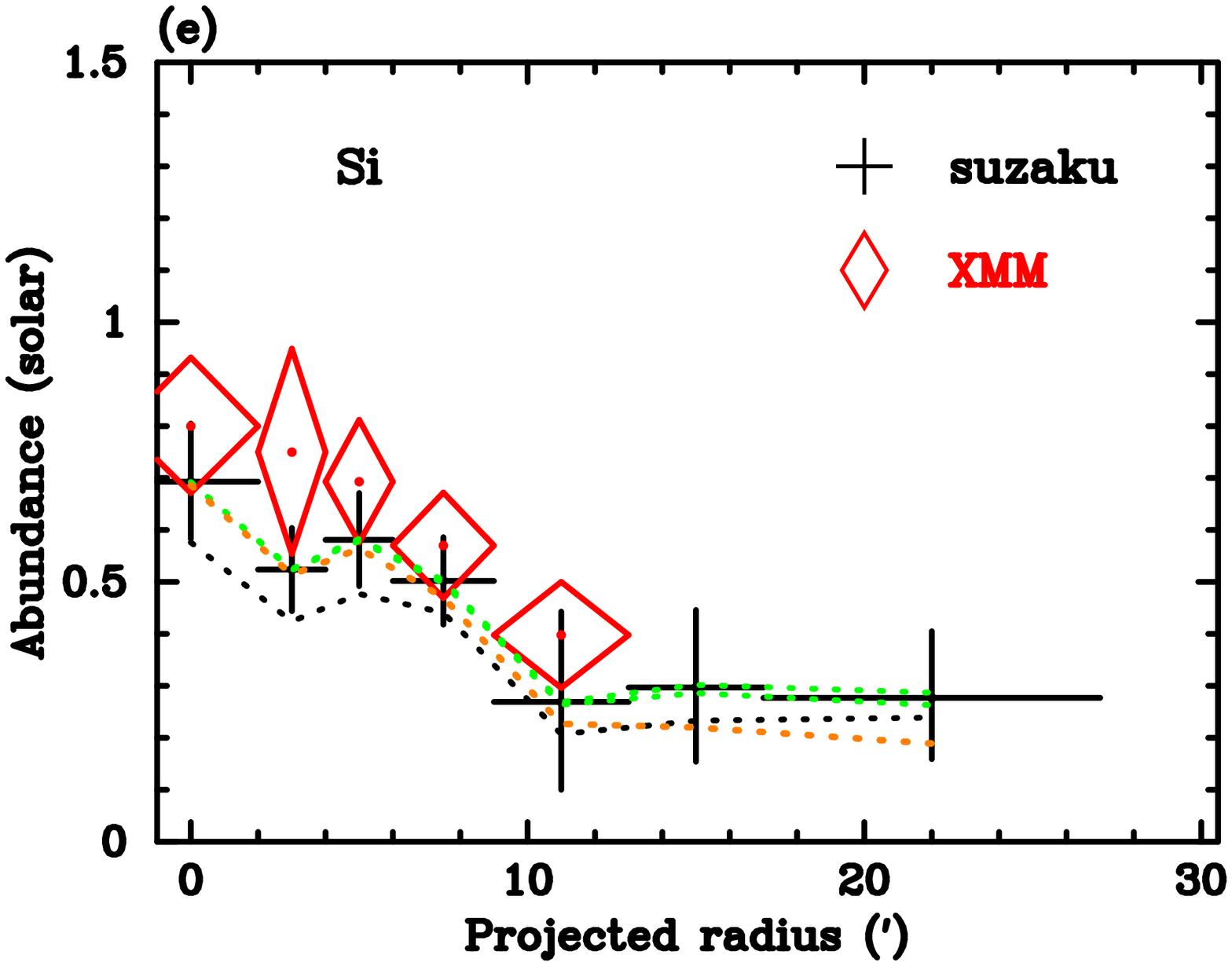}
\end{minipage}%
\begin{minipage}{0.33\textwidth}
\FigureFile(\textwidth,\textwidth){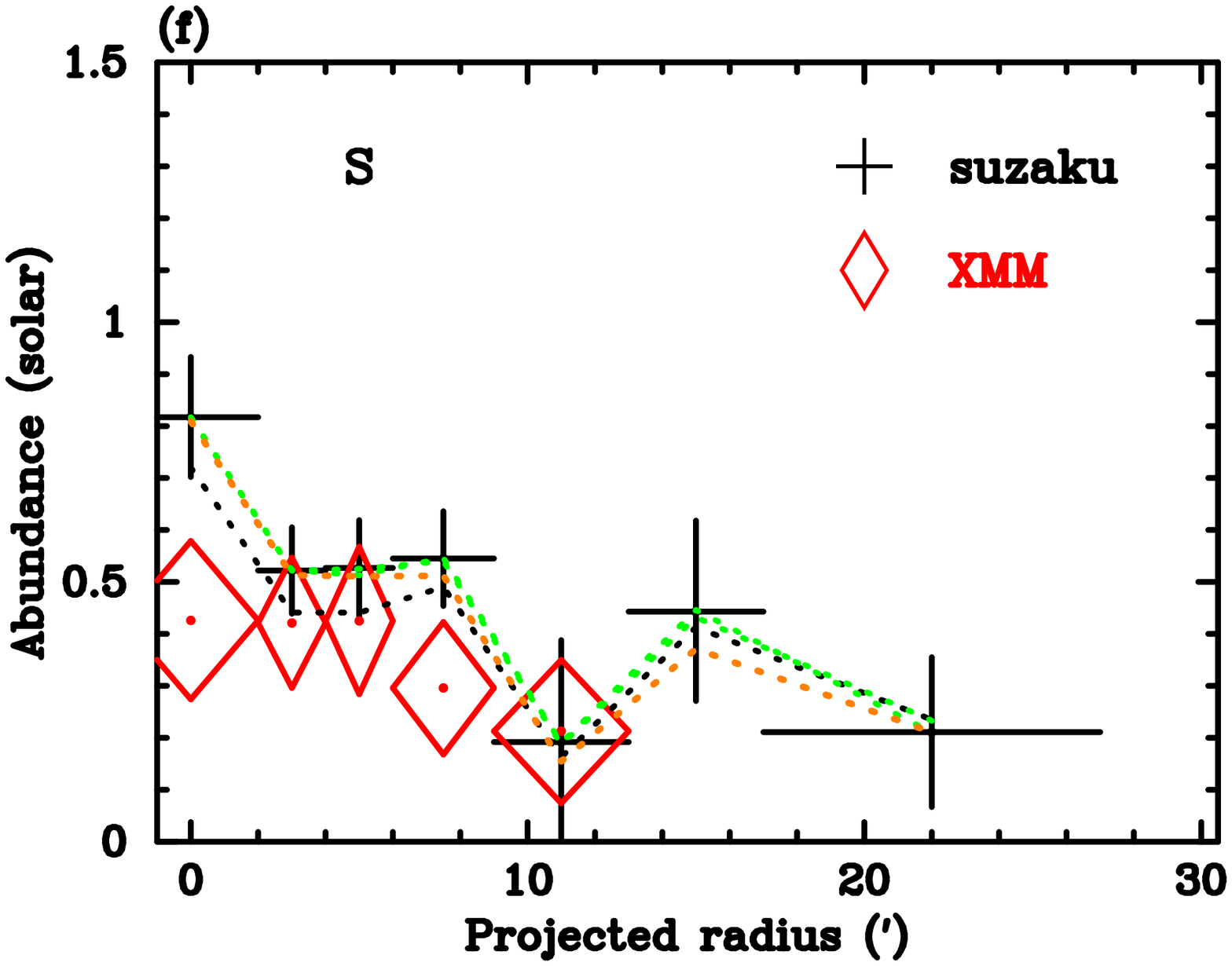}
\end{minipage}

\begin{minipage}{0.33\textwidth}
\FigureFile(\textwidth,\textwidth){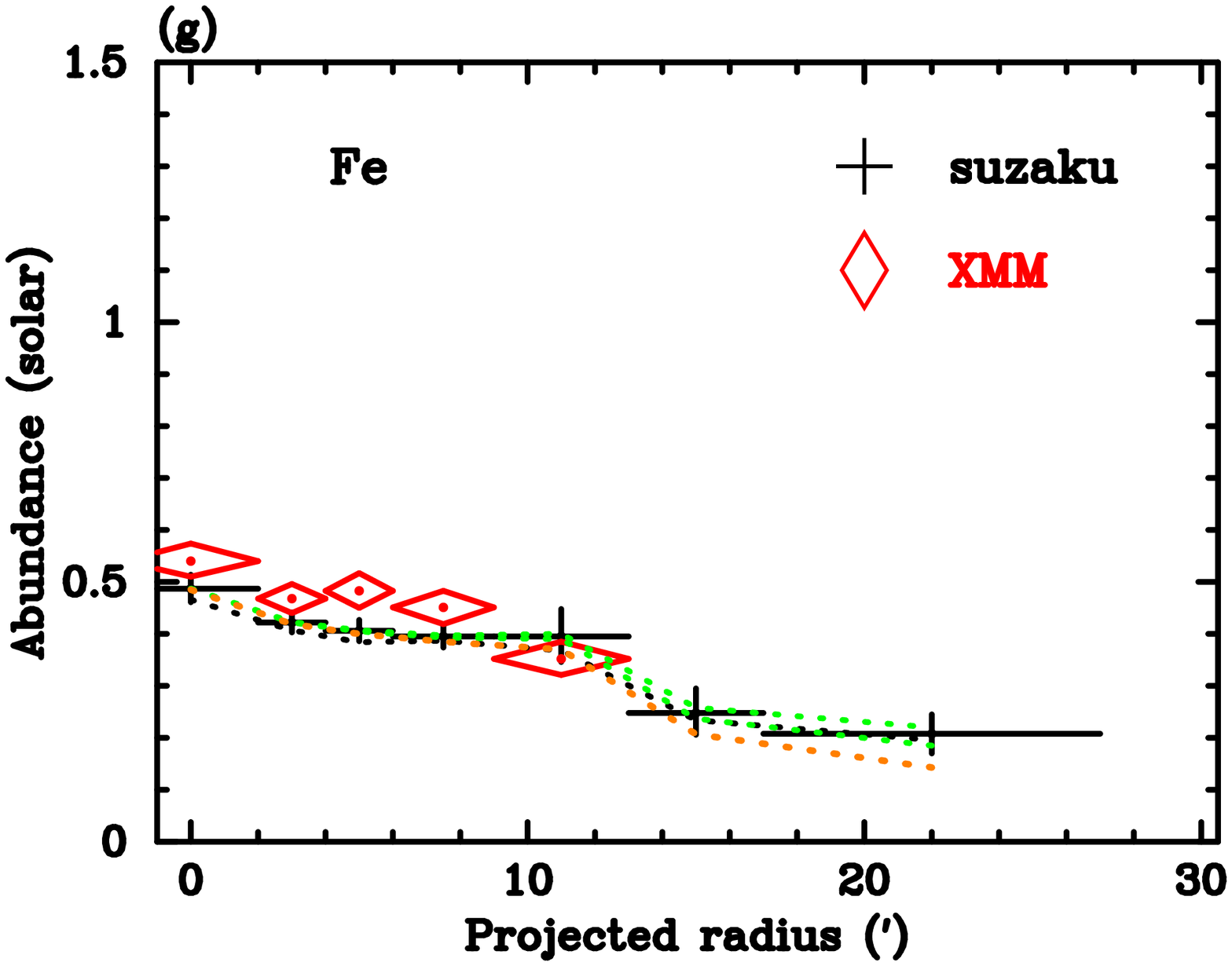}
\end{minipage}%
\begin{minipage}{0.33\textwidth}
\FigureFile(\textwidth,\textwidth){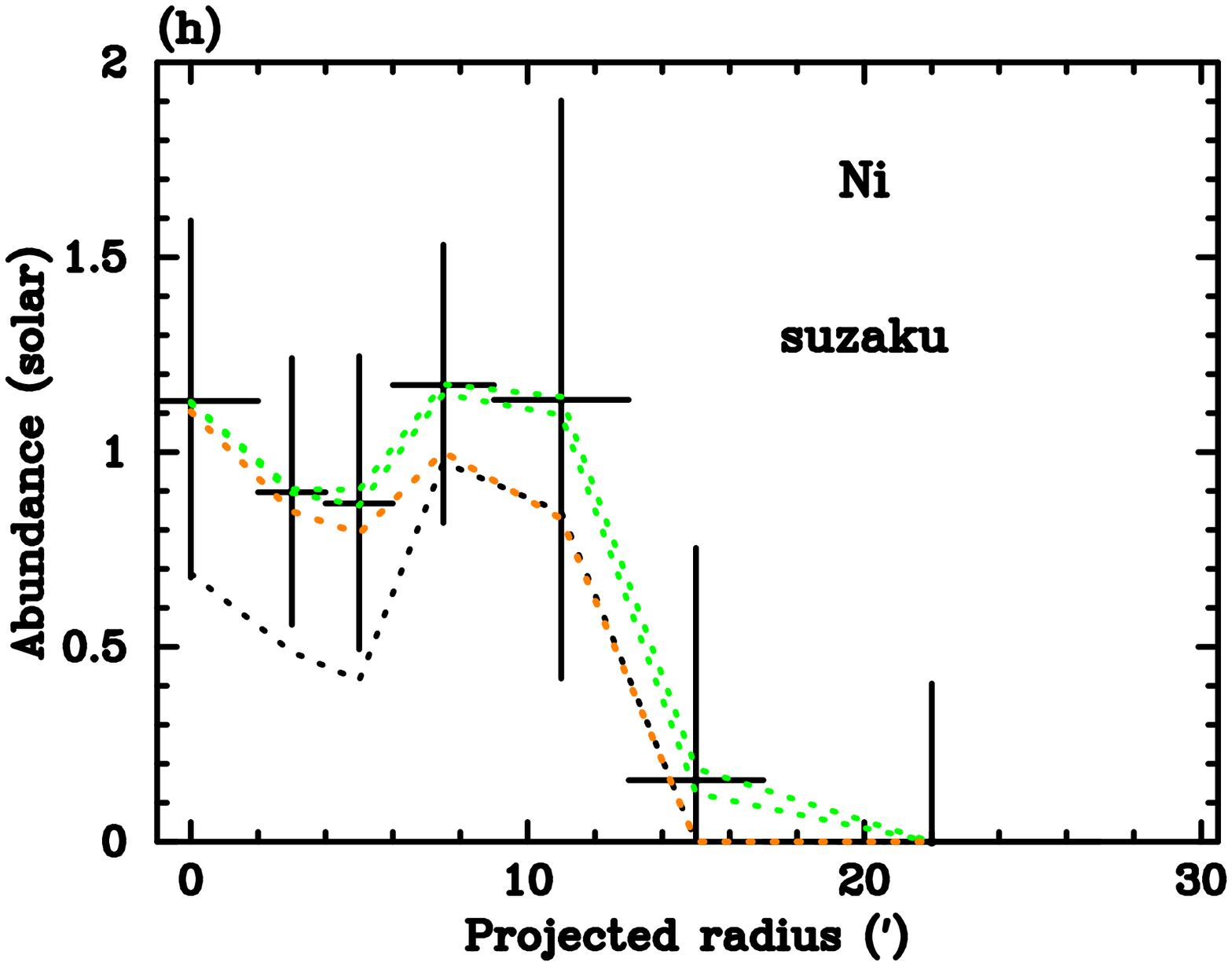}
\end{minipage}

\vspace*{-0.5ex}
\caption{
(a) Radial temperature profiles derived from spectral fit of
the Suzaku (black) and XMM-Newton (red) spectra at each annulus. 
The horizontal axis denotes the projected radius
and deprojection are not conducted.
The same data with table~\ref{tab:fit_icm} are used for Suzaku.
The black dotted lines correspond to shifts of the best-fit values
by changing thickness of the OBF contaminant by $+ 20$\%.  
The green dotted lines denote those when the estimated CXB and NXB
levels are changed by $\pm 10$\%.
The orange dotted line shows the best-fit value when the Galactic
component is modeled by a single {\it apec}.
Regarding XMM-Newton, the ${\it phabs}\times vapec$ model
are used for the spectral fit, and the O, Ne, Mg, and Ni abundances
are fixed to 1 solar in the assumed abundance table of {\it angr}.
(b)--(h) Radial abundance profiles derived and plotted in the
same way as (a).
}\label{fig:result}
\end{figure*}

Before entering the spectral fit at each annulus,
we investigate the fit result in the central 0--6$'$ region
to test the capability in the abundance determination
with the Suzaku XIS\@.
This region exhibits nearly constant temperature
and metal abundances (figure~\ref{fig:result}),
and the contribution of the Galactic component is almost
negligible (figure~\ref{fig:spec}).
The fit result is presented in table~\ref{tab:fit_center},
in which most of element abundances
are allowed to be free in the spectral fit,
except for He, C, and N, which are fixed to 1~solar
with the assumed abundance ratio of {\it angr}.

Although the fit was not acceptable due mainly to the very
high photon statistics than the systematic errors
in the instrumental response,
this result was useful to assess whether each element abundance
was reasonably (sub-solar to $\sim 2$~solar) determined or not.
The Ar and Ca abundances were reasonably determined
although the errors are larger than other elements.
We therefore decided to link the S, Ar, and Ca abundances
to be the same, because they showed similar values.
The Al abundance became 0.0, which is physically strange,
therefore we fixed the Al abundance to 1 solar.
The $\chi^2$/dof was increased to 1898/988 with this treatment,
however we confirmed that other parameters did not change
beyond the quoted error range in table~\ref{tab:fit_center}.
Somehow, the Ne and Ni abundances were larger than other elements.
This might be due to these element lines could not
be resolved from the Fe-L line complex.
Note that the Ni abundance was also determined by Ni-L lines
because we ignored energy range above 7.1~keV,
while the Fe abundance was determined by both Fe-L and Fe-K lines.
Anyway, we left these Ne and Ni abundances to move freely
during the spectral fit. We also present results when
the O, Ne, and Mg abundances are linked to have the same value
in Appendix~\ref{app:lodd}.

We also found that the assumed abundance table, which defines
the ``1 solar'' of each element relative to H,
significantly affects the determination of temperature and abundance.
In fact, difference in the abundance table affects the fit results
in three ways. Firstly, we assume the Galactic component to have the
${\it apec}_1 + {\it apec}_2$ model with ``1 solar''.
Secondly, the Galactic absorption model of {\it phabs} is also
changed by the assumed abundance table.
Lastly, the derived abundances are given in unit of ``solar''
by the {\it vapec} model.
The first two effects are rather complicated,
and details are investigated in Appendix~\ref{app:lodd}.
In this section, we treat only the solar abundance ratio of
{\it angr} \citep{anders89} with the {\it phabs} absorption model,
simply because they are the {\em default standard} of XSPEC.
In terms of the $\chi^2$, it appears to give the minimum $\chi^2$
with the {\it lodd} \citep{lodders03} abundance table
in combination with the {\it wabs} \citep{morrison83} absorption model
which utilizes the abundance table of {\it aneb} \citep{anders82}
built-in the code (table~\ref{tab:chi}).

We then examined the influence of uncertainty in the OBF contaminant
using the spectrum within 6$'$.
Figure~\ref{fig:contami_cen}(a) shows a magnification of
the best-fit spectrum in the energy range of 0.4--1.0~keV
with the ``nominal'' ARF response, in which transmission of
the OBF contaminant was estimated according to the calibration files
{\tt ae\_xi$N$\_contami\_20060525.fits} ($N=0,1,2,3$), as shown
in figures~\ref{fig:trans} and \ref{fig:xis}(b).
As described in subsection~\ref{subsec:strategy},
the spectrum was simultaneously fitted with the 17--27$'$ annulus
shown in figure~\ref{fig:contami_cen}(d).
In this figure, it is suggested that the absorption in the
low energy band is slightly inconsistent between BI and FI sensors,
namely, the BI spectrum appears to need more absorption in 0--6$'$
where the amount of contamination is larger than 17--27$'$
(see figure~\ref{fig:trans}).
We also found that the contamination measurement by RXJ1856.5$-$3754
on October 25, 2005 in the XIS hardware paper (figure~14 of \cite{koyama06})
indicates by about 20\% larger amount of the OBF contaminant than
the ``nominal'' value for BI (XIS1).

We therefore generated ARF responses changing the amount of
the OBF contaminant by $+20$\%. The fit result is shown in
figure~\ref{fig:contami_cen}(b), (e) and
table~\ref{tab:contami_cen} (``+20\%'' row).
Note that the CXB background to subtract was also modified in this fit.
The fit residual for 0--6$'$ was improved,
and the $\chi^2$ was decreased by $\Delta\chi^2=132$.
We further tested the fit by adding an absorption of the OBF
contaminant with chemical composition of C\,O$_{1/6}$ to the fit model
using the XSPEC {\it varabs} model.\footnote{
In this method, we used the same CXB background with the ``nominal'' ARF,
and the surface brightness of the Galactic component
was slightly different between 0--6$'$ and 17--27$'$.}
The fit result is shown in figure~\ref{fig:contami_cen}(c) and (f).
The best-fit values are summarized in table~\ref{tab:contami_cen}
(``free'' row), and the $\chi^2$ was improved by $\Delta\chi^2=197$.
The derived amount of contaminant for BI was by 25\% larger than
the ``nominal'' value, and by 3\% larger for FI\@.
We think that the deviation from the ``nominal'' value for BI is
beyond the calibration uncertainty, however, almost all of the best-fit
values at the ``free'' row in table~\ref{tab:contami_cen} are
between values at ``nominal'' and ``+20\%''.
We therefore consider the $+20$\% result as the systematic error range
due to the uncertainty in the OBF contaminant.

Considering these systematics,
results of the spectral fit at each annulus are summarized in
table~\ref{tab:fit_icm} and figure~\ref{fig:result}.
We tested the results by changing the background normalization 
by $\pm 10$\%, and they are plotted in green dotted lines
in figure~\ref{fig:result}. The systematic error due to
the background estimation is almost negligible.
Difference in the best-fit values by modeling the Galactic component
with a single {\it apec} was investigated, and they were indicated
by orange dotted lines.
The differences are within the statistical error for the
inner five annuli, however, O abundance becomes lower than
the 90\% confidence error at the outer two annuli as indicated in 
figure~\ref{fig:fe-l},
and temperature and Fe abundance become lower at the outermost annulus.
This is due mainly to the fact that the XIS cannot resolve
the ICM O\emissiontype{VIII} line from the Galactic O\emissiontype{VIII}
by redshift.
The systematic error range due to the uncertainty in the OBF contaminant
is indicated by black dotted lines.
It is sometimes larger than the statistical errors at small radii
($r\lesssim 6'$), particularly for $kT$, O, and Mg abundances.
Though Ne abundance is significantly larger than other elements
at small radii ($r < 9'$), it is probably not reliable
because the Suzaku XIS cannot resolve the ionized Ne lines from
the Fe-L line complex.

We also plot the XMM-Newton results for the inner five annuli
with red diamonds, in which the ${\it phabs}\times vapec$ model
are used for the spectral fit, and the Ne, Ar, Ca and Ni abundances
are fixed to 1 solar. Namely, the Galactic component is ignored.
They are almost consistent with the Suzaku results,
although XMM-Newton appears to give slightly higher
abundance for Fe and Si, while lower abundance for S\@.

\subsection{Direct Comparison of O\emissiontype{VII} and O\emissiontype{VIII} Intensities}
\label{subsec:direct}

\begin{table}[t]
\caption{
Line intensities of O\emissiontype{VII} and O\emissiontype{VIII}
at each annulus of A~1060 and the NGC~2992 field
in unit of photons~cm$^{-2}$~s$^{-1}$~sr$^{-1}$.
These intensities are derived from the spectral fit
with {\it power-law} + {\it gaussian} + {\it gaussian} model
(see figure~\ref{fig:ngc2992} for the NGC~2992 field),
assuming the uniform-sky ARF response.
Intensities by \citet{mccammon02}
measured with a high resolution microcalorimeter array
for a large sky area of $\sim 1$~sr are also presented.
}\label{tab:sb}
\begin{center}
\begin{tabular}{llr} 
\hline \hline
\makebox[14em][l]{Region} & \multicolumn{1}{c}{O\emissiontype{VII}} & \multicolumn{1}{c}{O\emissiontype{VIII}} \\
\hline
\makebox[6em][l]{A~1060 center} (0--2$'$)   $\dotfill$ & \makebox[0in][r]{1}1.9$^{+15.5}_{-11.9}$ & 17.1$\pm$9.2 \\
\makebox[6em][l]{A~1060 center} (2--4$'$)   $\dotfill$ & 7.2$\pm$7.2  & 12.1$\pm$5.0 \\
\makebox[6em][l]{A~1060 center} (4--6$'$)   $\dotfill$ & \makebox[0in][r]{1}0.0$\pm$4.4  & 8.6$\pm$3.3 \\
\makebox[6em][l]{A~1060 center} (6--9$'$)   $\dotfill$ & 7.3$\pm$2.0  & 6.9$\pm$1.4 \\
\makebox[6em][l]{A~1060 offset} (9--13$'$)  $\dotfill$ & 8.3$\pm$2.8  & 5.7$\pm$1.8 \\
\makebox[6em][l]{A~1060 offset} (13--17$'$) $\dotfill$ & 8.6$\pm$1.7  & 3.5$\pm$0.9 \\
\makebox[6em][l]{A~1060 offset} (17--27$'$) $\dotfill$ & 8.6$\pm$1.2  & 2.0$\pm$0.5 \\
NGC~2992 field                              $\dotfill$ & 4.1$\pm$0.6  & 0.6$\pm$0.2 \\
\citet{mccammon02}                          $\dotfill$ & 4.8$\pm$0.8 & 1.6$\pm$0.4 \\
\hline 
\end{tabular}
\end{center}
\end{table}

\begin{figure*}[t]
\begin{minipage}{0.47\textwidth}
\FigureFile(\textwidth,\textwidth){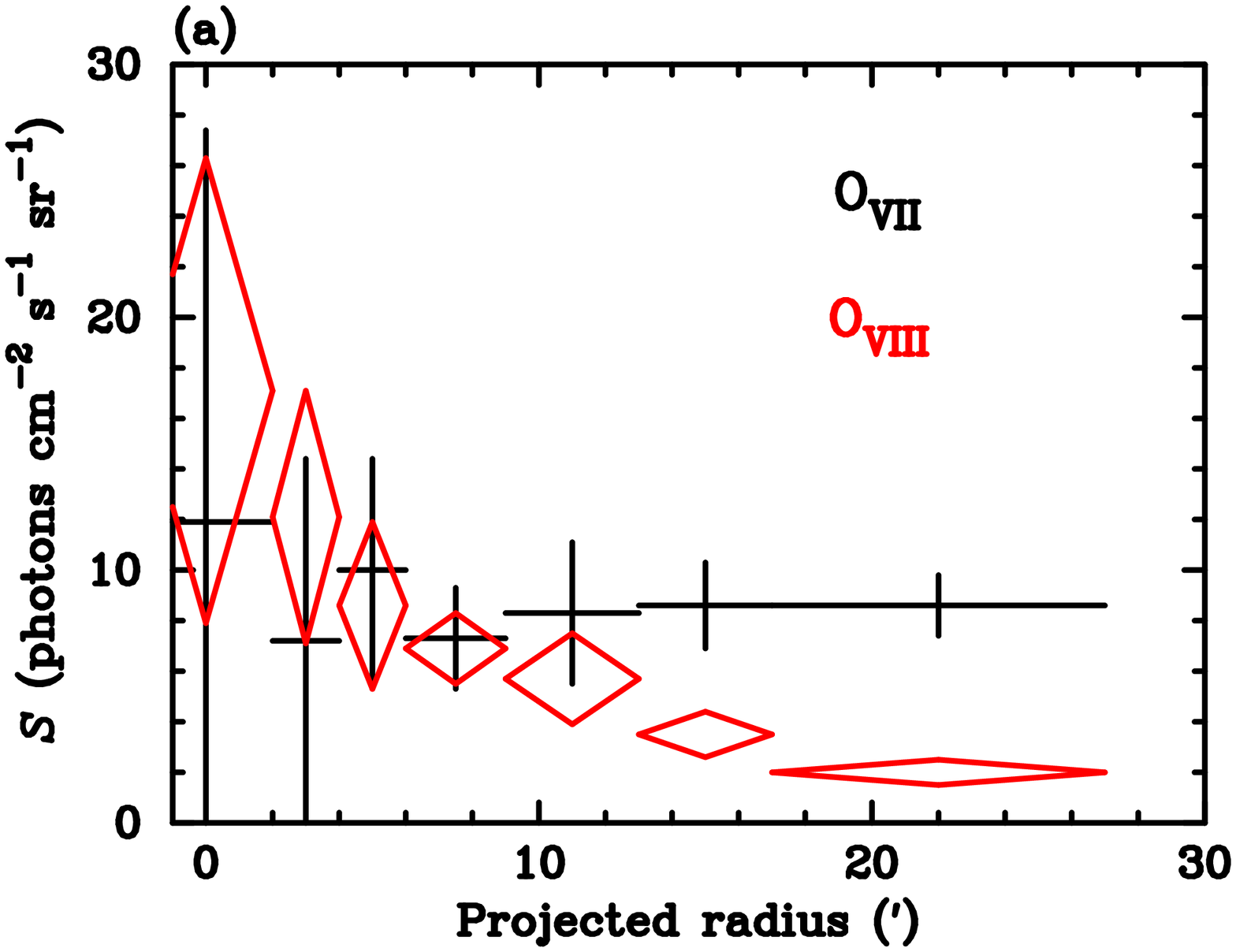}
\end{minipage}\hfill
\begin{minipage}{0.5\textwidth}
\FigureFile(\textwidth,\textwidth){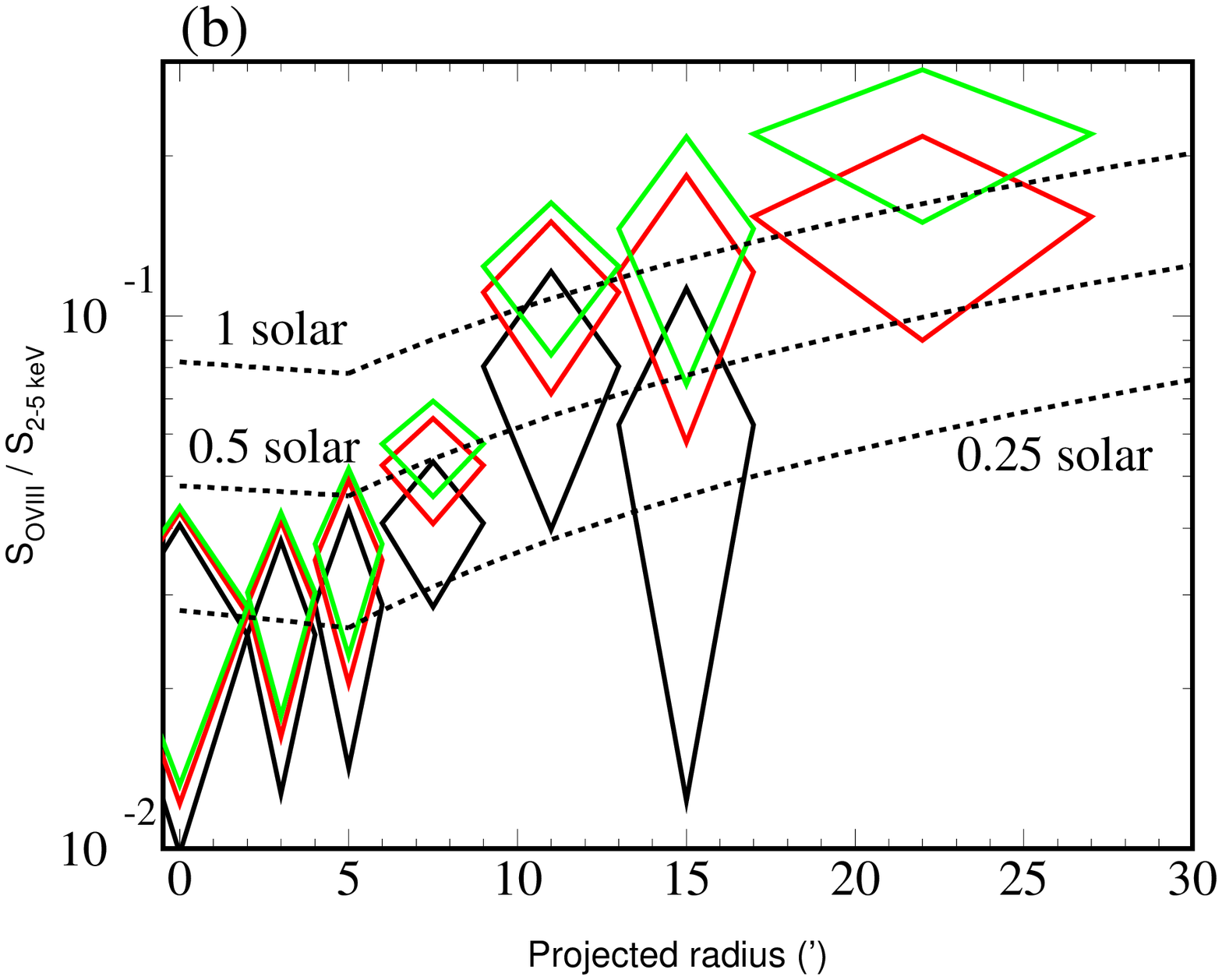}
\end{minipage}
\caption{
(a) Line intensities of O\emissiontype{VII} and O\emissiontype{VIII}
at each annulus of A~1060 in unit of photons~cm$^{-2}$~s$^{-1}$~sr$^{-1}$.
Each value is shown in table~\ref{tab:sb}.
(b) The O\emissiontype{VIII} emission line intensities in table~\ref{tab:sb}
divided by the surface brightness of the 2--5~keV continuum
are plotted in green diamonds against the radius of each annulus.
The red or black diamonds correspond to the ratios
when the O\emissiontype{VIII} intensities of the NGC~2992 field
or the outermost annulus (17--27$'$) are subtracted as an offset.
Dashed lines represent expected ratios for the {\it vapec} model
with 0.25, 0.5, and 1 solar abundances, assuming the observed
temperature at each annulus (table~\ref{tab:fit_icm}).
}\label{fig:sb}
\end{figure*}

\begin{figure}[t]
\centerline{\FigureFile(0.48\textwidth,0.48\textwidth){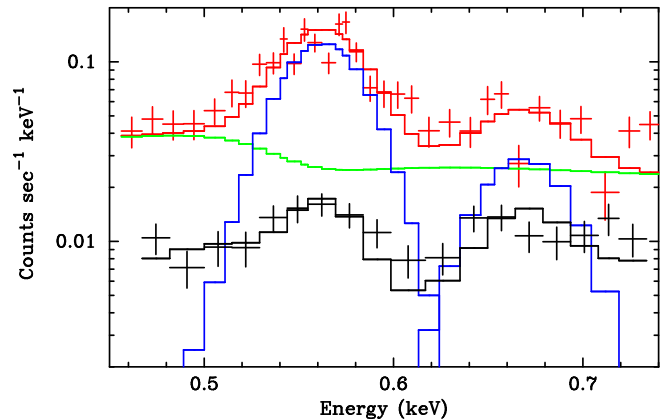}}
\caption{
The 0.46--0.74~keV spectra of the NGC~2992 field
with BI (red) and FI (black) are
fitted with the {\it power-law} + {\it gaussian} + {\it gaussian} model.
The red and black lines represent the best-fit model,
and the model components for BI are plotted by green ({\it power-law})
and blue (two {\it gaussian}) lines.
The central energies of two Gaussians are constrained to be within
$\pm 5$~eV range of O\emissiontype{VII} and O\emissiontype{VIII} lines,
and $\sigma$ of Gaussians are fixed to 0.0.
}\label{fig:ngc2992}
\end{figure}

We also investigated the surface brightness of
the O\emissiontype{VII} and O\emissiontype{VIII} emission lines
in order to estimate the O abundance of the ICM
directly from the line intensities.
The surface brightness of the O\emissiontype{VII}
and O\emissiontype{VIII} were derived by fitting the annular
spectrum with a {\it power-law} + {\it gaussian} + {\it gaussian} model.
In the fitting, we fixed the Gaussian $\sigma$ to be 0,
and allowed the energy center of the two Gaussians
to vary within 555--573~eV or 648--658~eV for
O\emissiontype{VII} or O\emissiontype{VIII}, respectively.
The derived line intensities are summarized in table~\ref{tab:sb}
and figure~\ref{fig:sb}(a).
There is a clear excess of the O\emissiontype{VIII} intensity
towards the cluster center,
while O\emissiontype{VII} is consistent with constant.
This is a strong evidence that the O\emissiontype{VIII} emission
is associated with the ICM itself, on the other hand,
O\emissiontype{VII} might be due mainly to the Galactic origin.

We compared the surface brightness of the oxygen lines
with that at a neighbor of A~1060, the NGC~2992 field.
Though NGC~2992 is located at 18$^\circ$ offset from A~1060, 
the Galactic latitude of A~1060 and NGC~2992 is similar (table\ref{tab:ob}),
and the neutral hydrogen column density
($N_{\rm H} = 5.1\times 10^{20}$~cm$^{-2}$; \cite{dickey90})
is comparable to A~1060.
NGC~2992 ($z=0.007710$) has an active galactic nuclei,
which is expected to emit no oxygen lines.
In addition, we excluded the main target with a radius of 5$'$ 
centered on NGC~2992, and utilized the outer region
for the spectral analysis in figure~\ref{fig:ngc2992}.

The surface brightness of the O\emissiontype{VII} 
and O\emissiontype{VIII} emissions 
in the outermost annulus of the A~1060 observation
was roughly twice larger than the NGC~2992 field.
Line intensities by \citet{mccammon02}
measured with a high resolution microcalorimeter array
for a large sky area of $\sim 1$~sr are also presented
in table~\ref{tab:sb}, and the O\emissiontype{VIII} intensity
was consistent with the outermost annulus (17--27$'$) of the A~1060.
It is suggested that both the O\emissiontype{VII} and
and O\emissiontype{VIII} lines in our observation are
somewhat mixture of the ICM and the Galactic origin.
However, we could not confirm excess O\emissiontype{VII} emission
towards the cluster center, we concentrate on
the O\emissiontype{VIII} emission.

Then we calculated the surface brightness ratio of the
O\emissiontype{VIII} line divided by the 2--5~keV continuum
at each annulus, as presented in figure~\ref{fig:sb}(b) by green diamonds.
Here, we subtracted the O\emissiontype{VIII} intensity
at the NGC~2992 field (red diamonds)
or the intensity at the outermost annulus (black diamonds),
as an ``offset'' due to the Galactic component.
The dashed lines represent the calculated ratio by the XSPEC simulation
fixing the O abundance of ICM to 0.25, 0.5, and 1 solar.
The ICM temperature at each annulus is assumed to have
the measured value in figure~\ref{fig:result}(a).
Actual value of the O abundance is supposed to lie
between the black and green diamonds.
The radial O abundance profile obtained in figure~\ref{fig:result}(b)
is mostly consistent with this plot.

\subsection{Central Cool Component of A~1060}
\label{subsec:central}

Chandra observation resolved two central elliptical galaxies
of A~1060, NGC~3311 and NGC~3309 \citep{yamasaki02}.
We estimated the flux of an additional cool component for the central 
galaxies with the {\it vapec} model of $kT=0.8$~keV and 0.5 solar abundance
at $z=0.0114$, and obtained an upper limit (90\% confidence level) to be 
$L_{\rm X}= 5.2\times 10^{40}$ erg~s$^{-1}$ (0.4--4.5~keV)
in the central region within $r<2'$\@.
In this fit, temperatures of the Galactic components (two {\it apec})
were fixed at 0.146 and 0.662~keV, 
and only the ICM component was allowed to be free.
This upper limit is consistent with the sum of the ISM flux
of the two elliptical galaxies by \citet{yamasaki02},
$L_{\rm X}= (1.8\pm 0.6)\times 10^{40}$ erg~s$^{-1}$ (0.4--4.5~keV)\@.

\section{Discussion}\label{sec:discuss}

\subsection{Temperature Profile}
\label{subsec:temp}

The temperature in A1060 observed with Suzaku shows a decline from
$\sim 3.4$~keV in the central region ($r\lesssim 5'\sim 70$~kpc) to
$\sim 2.2$~keV in the outermost annulus of
17--27$'$ $\sim 240$--380~kpc $\sim 0.16$--0.25$\;r_{180}$.
This feature is roughly consistent with the previous
result for the inner region ($r < 13'$) measured with XMM-Newton
\citep{hayakawa06}, 
although our Suzaku result indicate slight flattening of
the temperature profile in the outer annuli.
\citet{hayakawa06} suggested 
a faster decline of the temperature profile of A~1060
than the average $T/\langle T\rangle$ curve by \citet{markevitch98}
obtained from 30 nearby clusters (excluding A~1060) with ASCA observations.
However, now it appears to become consistent with them by taking
the emission-weighted average temperature, $k\langle T\rangle = 2.5$~keV\@.

Recent measurements of 13 nearby relaxed clusters (excluding A~1060)
with Chandra by \citet{Vikhlinin2005} showed that the temperature
reaches at a peak at $r\sim 0.15\;r_{180}$ and then declines to $\sim 0.5$
of its peak value at $r\simeq 0.5\;r_{180}$, in good agreement with
\citet{markevitch98}.
They also found that clusters whose temperature profiles peak at
$r<70$~kpc $\ll 0.15\; r_{180}$ (MKW4 and RX J1159+5531) shows 
a larger peak-temperature-ratio of $T_{\rm p}/\langle T\rangle\sim 1.35$.
This is exactly the case of A~1060, which shows the peak-temperature
of $kT_{\rm p}= 3.42\pm 0.04$~keV at $r = 4$--6$'\sim 70$~kpc
with our Suzaku measurement. Using the relation of
$T_{\rm p}/\langle T\rangle\sim 1.35$,
the emission-weighted average temperature
is calculated to be $k\langle T\rangle = 2.5$~keV,
as suggested in the previous paragraph.
These three correspond to clusters without cooling core,
hence this apparent steep temperature decline at $r\lesssim 0.2\;r_{180}$
might be a common feature for the non-cooling flow clusters.

In order to determine actual value of $k\langle T\rangle$ for A~1060,
further offset observations toward $r_{\rm 180}\sim 110'$ are required.
Since the wide and energy-dependent
point-spread function of ASCA tended to give a flatter temperature
profile for A~1060 \citep{tamura96,furusho01},
studies of the $T/\langle T\rangle$ curve in cluster outskirts
with Suzaku will be useful to look at the true temperature features.

\citet{Dolag2004} conducted a smoothed particle hydrodynamics (SPH)
simulation to reconstruct the structure of the Local Universe,
which well reproduces the position, mass, and temperature of A~1060
(``Hydra'' of table~1 in \cite{Yoshikawa2004}).
The simulated temperature of A~1060 was 3.4~keV within
$r<20'\sim 0.18\; r_{180}$.
The radial temperature profile of simulated A~1060 is presented
by \citet{Kawahara2006}, based on the updated version
of the SPH simulation by \citet{Dolag2005}.
It reproduces the shape of the temperature profile obtained
with our Suzaku observation pretty well, although the temperature
in the simulation is higher than our observation.
This is partly because our temperature profile in
figure~\ref{fig:result}(a) is plotted against the projected radius,
in which the ICM emission is integrated along the line of sight,
while the simulated temperature is plotted against the spherical radius.
It is pointed out that the spectroscopic temperature in the real X-ray
observation tend to give lower temperature than the emission-weighted
temperature calculated in the simulation due to the multi-phase nature
of the ICM not only from the radial profiles but also from the local
inhomogeneities \citep{Mazzotta2004,Kawahara2006}.

There is a small flat-top in the observed temperature profile
within $r\lesssim 5'\sim 70$~kpc in figure~\ref{fig:result}(a).
This feature might indicate an initial phase in the formation
of cooling core. The SPH simulation predicts that the period
of the last major merger occurred at an age of $\sim 3$~Gyr ago ($z\sim 0.3$).
This implies that the temperature flat-top has been made
within recent $\sim 3$~Gyr.
The thermal conduction length scale is calculated as
\begin{eqnarray}
\nonumber
r_{\rm cond} = 155\;
        \biggl(\frac{t_{\rm cond}}{\rm 3\;Gyr} \biggr)^\frac{1}{2}
        \!\biggl(\frac{n_{\rm e}}{\rm 0.01\;cm^{-3}} \biggr)^{-\frac{1}{2}}
        \!\biggl(\frac{kT}{\rm 3\;keV}\biggr)^\frac{5}{4}
        \;\makebox{kpc},
\end{eqnarray}
where $t_{\rm cond}$ is the thermal conduction time,
$n_{\rm e}$ is the electron density,
assuming the classical Spitzer conductivity.
This length scale is similar to the size of the flat-top.

In subsection \ref{subsec:central}, only an upper limit of $5\times 10^{40}$
erg~s$^{-1}$ for the cool component in the ICM was obtained.  It
confirms that the sharp cusp structure in the Chandra image of A1060
\citep{hayakawa04} really reflects the gravitational mass
concentration and does not come from cool gas of high emissivity.

\begin{figure*}[t]
\begin{minipage}{0.46\textwidth}
\centerline{
\FigureFile(\textwidth,\textwidth){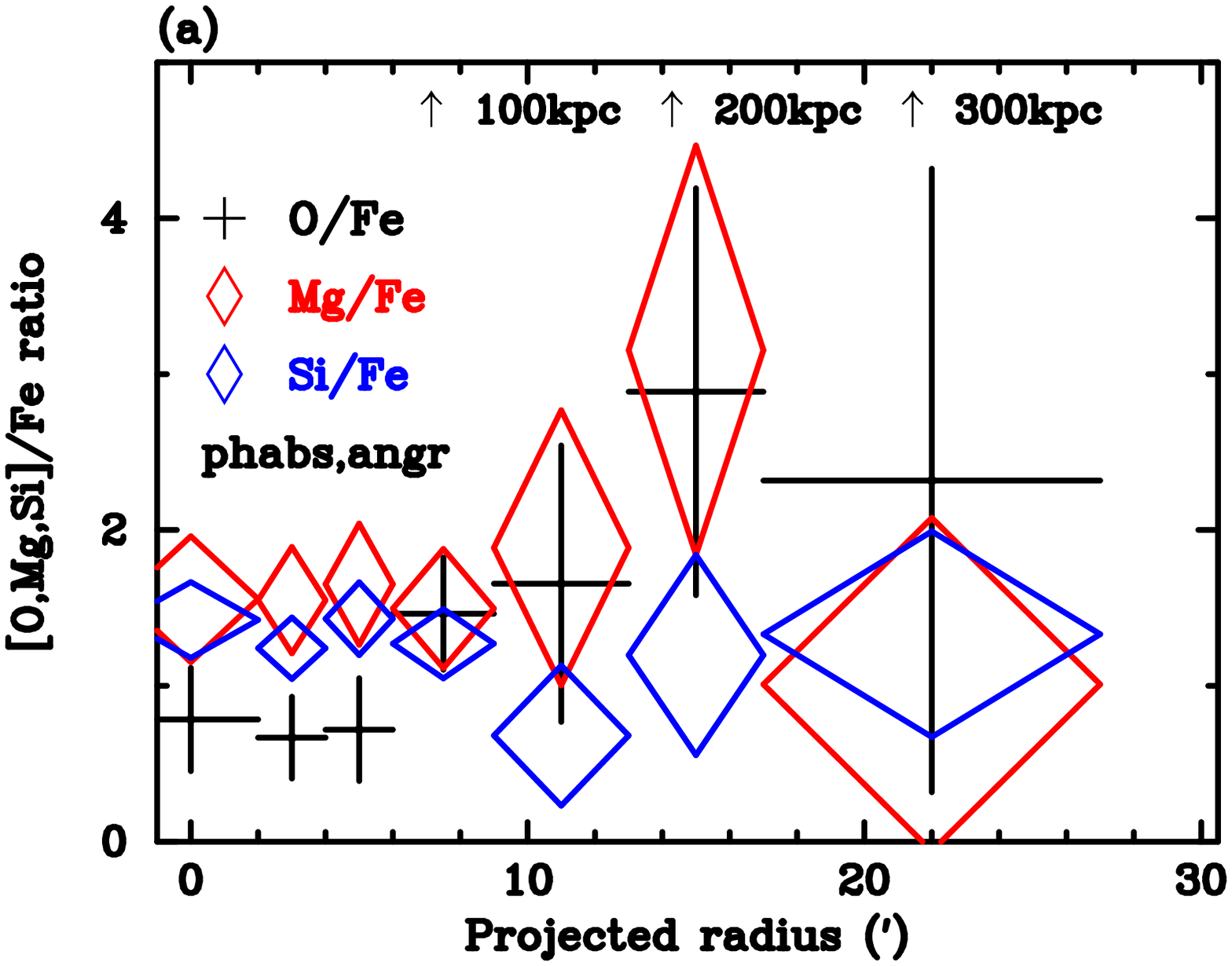}}
\end{minipage}\hfill
\begin{minipage}{0.52\textwidth}
\centerline{
\FigureFile(\textwidth,\textwidth){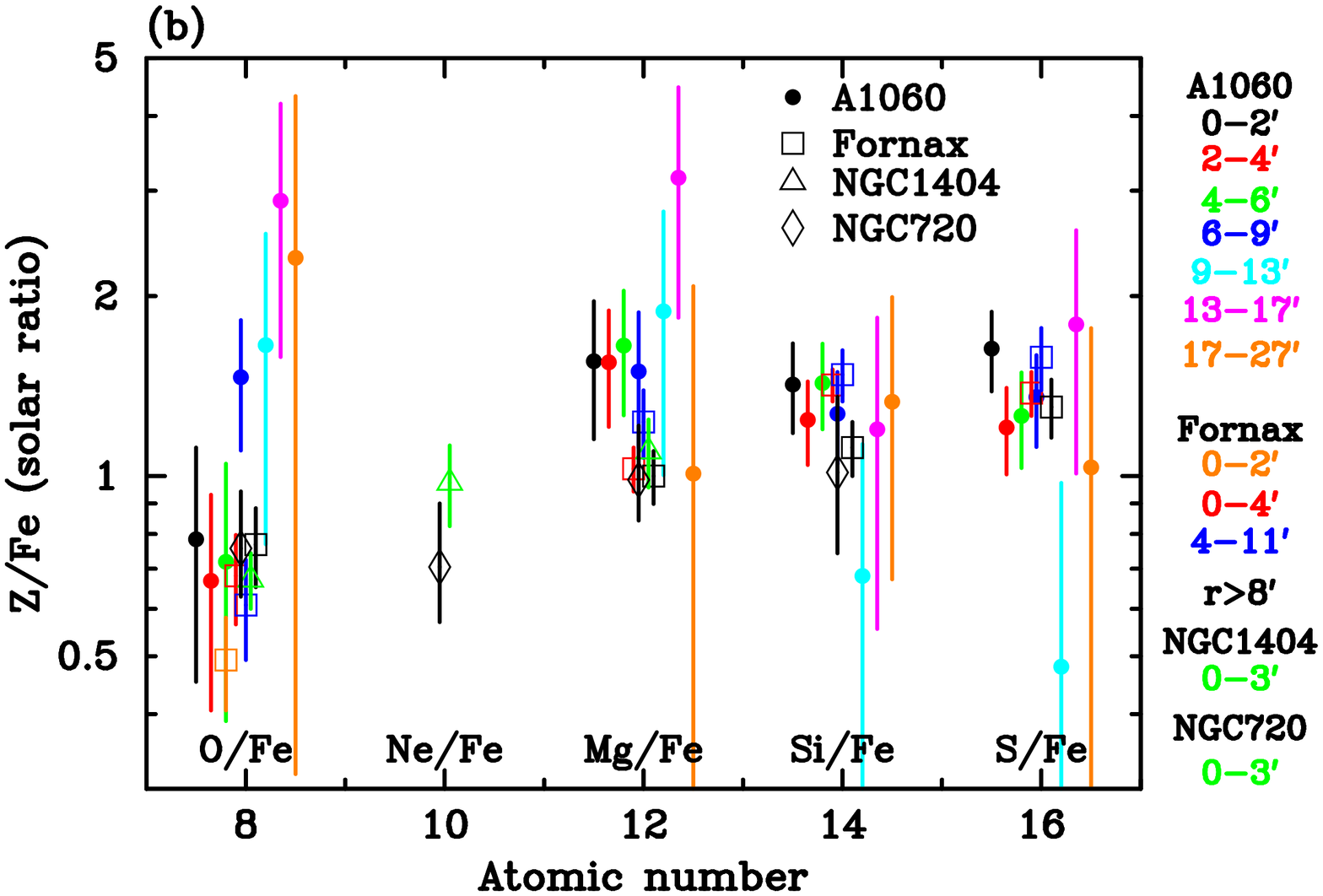}}
\end{minipage}
\caption{
(a) Abundance ratios of O (black), Mg (red) and Si (blue) divided by Fe.
(b) Abundance ratios of O, Ne, Mg, Si, and S, divided by Fe
for Fornax cluster, NGC~1404 \citep{matsushita06a},
NGC~720 \citep{tawara06}, and A~1060.
In both plots, the solar abundance ratio of {\it angr} \citep{anders89}
is assumed, and only the statistical errors are plotted.
}\label{fig:ratio}
\end{figure*}

\begin{figure*}[t]
\begin{minipage}{0.32\textwidth}
\centerline{\FigureFile(\textwidth,\textwidth){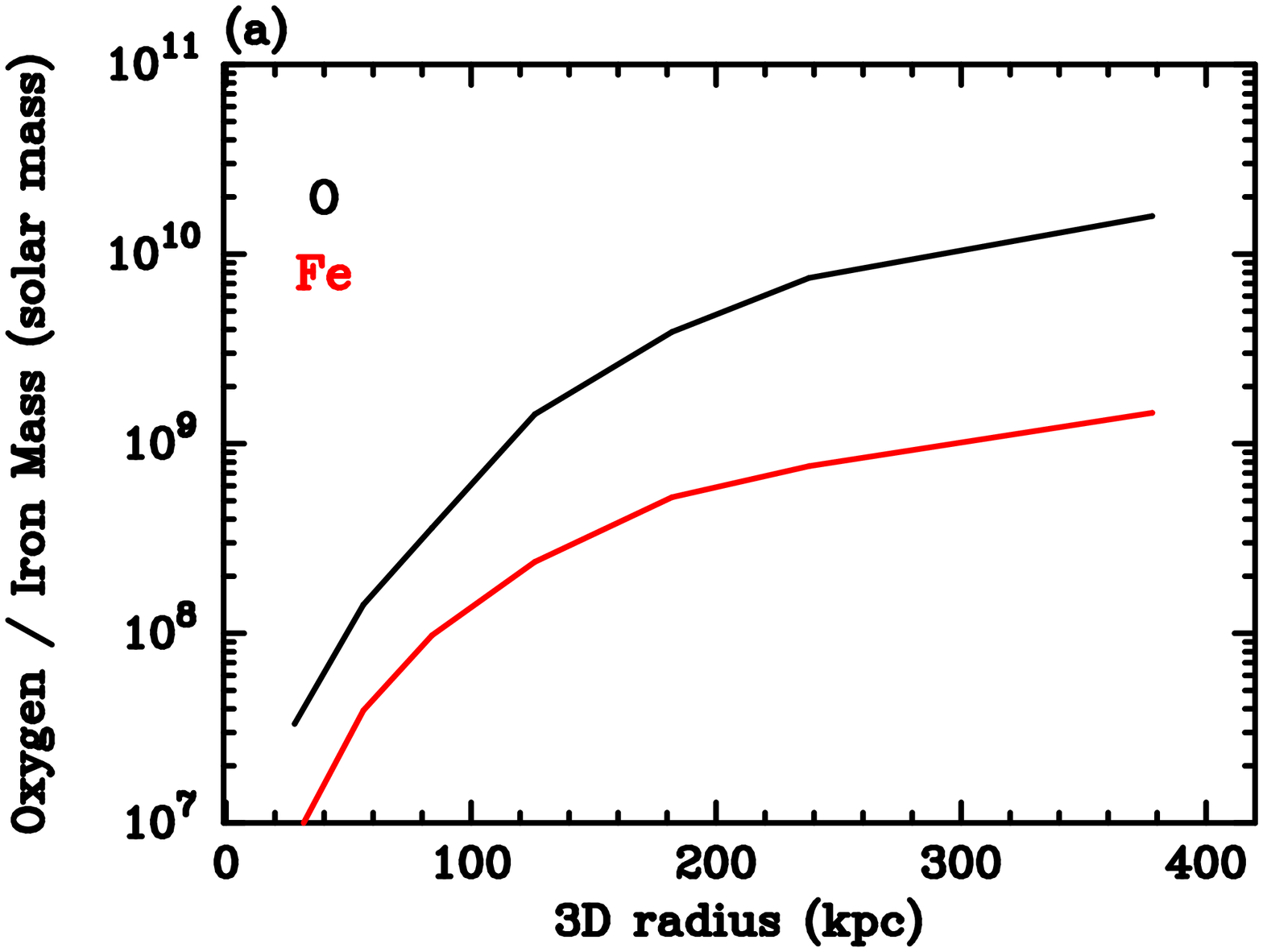}}
\end{minipage}\hfill
\begin{minipage}{0.32\textwidth}
\centerline{\FigureFile(\textwidth,\textwidth){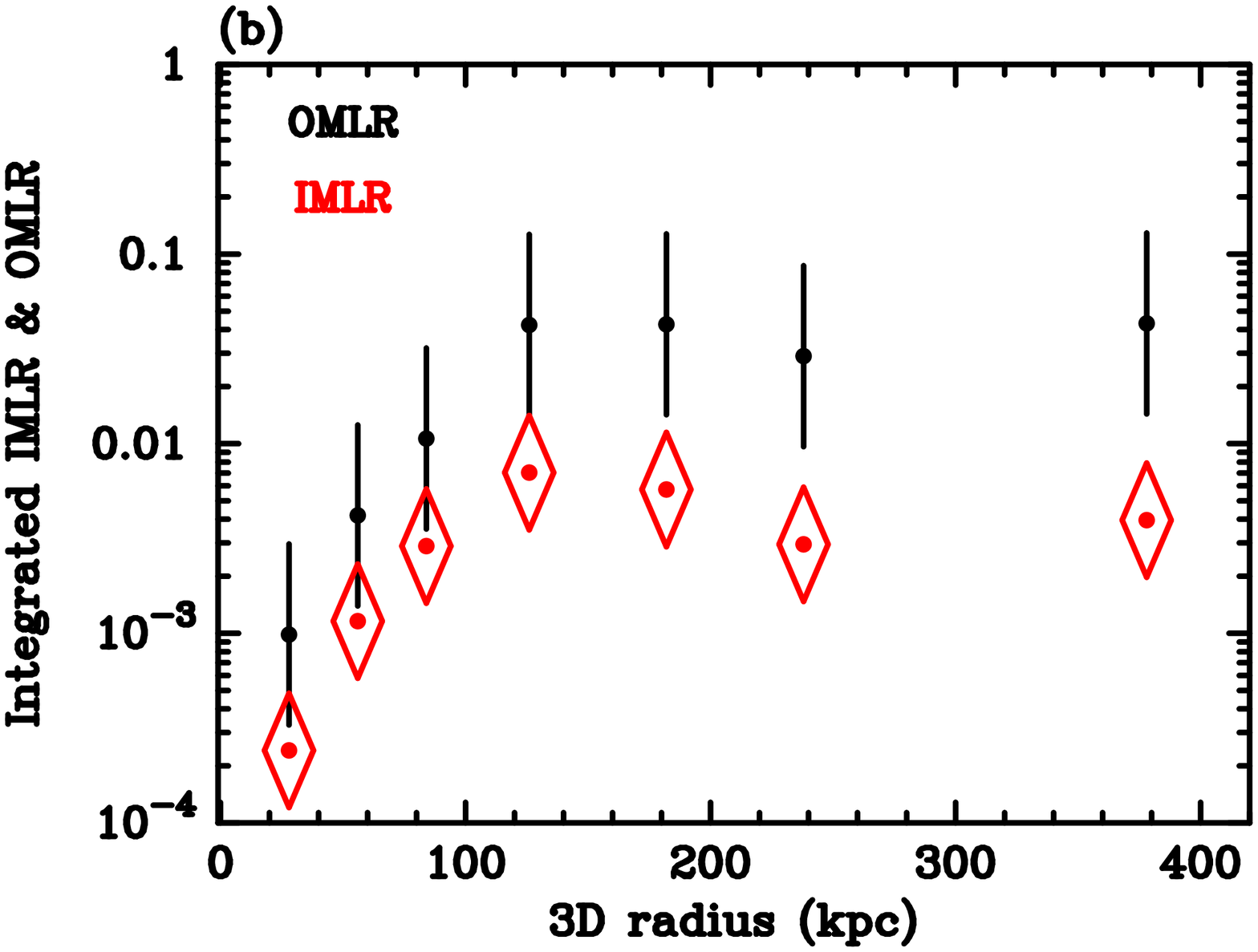}}
\end{minipage}\hfill
\begin{minipage}{0.32\textwidth}
\centerline{\FigureFile(\textwidth,\textwidth){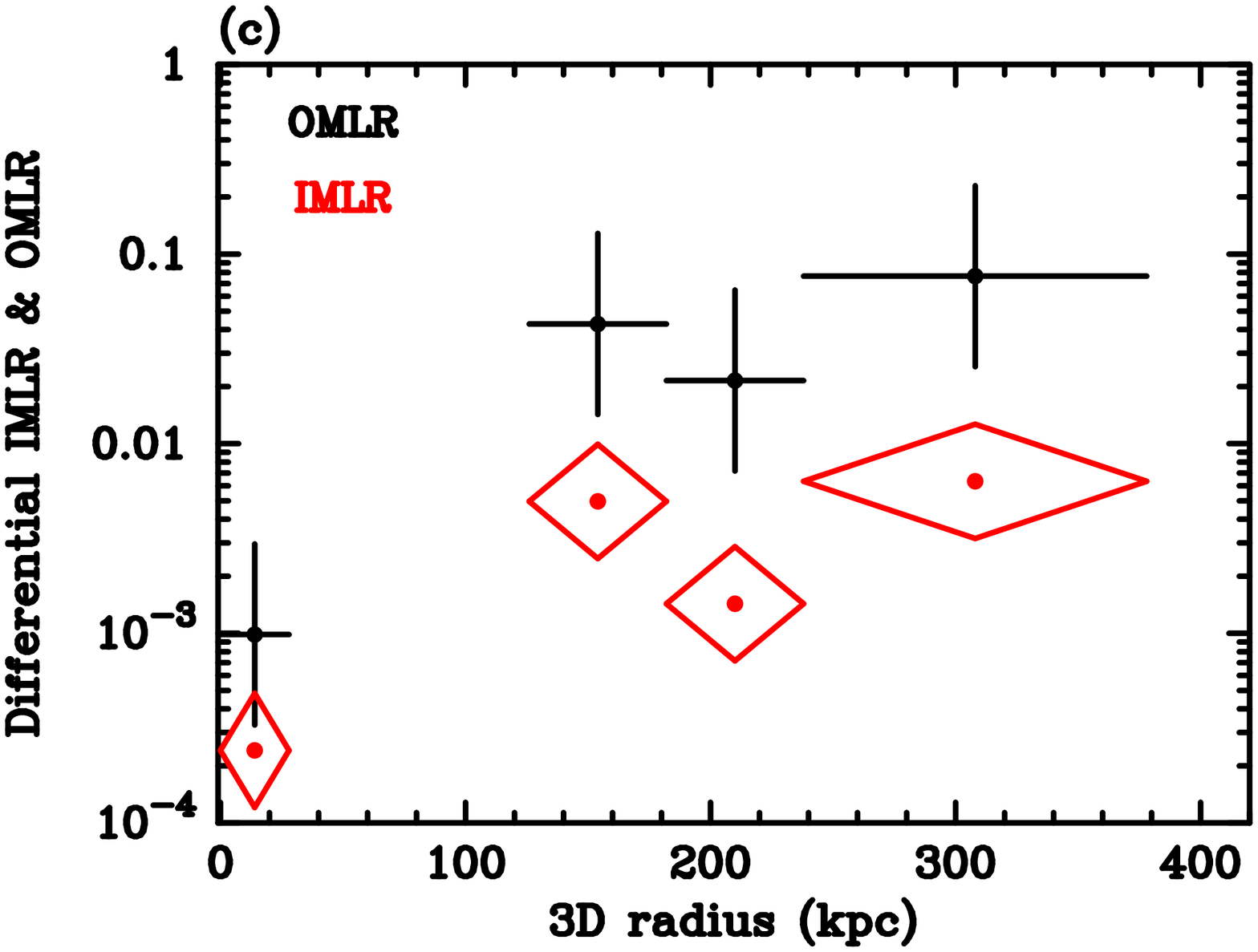}}
\end{minipage}\hfill
\caption{
(a) Cumulative Fe (red) and O (black) mass profiles of A~1060.
(b) Integrated iron mass-to-light ratio (IMLR) and
oxygen mass-to-light ratio (OMLR) in unit of
$M_\odot/L_\odot$ in the blue band.
(c) Differential IMLR and OMLR calculated at each annulus.
The annular range of 2--9$'$ are not plotted
because there are no member galaxies cataloged by \citet{christlein03}.
See text subsection \ref{subsec:omlr} for details.
}\label{fig:omlr}
\end{figure*}

\begin{table*}
\caption{Comparison of IMLR and OMLR with other systems.
}\label{tab:omlr}
\begin{center}
\begin{tabular}{lrrrrl}
\hline\hline
\hspace*{12em} & IMLR & OMLR & \multicolumn{1}{c}{$< r$} & \multicolumn{1}{c}{$kT$} & Reference \\
\hline
NGC~720 $\dotfill$
        & $1\times 10^{-4}$ & $4\times 10^{-4}$ & 25~kpc & $\sim 0.56$ keV &
        \citet{tawara06} \\
Fornax  $\dotfill$
        & $4\times 10^{-4}$ & $2\times 10^{-3}$ & 130~kpc & $\sim 1.3$ keV & 
        \citet{matsushita06a} \\
Centaurus $\dotfill$
        & $4\times 10^{-3}$ & $3\times 10^{-2}$ & 190~kpc & $\sim 4$ keV &
        \citet{matsushita06b} \\
A~1060  $\dotfill$
        & $4.0\times 10^{-3}$ & $4.3\times 10^{-2}$ & 380~kpc & $\sim 3$ keV &
        This work \\
\hline
\end{tabular}
\end{center}
\end{table*}

\subsection{Abundance Profiles}
\label{subsec:abun}

In this paper, we obtained emissivity weighted abundance profiles of
Ni, Fe, Si, S, Mg, Ne and O in the ICM of A 1060 up to a radius of
$27'\simeq 380$~kpc, although Ni and Ne abundances are not reliable
due to the strong and complex Fe-L line emissions.
Abundances of Si, S, and Fe decrease from $\sim 0.7$,
0.8 and 0.5 solar in the central region, to $\sim 0.3$, 0.2 and 0.2 solar,
respectively, in the outskirts of the cluster.
\citet{degrandi01} showed that non-cooling flow clusters
do not exhibit a steep abundance gradient due possibly
to disruption of the central cool cores after major merger events.
A~1060 shows clear gradients for Fe, Si, and S,
although it is a very relaxed system without the cD galaxy and cooling core.

The other elements, Mg and O, show somewhat flatter radial distributions.
The measured Mg abundance is consistent with constant up to the 13--17$'$
annulus, namely out to 240~kpc from the cluster center.
Oxygen also exhibits a flat distribution or a slight increase at $r\simeq10'$.
We have to note that the O abundances depend on the estimation of
the Galactic component as described in subsection \ref{subsec:galactic}.
The O abundance in the outer annuli is particularly affected by the surface
brightness and temperature of the cooler Galactic component
($kT_1\sim 0.15$~keV)\@. The O and Mg abundances at $r\lesssim 6'$
are also affected by the uncertainties in the OBF contaminant
(subsection \ref{subsec:radial}).
We have tried to take into account all the possible systematic errors,
and the results are presented in figure~\ref{fig:result}.
In addition, systematic effects caused by the definition of the solar
abundances were addressed and described in appendix~\ref{app:lodd}.

In order to compare the relative variation in the abundance profiles,
we show abundance ratios of O, Mg, and Si divided by Fe
as a function of projected radius in figure~\ref{fig:ratio}(a).
Apparently, the profiles of Si/Fe, O/Fe, and Mg/Fe show different gradients.
The Si/Fe ratio is consistent with a constant value,
while the O/Fe and Mg/Fe ratios seem to increase with radius
excluding the outermost annulus (17--27$'$).
This trend is similar when the O, Ne, and Mg abundances
are constrained to have the same values in different solar units
in figure~\ref{fig:lodd-onemg}(c).

Recent Suzaku observations also showed the O and Mg abundances in
poorer systems: an elliptical galaxy NGC~720 \citep{tawara06},
the Fornax cluster, and NGC~1404 \citep{matsushita06a}.
The Fe ratios of elements for these systems are compared in
figure~\ref{fig:ratio}(b).
The ratios Si/Fe and S/Fe are almost the same with A~1060 in other systems.
Somehow, the Mg/Fe ratio in A~1060 is $\sim 50$\% higher
than those in other systems.
However, if one considers the influence of the OBF contaminant,
this ratio can drop and the difference becomes insignificant.
Abundance ratio of O/Fe shows almost the same values with A~1060
at their centers, and in the outer regions A~1060 indicates
a higher value than others.

This peculiar behavior of the O/Fe ratio has been reported by
\citet{tamura04} with XMM-Newton as an average of 19 clusters
which are mainly composed of X-ray bright and relaxed clusters
with a cD galaxy (excluding A~1060).
The O/Fe ratio is $\sim 0.7$ solar within $r\lesssim 50$~kpc,
which increases to $\sim 1.5$ solar at $r\sim 100$~kpc,
and more at $r\gtrsim 200$~kpc, in figure~5(b) of \citet{tamura04},
assuming the solar abundance ratio of {\it angr} \citep{anders89}.
This is much alike the O/Fe plot in figure~\ref{fig:ratio}(a).

Since these differences are not simply explained by the ion mass,
we need to consider that the enrichment processes have a
significant difference between O and Fe, for example.
It is generally considered that enrichment of O and other SN~II
originated metals has occurred in the early stage of cluster formation,
certainly before the last merger epoch.
The relatively flat distributions of O and Mg indicate
that the early metal enrichment has caused these features.
One plausible explanation is the enrichment in the form of
starburst-driven galactic winds (e.g.\ \cite{strickland00}).
Recent numerical simulation of the cluster metal enrichment indicates
that even a metallicity peak in an intermediate radius can be created
by enrichment through galactic winds (Kepferer et al.\ 2006, priv.\ comm.),
which resembles our oxygen feature.

The other possibility is the early metal enrichment
by massive Population III stars (e.g.\ \cite{Matteucci2006})
In this case, a large fraction of the intergalactic space
would be enriched with metals.
One big riddle in the obtained abundance profiles is that
the Mg abundance at the central region appears to be
significantly higher than the O abundance,
although both O and Mg are mainly produced by SN~II\@.
\citet{Loewenstein2001} suggests that preenrichment by
a generation of massive Population III stars may account for
the abundance anomalies of O, Si, and Fe observed in the ICM\@.
However, \citet{Yoshida2004} concluded that metals originating from
the earliest generation of stars cannot be responsible for the
observed abundance anomalies in the ICM\@.
Future quantitative studies of metal
abundances in the outer regions ($\sim r_{\rm 180}$) or even outside
of clusters will be important to locate the origin of
metals. Instruments with much higher energy resolution such as
microcalorimeters will be desired for these studies.

\subsection{IMLR and OMLR}\label{subsec:omlr}

Combining the X-ray luminous gas mass profile by \citet{hayakawa06}
with XMM-Newton and the abundance profiles obtained with Suzaku,
we calculated the cumulative iron or oxygen mass profiles
in figure~\ref{fig:omlr}(a). Strictly speaking, the abundance
profile is derived from the projected spectrum along the line of sight,
we approximated it to the spherical distribution.
The iron and oxygen mass within $r\lesssim 380$~kpc is calculated
to be $\sim 1.4\times 10^9\; M_\odot$ and $1.6\times 10^{10}\; M_\odot$,
respectively.

Based on the member galaxy catalog
(69 galaxies within the projected radius, $r < 27'\sim 380$~kpc)
by \citet{christlein03},
we also calculated the iron mass-to-light ratio (IMLR) and
oxygen mass-to-light ratio (OMLR) in figure~\ref{fig:omlr}(b) and (c).
We utilized the observed redshift to estimated the 3-dimensional
distribution of the galaxy, assuming the spherical symmetry.
The IMLR and OMLR are calculated to be
$\sim 4.0\times10^{-3}\; M_{\odot}/L_{\odot}$ and
$4.3\times10^{-2}\; M_{\odot}/L_{\odot}$ within $r\lesssim 380$~kpc.
Both are subjected to large errors,
and we tentatively adopted factors of two and three, respectively,
in figure~\ref{fig:omlr}.

We summarize recent measurements of IMLR and OMLR in table~\ref{tab:omlr}.
NGC~720 and the Centaurus cluster exhibit similar values with A~1060
when compared at the same radius, while the Fornax cluster shows
significantly smaller values by an order of magnitude.
It is suggested that the smaller systems with lower gas temperature
tend to show lower IMLR in \citet{makishima01},
however its deviation is outstanding.

Both the integrated IMLR and OMLR steeply increase up to $\sim 100$~kpc
and seem to reach almost maximum at 100--200~kpc.
This trend is similar to the IMLR obtained with XMM-Newton
for M~87 and the Centaurus cluster by \citet{matsushita06b}
within $r\lesssim 100$~kpc.
The steep increase in the $r\lesssim 100$~kpc region
suggests that the Fe and O ions which was synthesized in
the central galaxies have diffused to the ICM\@.
On the other hand, as pointed out in \citet{ezawa97},
abundance gradient over a few hundred kpc scale
in clusters of galaxies should follow the mass ratio
between the galaxies and the ICM gas.
This is because heavy ions released from galaxies without
much kinetic energy would not diffuse out more than 10~kpc
over the Hubble time.
\citep{Bohringer2004} indicated that it takes about $10^{10}$~yr
to synthesize Fe mass within $r\lesssim 100$~kpc range.

These considerations suggest that both Fe and O enriched gas
was released to the ICM space with significant kinetic energy
and/or the gas stripping was efficiently occurred in
the $\sim 100$~kpc range due to the galactic motion.
We also note that two central galaxies (NGC~3311 and NGC~3309)
of A~1060 are slightly off-center ($\sim 0.3'$ and $2'$)
from the X-ray peak of the cluster (see figure~1(b) of \cite{hayakawa06}),
and the line of sight velocity is calculated to be 180~km~s$^{-1}$
and 660~km~s$^{-1}$. Ram-pressure stripping is also likely to be
the reason why these central galaxies show very compact X-ray halo
\citep{yamasaki02}.

The obtained O mass in figure~\ref{fig:omlr}(a) provides
the rough estimate of the SN~II rate.
Theoretical calculations \citep{Nomoto2006,Thielemann1996,Tsujimoto1995}
predict the O product of $\sim 1.5\;M_\odot$ for $20\;M_\odot$
progenitor mass and $\sim 9\;M_\odot$ for $40\;M_\odot$ progenitor.
This requires at least $\sim 3\times 10^{9}$ SN~II in the whole cluster
since the formation.

\subsection{Warm Component in the Galactic Emission}
\label{subsec:warm}

As shown in figure~\ref{fig:sb}(a), the association of
O\emissiontype{VIII} line and the ICM is evident.  The surface
brightness of O\emissiontype{VII} lines are brighter than the level of
NGC~2992 region or than the average of wider sky region
\citep{mccammon02}, but consistent with the uniform distribution over
the cluster. A large fraction of O\emissiontype{VII} photons is
considered to be originated from the Milky Way Halo, whose typical
temperature measured with XMM-Newton is $kT\sim$ 0.2 keV
\citep{lumb02}.  If there is a cooler component like Local Hot Bubble,
$kT \sim$ 0.08~keV by \citet{lumb02,snowden98}, or
Warm-Hot Intergalactic Medium, it is hard to detect it under the
strong emission from the ICM\@.  Our spectral fit showed that a
$kT\sim 0.7$ keV component was required to reproduce the spectral
structure around the Fe L-line complex.  The surface brightness of
this component shows a flat distribution in our field of view
(subsection \ref{subsec:galactic}). The ROSAT 3/4 keV image shows an
enhancement with an angular extent of $\sim 10 \times 20$ deg$^{2}$,
which may be an extension of the North Polar Spur \citep{snowden97}.
These features suggest that the $\sim 0.7$~keV component is a Galactic
origin, but a further study is certainly needed and it is beyond the
scope of this paper.

\bigskip

Authors are greatful to K. Dolag for supplying the simulation data 
and S. Sasaki for discussions.  We also thank the referee for providing 
valuable comments. 
Part of this work was financially supported by the Ministry of
Education, Culture, Sports, Science and Technology of Japan,
Grant-in-Aid for Scientific Research No.\ 14079103, 15340088, 15001002.

\appendix

\section{Fraction from Corresponding Sky}
\label{app:frac}

\begin{figure*}
\begin{minipage}{0.25\textwidth}
\FigureFile(\textwidth,\textwidth){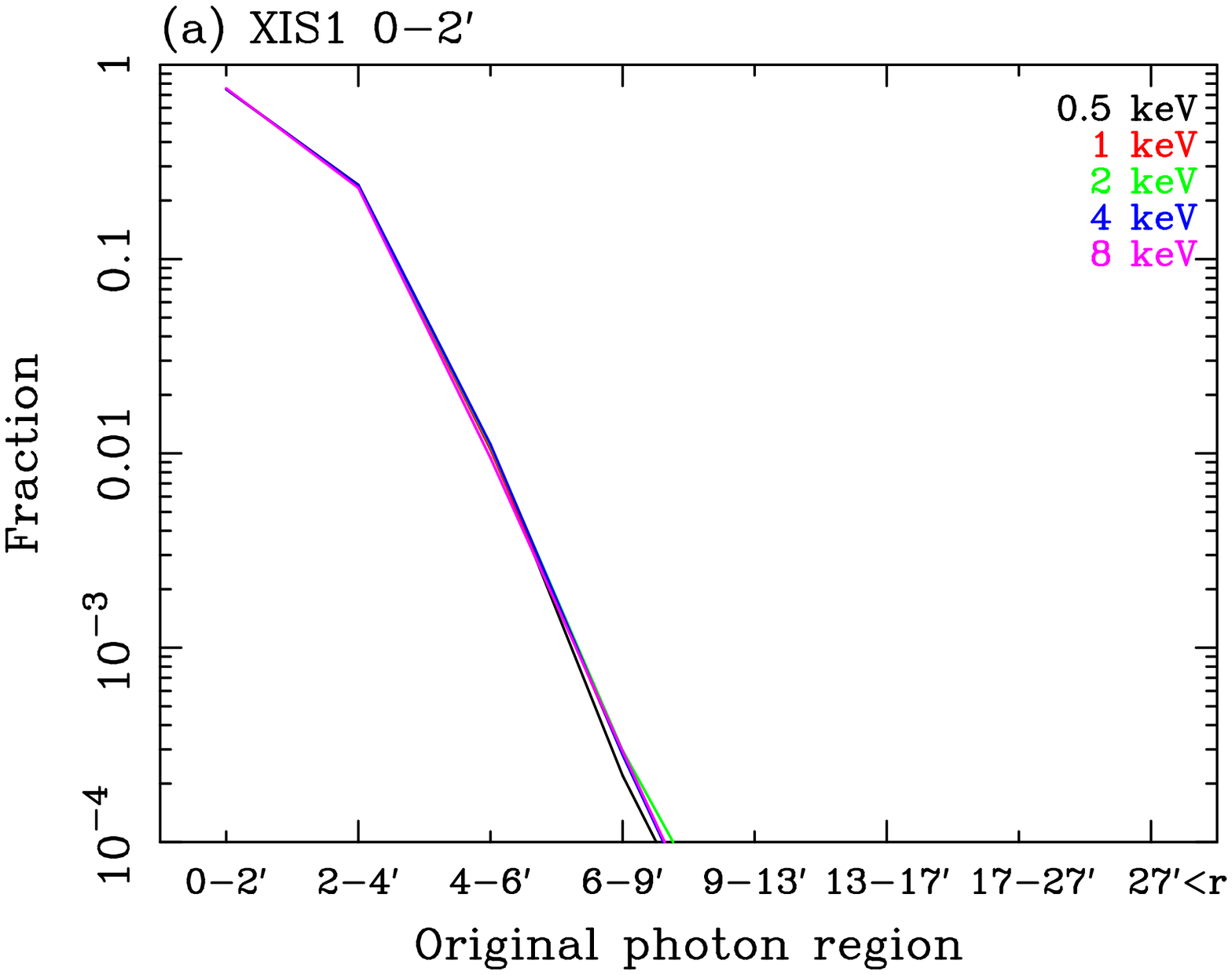}
\end{minipage}\hfill
\begin{minipage}{0.25\textwidth}
\FigureFile(\textwidth,\textwidth){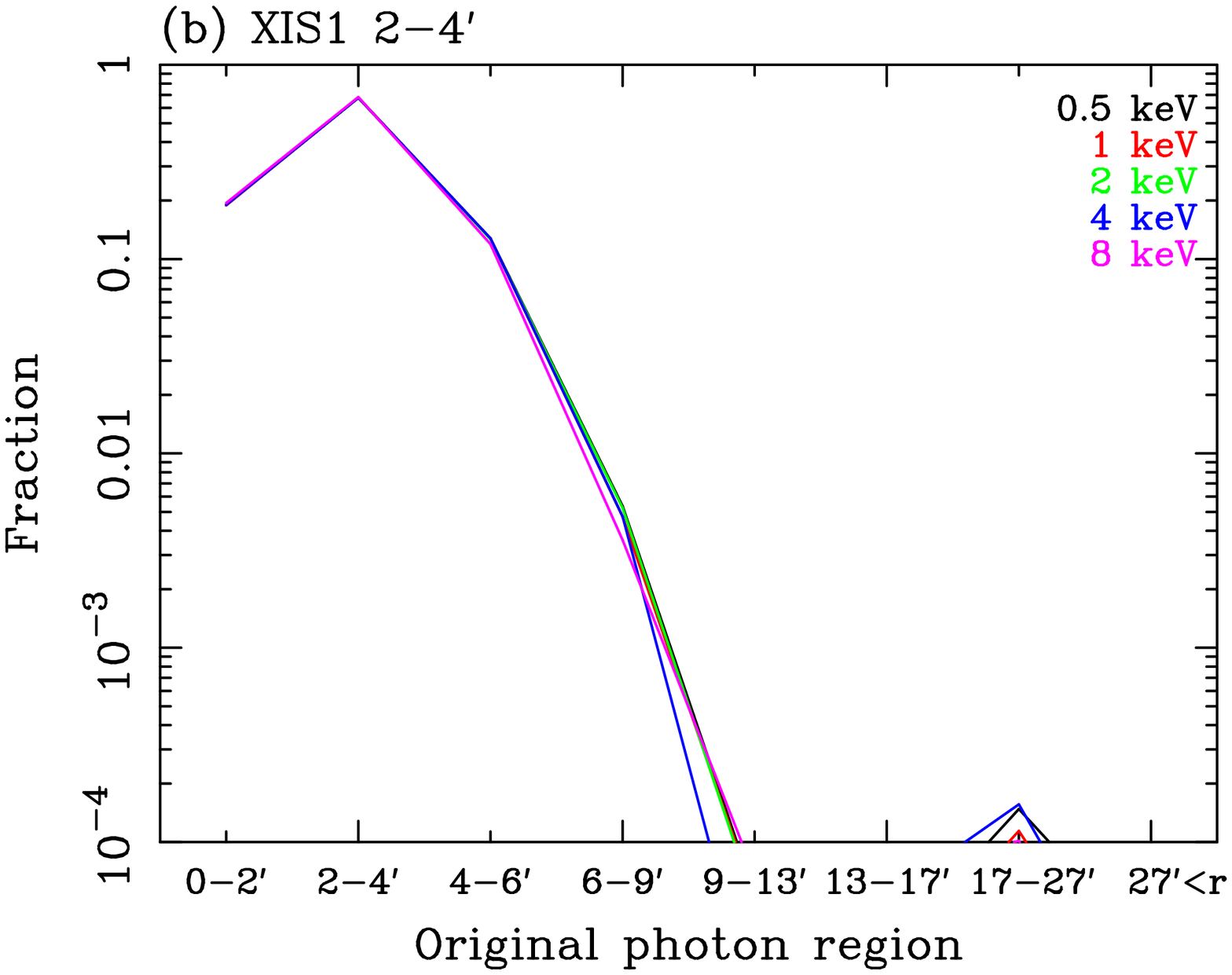}
\end{minipage}\hfill
\begin{minipage}{0.25\textwidth}
\FigureFile(\textwidth,\textwidth){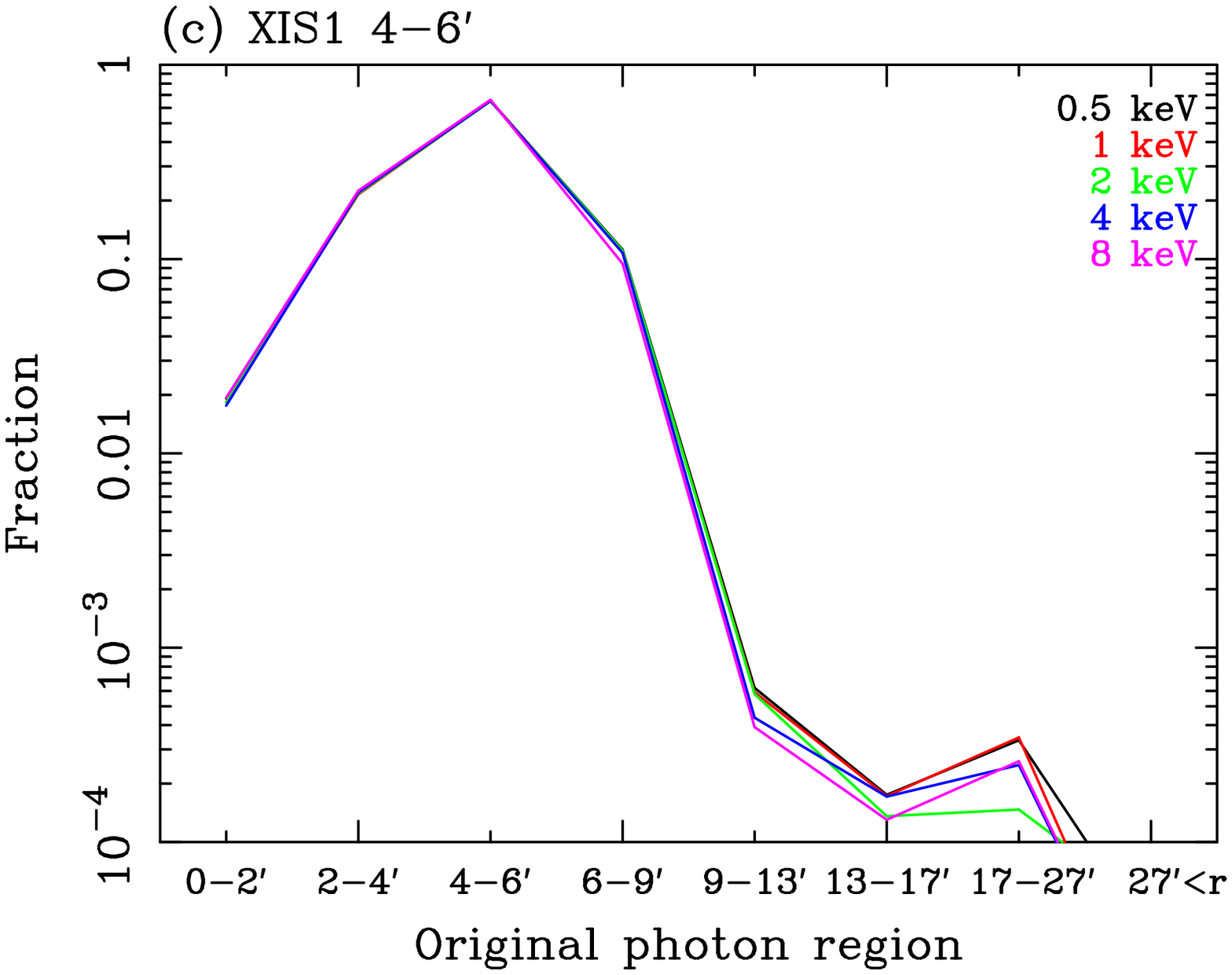}
\end{minipage}\hfill
\begin{minipage}{0.25\textwidth}
\FigureFile(\textwidth,\textwidth){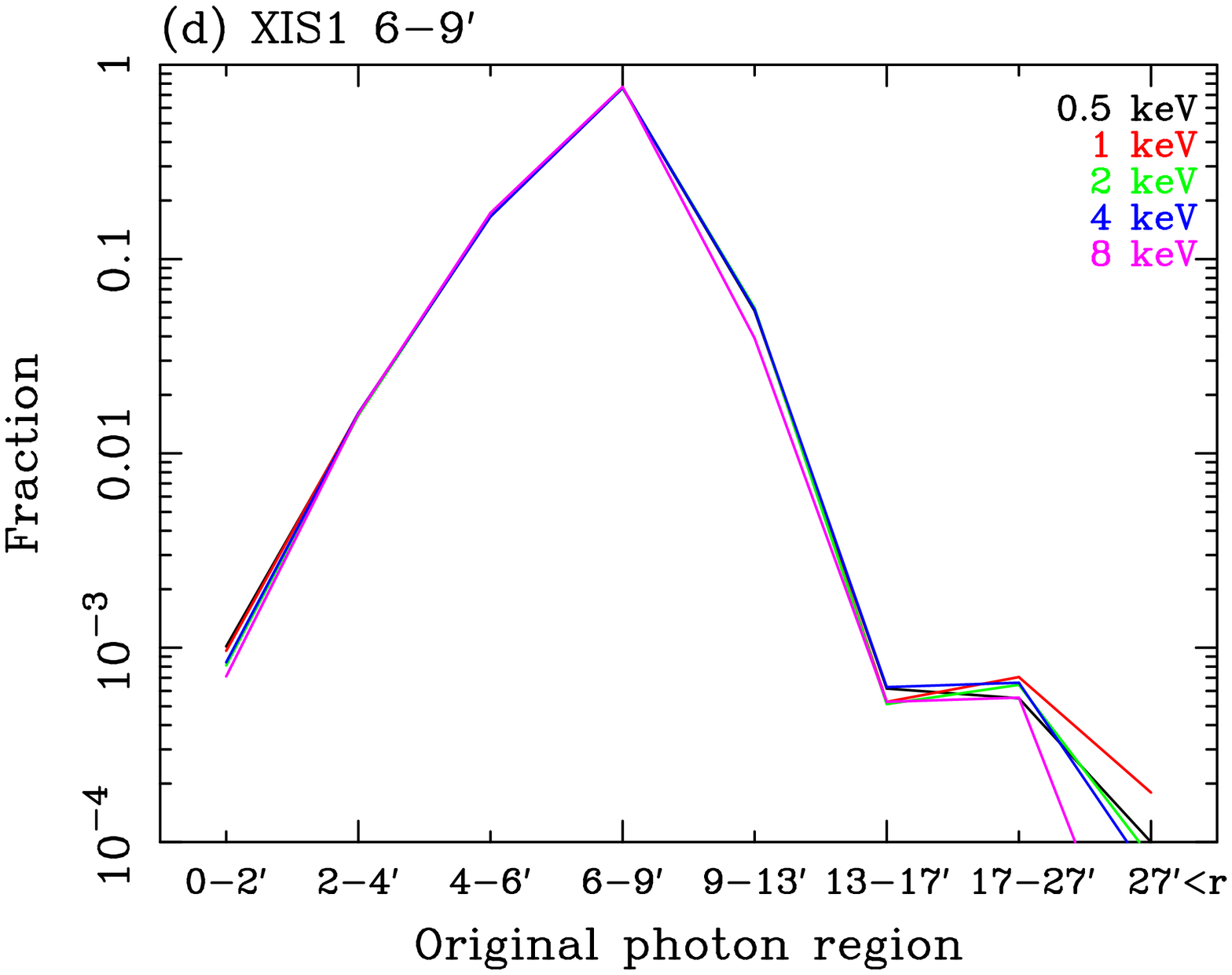}
\end{minipage}

\begin{minipage}{0.25\textwidth}
\FigureFile(\textwidth,\textwidth){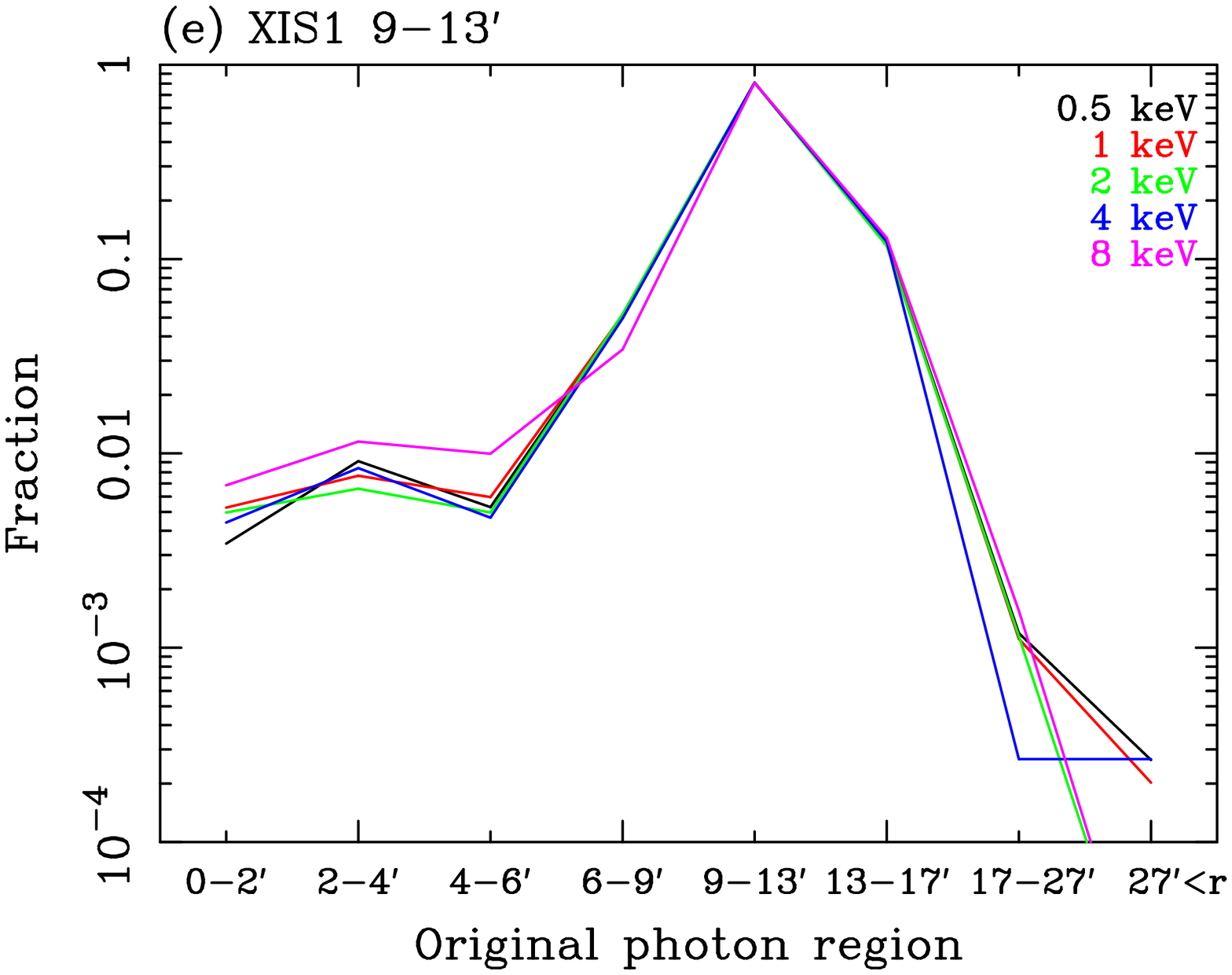}
\end{minipage}\hfill
\begin{minipage}{0.25\textwidth}
\FigureFile(\textwidth,\textwidth){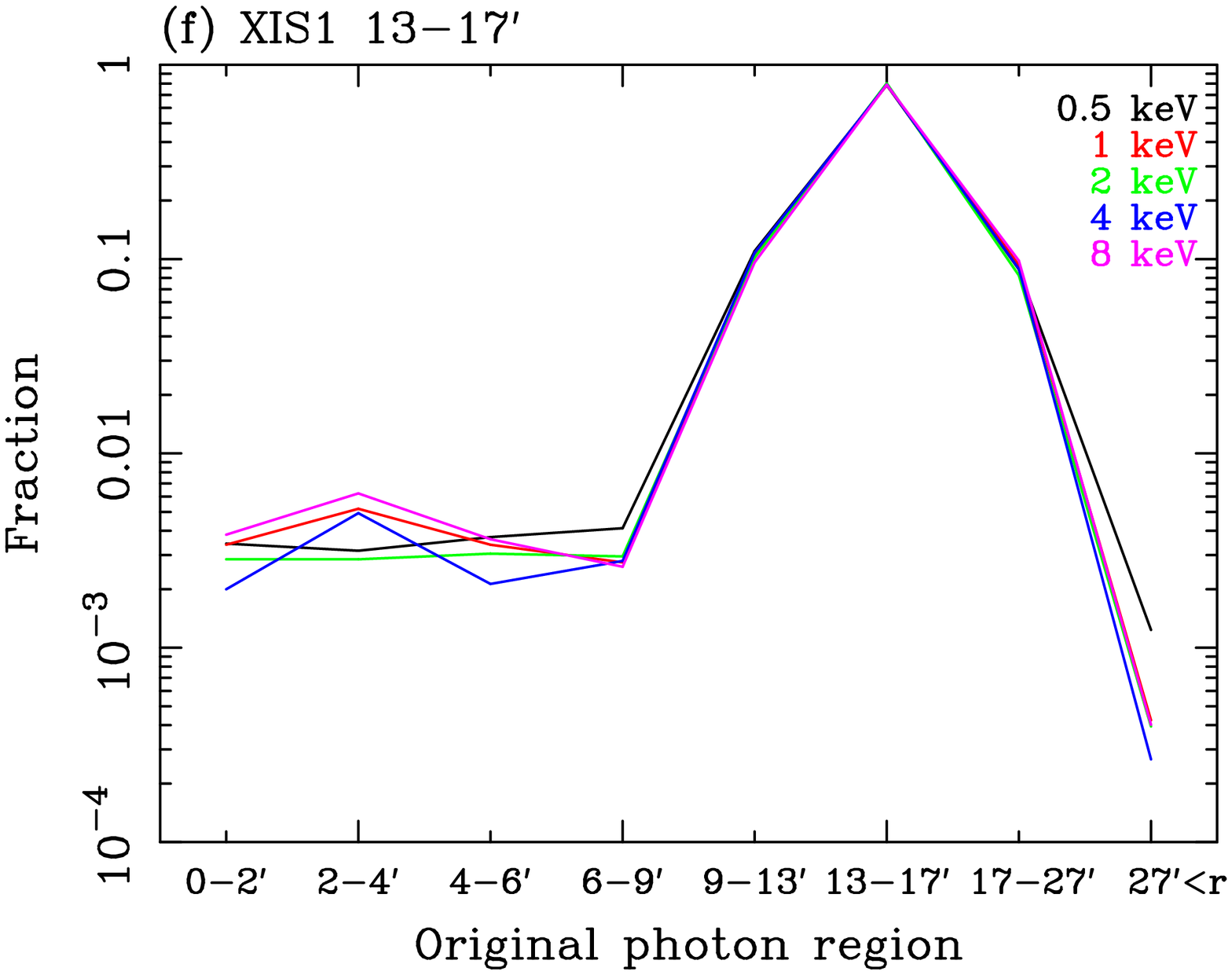}
\end{minipage}\hfill
\begin{minipage}{0.25\textwidth}
\FigureFile(\textwidth,\textwidth){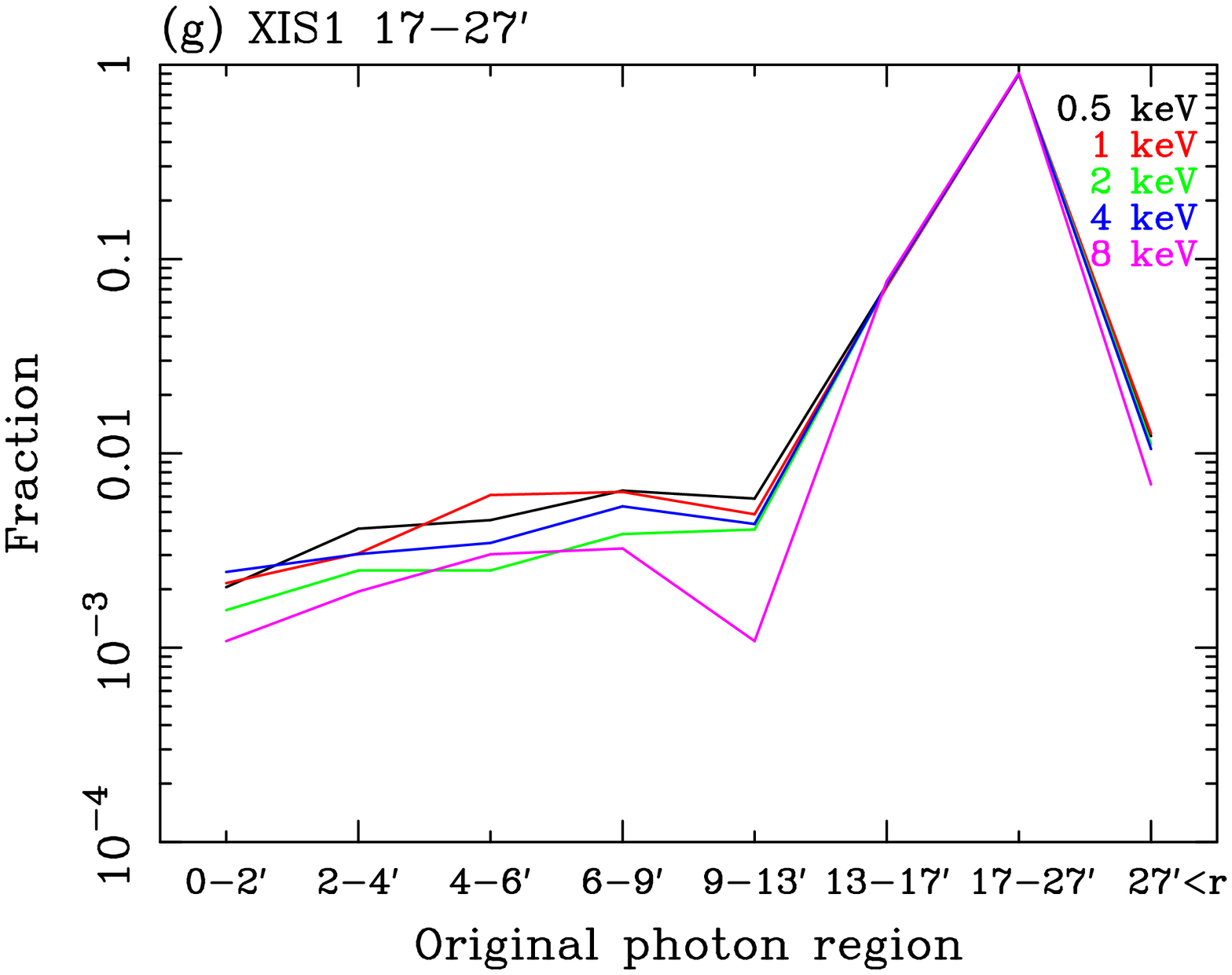}
\end{minipage}\hfill
\begin{minipage}{0.25\textwidth}
\FigureFile(\textwidth,\textwidth){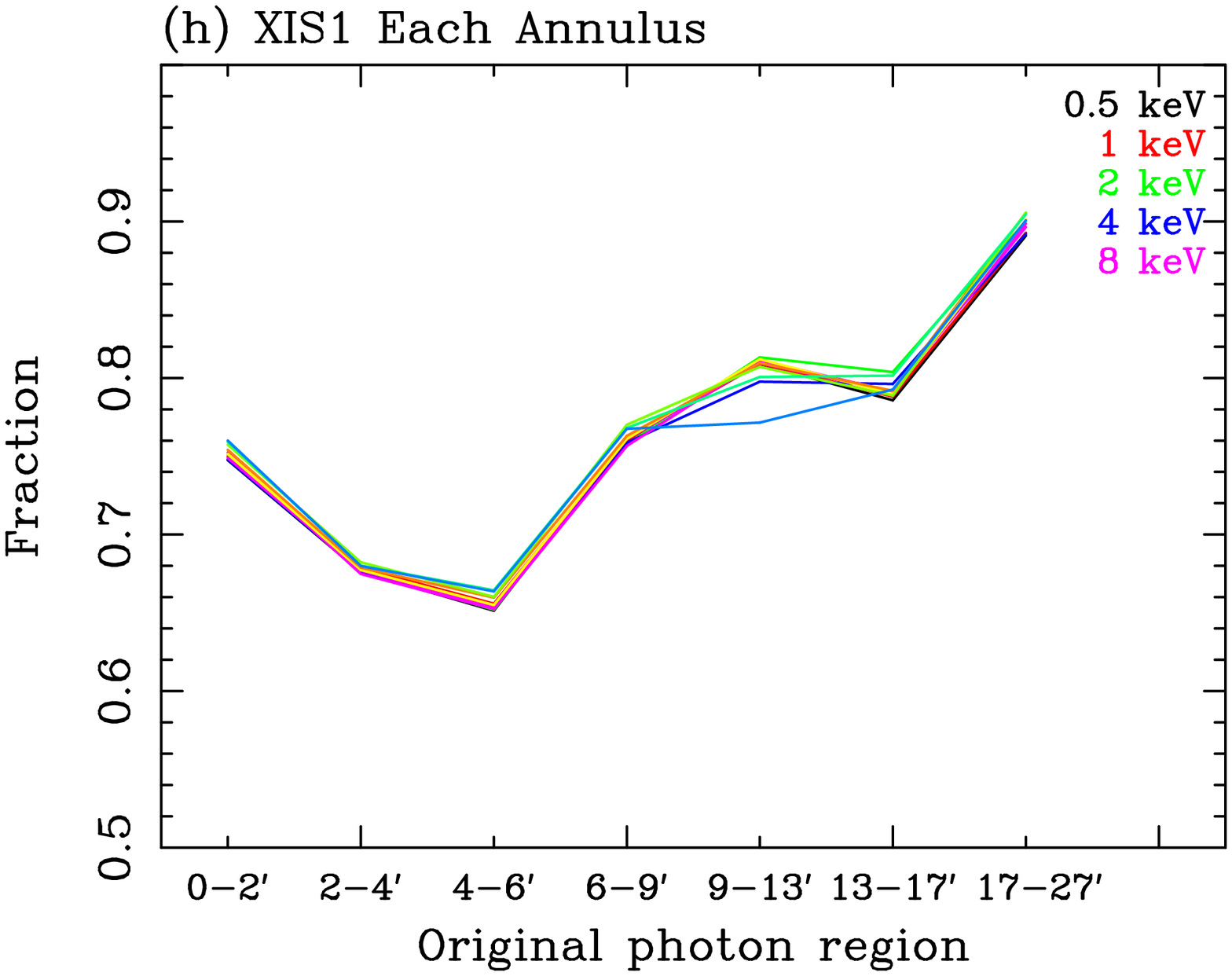}
\end{minipage}
\caption{
(a)--(g) Fraction of photons detected in the
0--2$'$ to 17--27$'$ extraction regions on the BI (XIS1) sensor
plotted against the original sky directions of incidence,
which is estimated by the ``xissim'' simulation.
Different energies are plotted in different colors.
(h)~Photon fraction of each annulus
coming from the corresponding sky region,
i.e., maximum values of panels (a)--(g) are plotted.
}\label{fig:frac}
\end{figure*}

Since angular resolution of X-Ray Telescopes (XRT; \cite{serlemitsos06})
of Suzaku ($\sim 2'$ in half power diameter; HPD) is not as good as
those of Chandra ($\sim 0.5''$) and XMM-Newton ($\sim 15''$),
fraction of photons may become significant
that come from outside of the sky direction
corresponding to the extraction region of spectra on the XIS detector plane.
This was a severe problem for the ASCA GIS
\citep{ohashi96,makishima96}, which employs similar type of
thin-foil-nested reflectors for the XRT \citep{serlemitsos95}
with $\sim 3.6'$ HPD,
and the fraction sometimes became more than half of total detected
photons as seen in figure~4 of \citet{kikuchi99}
for the MKW~3s and 2A~0335+096 clusters.
It made the cluster analysis further complicated that
energy dependence of the fraction could not be neglected
with the ASCA XRT/GIS system.

The situation has been much improved for Suzaku
than ASCA\@. The Suzaku XRT/XIS system adopts
longer focal length of 4.75~m than ASCA (3.5~m)
and the sky coverage of the XIS detector ($18'\times 18'$) is
smaller than the GIS ($\sim 22'$ radius), so that
the energy-dependent vignetting effect is much smaller.
Moreover, improvement of the mirror surface has reduced
scatters on the reflector and equipment of the pre-collimator
dramatically decreased the so-called ``stray-lights''
mainly from outside of the XIS field of view
\citep{serlemitsos06}.\footnote{
The ``stray-lights'' reach to the focal plain through abnormal paths,
e.g., reflection by only the secondary mirror,
or the mirror backside reflection \citep{mori05}.}

We therefore estimated the fraction of photons outside
the extraction region using a simulator of the Suzaku XRT/XIS system,
``xissim'' version 2006-08-26 \citep{ishisaki06}, in the following way:
{\it (1)}~Assuming the double-$\beta$ model surface brightness profile
(table~\ref{tab:2beta}), 2,000,000 count of monochromatic incident-photons
(in 0.5, 1, 2, 4, or 8~keV) were generated by a ``mkphlist'' task.
{\it (2)}~Simulated event files were created for both the central
and offset observations using the ``xissim'' task,
with the {\sf xis\_efficiency} parameter set to ``no''
to save photon statistics.
{\it (3)}~Using the {\sc ra} and {\sc dec} columns in the simulated
event files, the event files were splitted into 7+1 sky regions
corresponding to the extraction annuli of 0--2$'$,
2--4$'$, 4--6$'$, 6--9$'$, 9--13$'$, 13--17$'$, and 17--27$'$,
plus outside of them, $r>27'$.\footnote{
The ``fselect'' task with
the {\it angsep} function was utilized to calculated the angular separation
of events from the assumed center of the cluster,
(RA, Dec) = (\timeform{10h36m42.8s}, \timeform{-27D31'42''}).}
{\it (4)}~The detected photon count at each extraction region on the
XIS detector plane was calculated for each splitted event file
using the ``xselect'' task.

In this way, we estimated the fraction of photons for the 7 extraction annuli
from the 8 sky regions in 5 energies for the 4 XIS sensors.
We plot examples of BI (XIS1) in figure~\ref{fig:frac}.
It is confirmed that more than 65\% of photons are coming from the
corresponding sky directions by the ``xissim'' simulations.
The energy dependence is almost negligible within $\sim 5\%$
in figure~\ref{fig:frac}(h).
Fractions from the second next to the noticed annulus are less than
$\sim 1\%$ at the offset observation, even though the surface brightness
of the central part of the cluster is much brighter
by orders of magnitude (figure~\ref{fig:2beta}).
This is primarily owing to the existence of the pre-collimator for
the Suzaku XRT\@.
These results are almost the same with the other XIS sensors.
Note that this simulation does not include the CXB and
the Galactic background, which are almost uniform on the sky
so that the spatial response is different from figure~\ref{fig:frac}.

\section{Dependence on Abundance Tables}
\label{app:lodd}

\begin{figure*}[t]
\begin{minipage}{0.32\textwidth}
\FigureFile(\textwidth,\textwidth){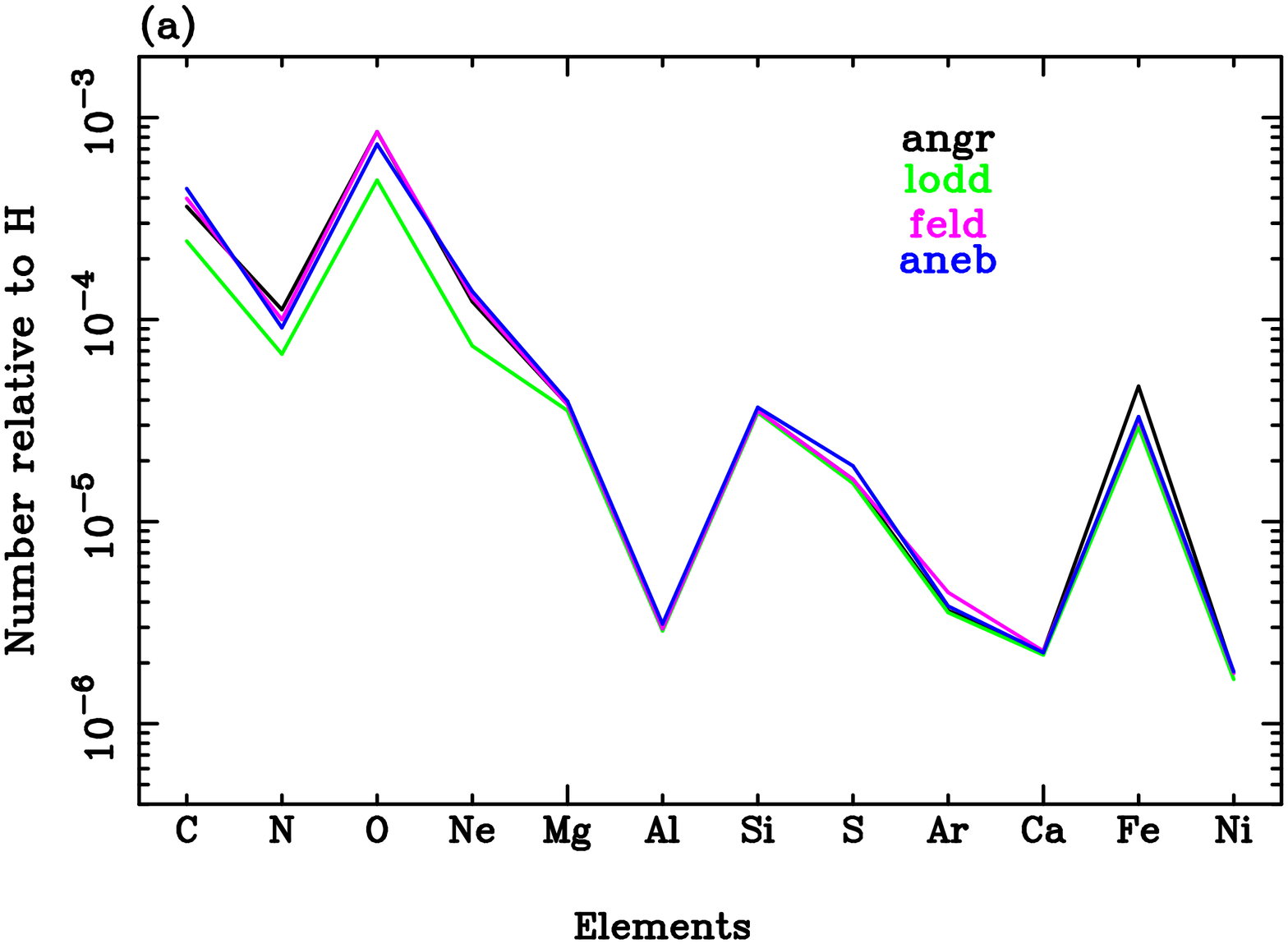}
\end{minipage}\hfill
\begin{minipage}{0.32\textwidth}
\FigureFile(\textwidth,\textwidth){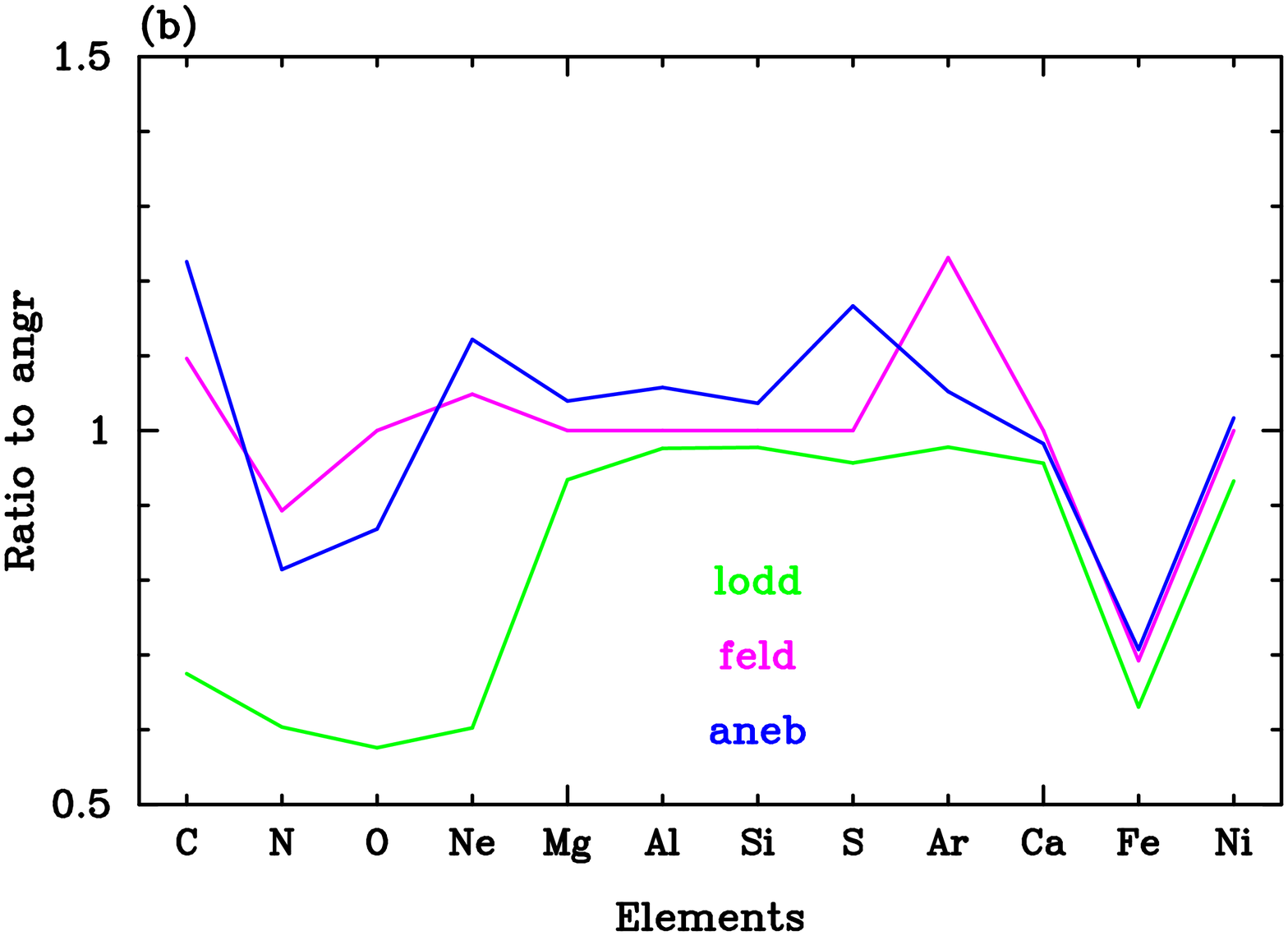}
\end{minipage}\hfill
\begin{minipage}{0.32\textwidth}
\FigureFile(\textwidth,\textwidth){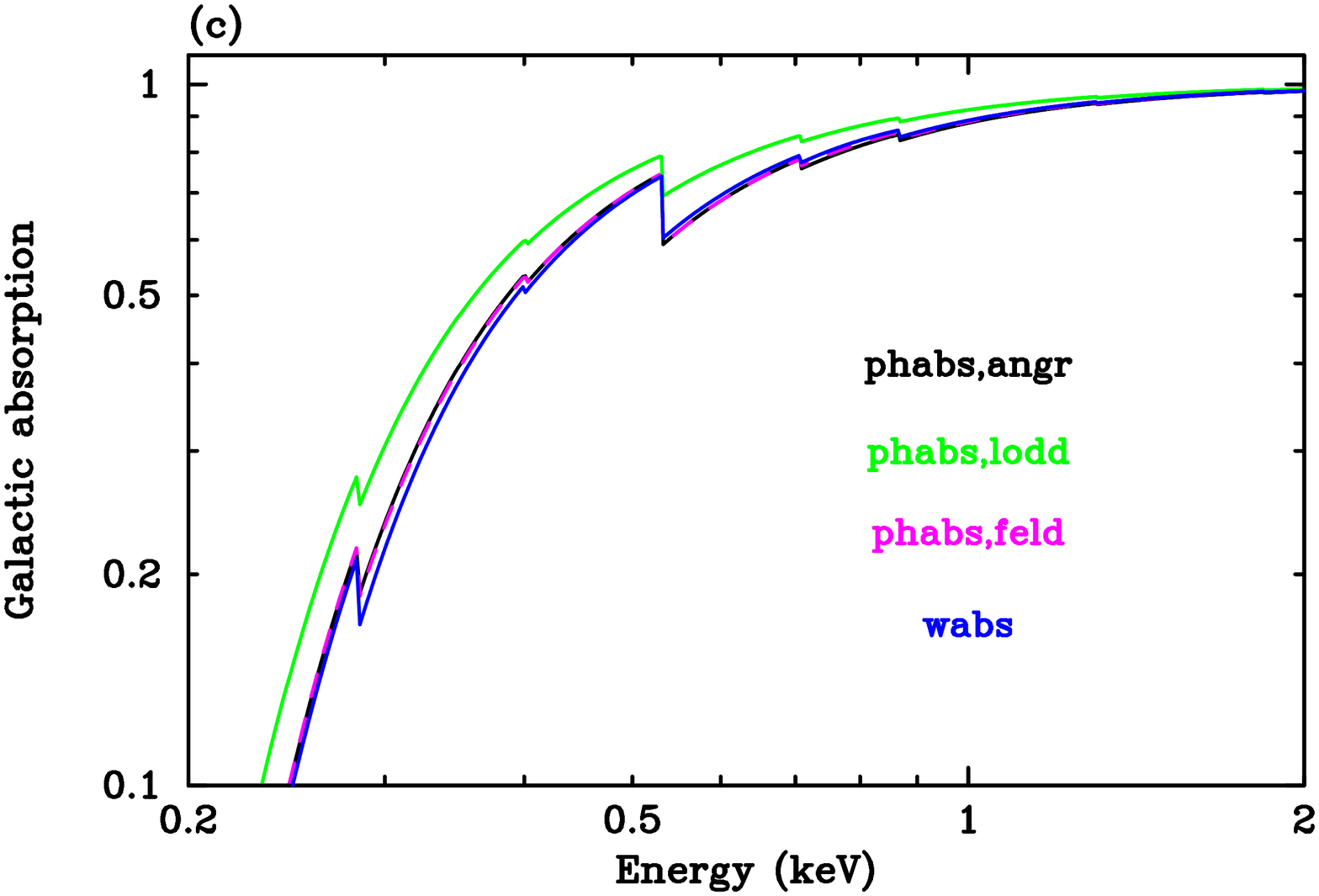}
\end{minipage}
\caption{
(a) The assumed metal abundances relative to H in number by the {\it angr},
{\it lodd}, {\it feld}, and {\it aneb} tables of XSPEC v11.3.0t.
(b) Same as (a) but normalized by the {\it angr} abundances.
(c) Comparison of absorption by the {\it phabs} model with
{\it angr} abundance table, {\it phabs} with {\it lodd},
and the {\it wabs} model in XSPEC\@.
The {\it wabs} model assumes the abundance ratio of {\it aneb}
built-in the code.
Neutral hydrogen column density of $N_{\rm H} = 4.9\times 10^{20}$~cm$^{-2}$
and ``1 solar'' abundance, which is different among the tables, are assumed.
The photoelectric absorption cross-section of {\it bcmc}
\citep{Balucinska-Church1992,Yan1998} is used.
}\label{fig:abund-abs}
\end{figure*}

\begin{figure*}[t]
\begin{minipage}{0.33\textwidth}
\FigureFile(\textwidth,\textwidth){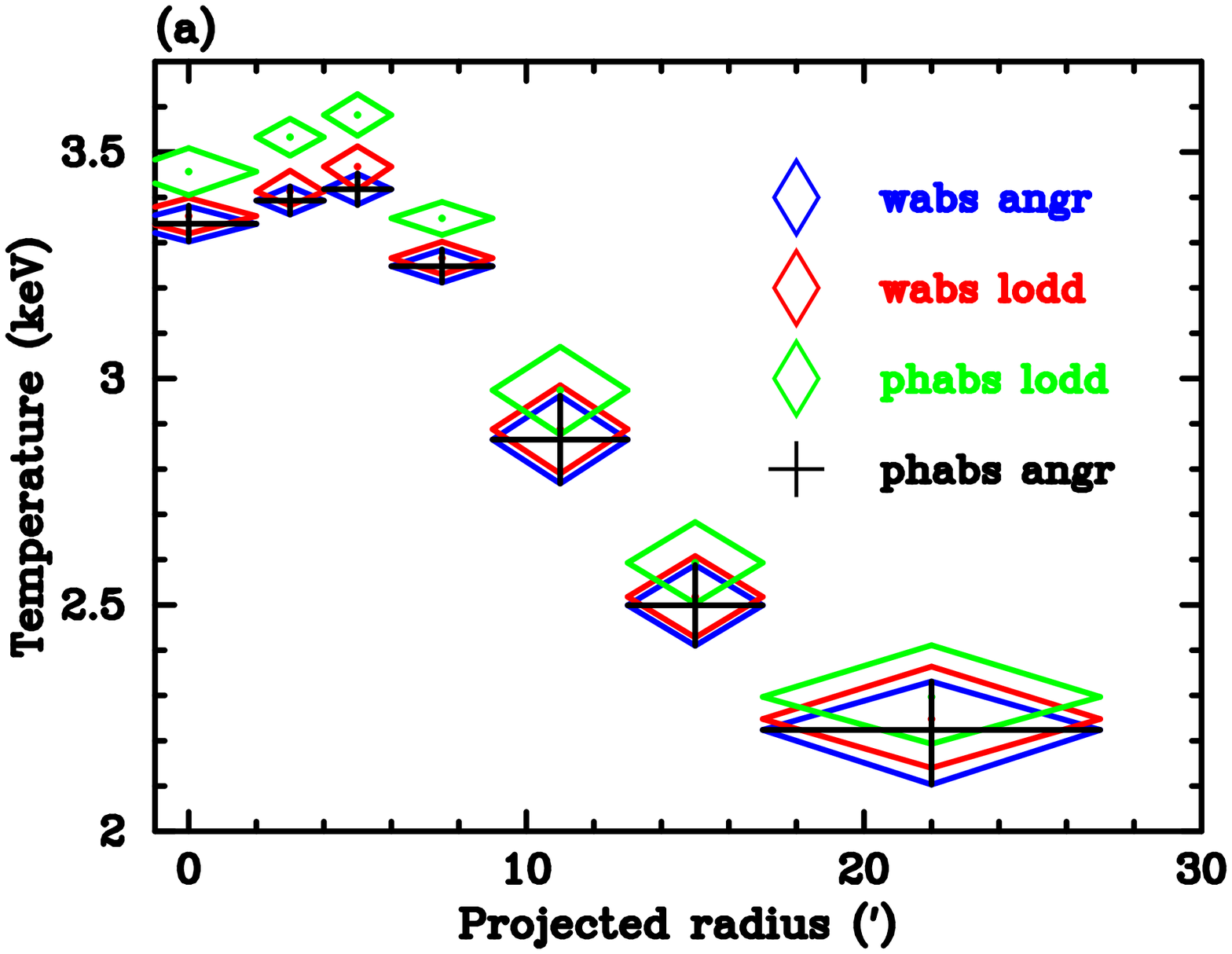}
\end{minipage}%
\begin{minipage}{0.33\textwidth}
\FigureFile(\textwidth,\textwidth){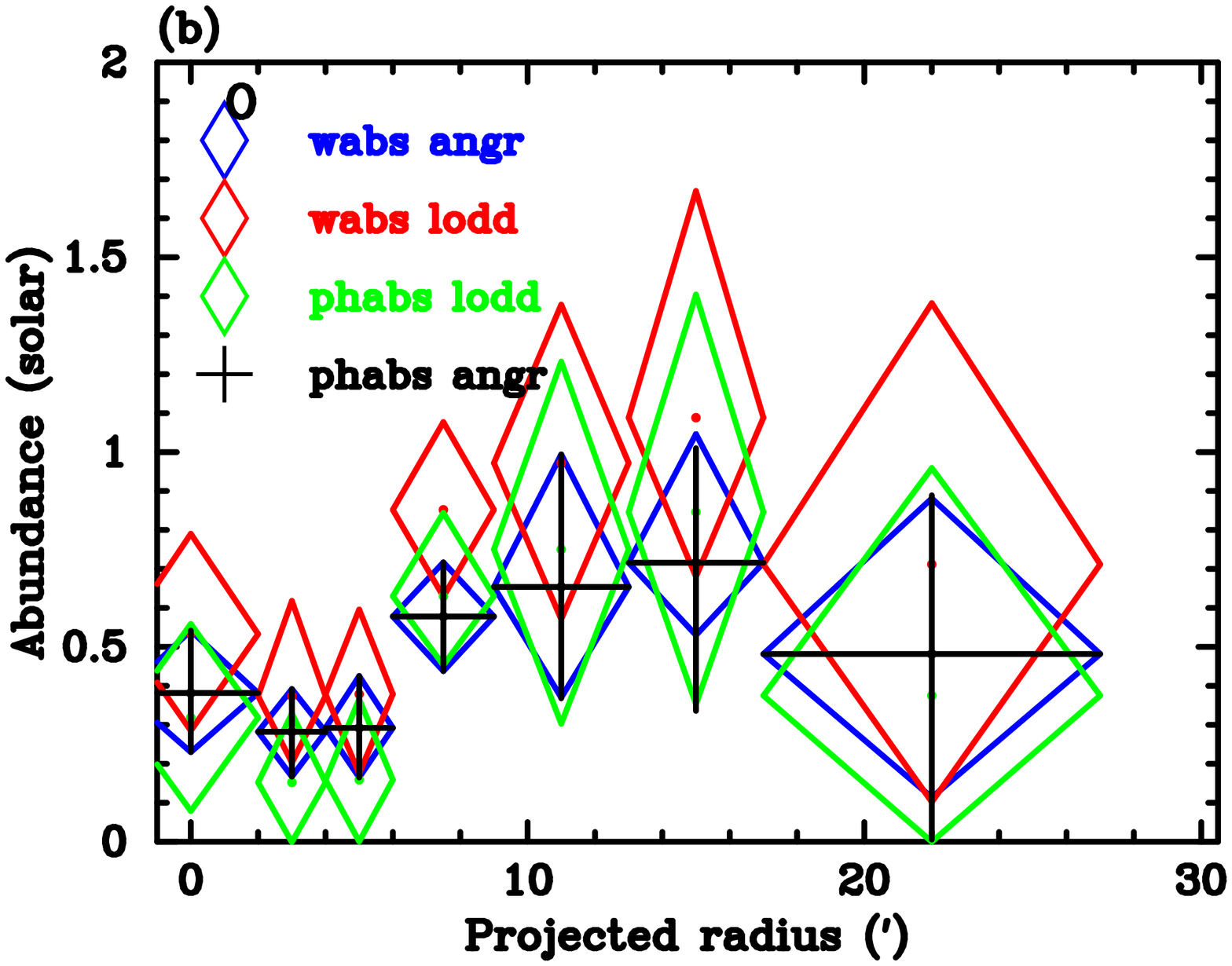}
\end{minipage}%
\begin{minipage}{0.33\textwidth}
\FigureFile(\textwidth,\textwidth){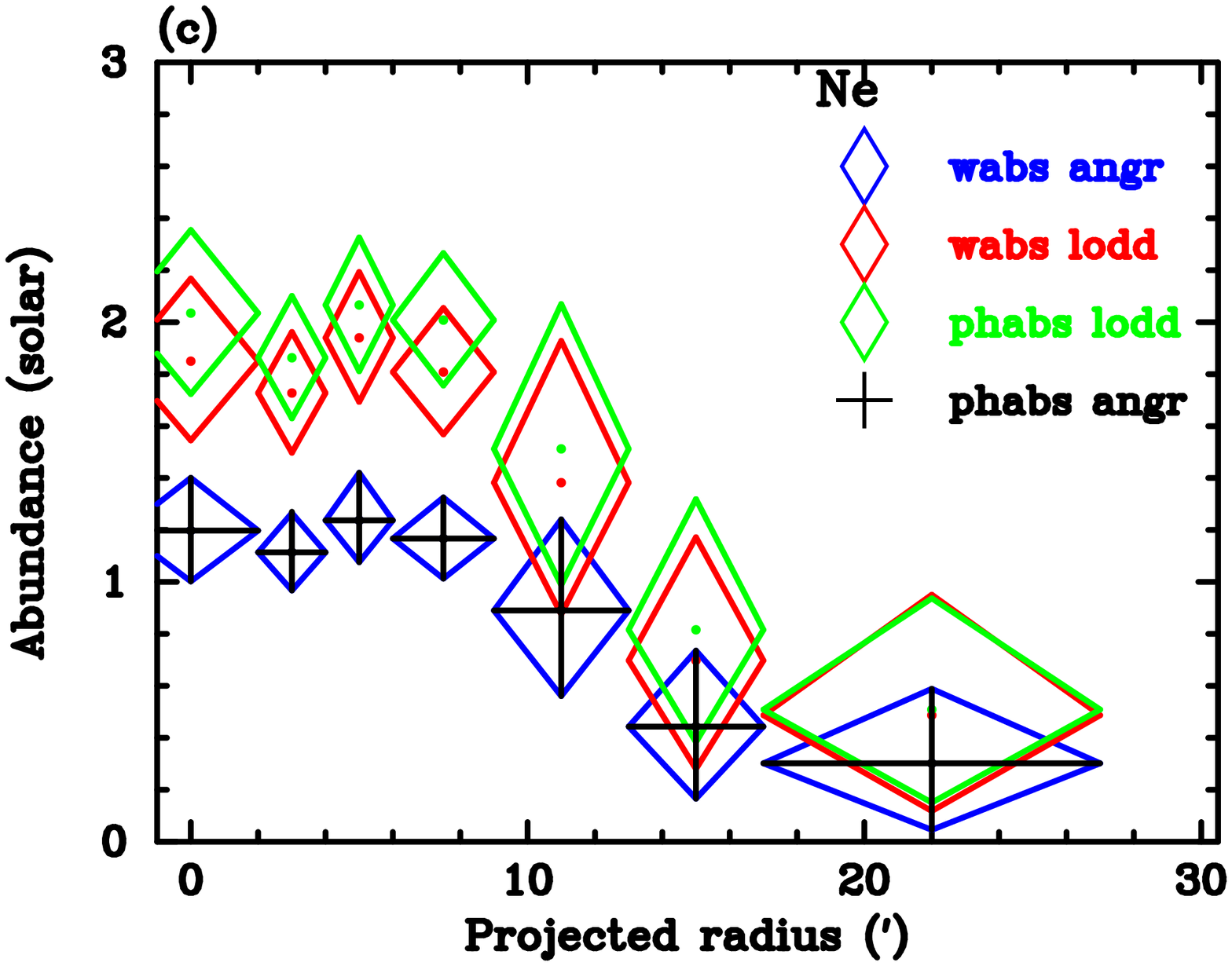}
\end{minipage}%

\begin{minipage}{0.33\textwidth}
\FigureFile(\textwidth,\textwidth){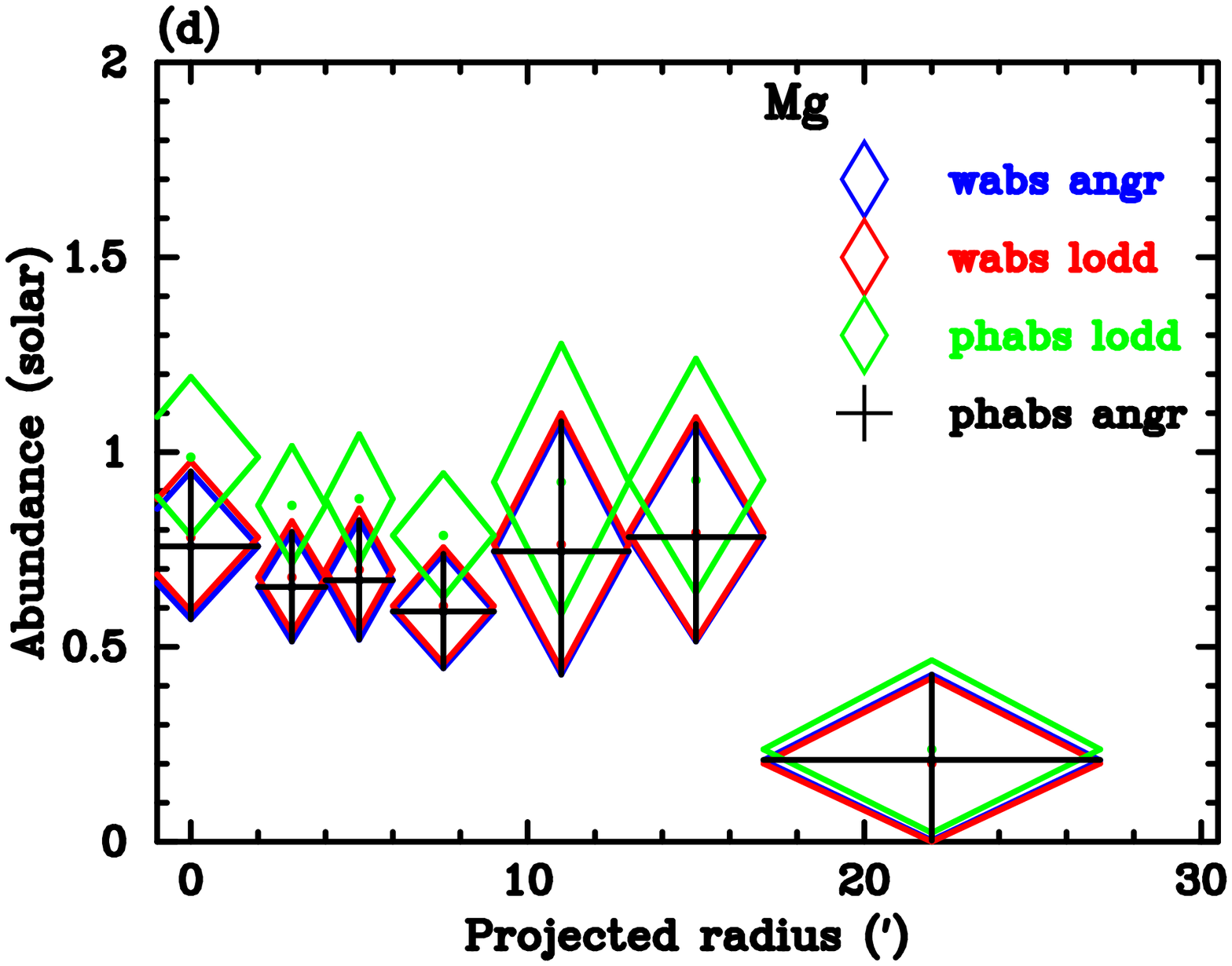}
\end{minipage}%
\begin{minipage}{0.33\textwidth}
\FigureFile(\textwidth,\textwidth){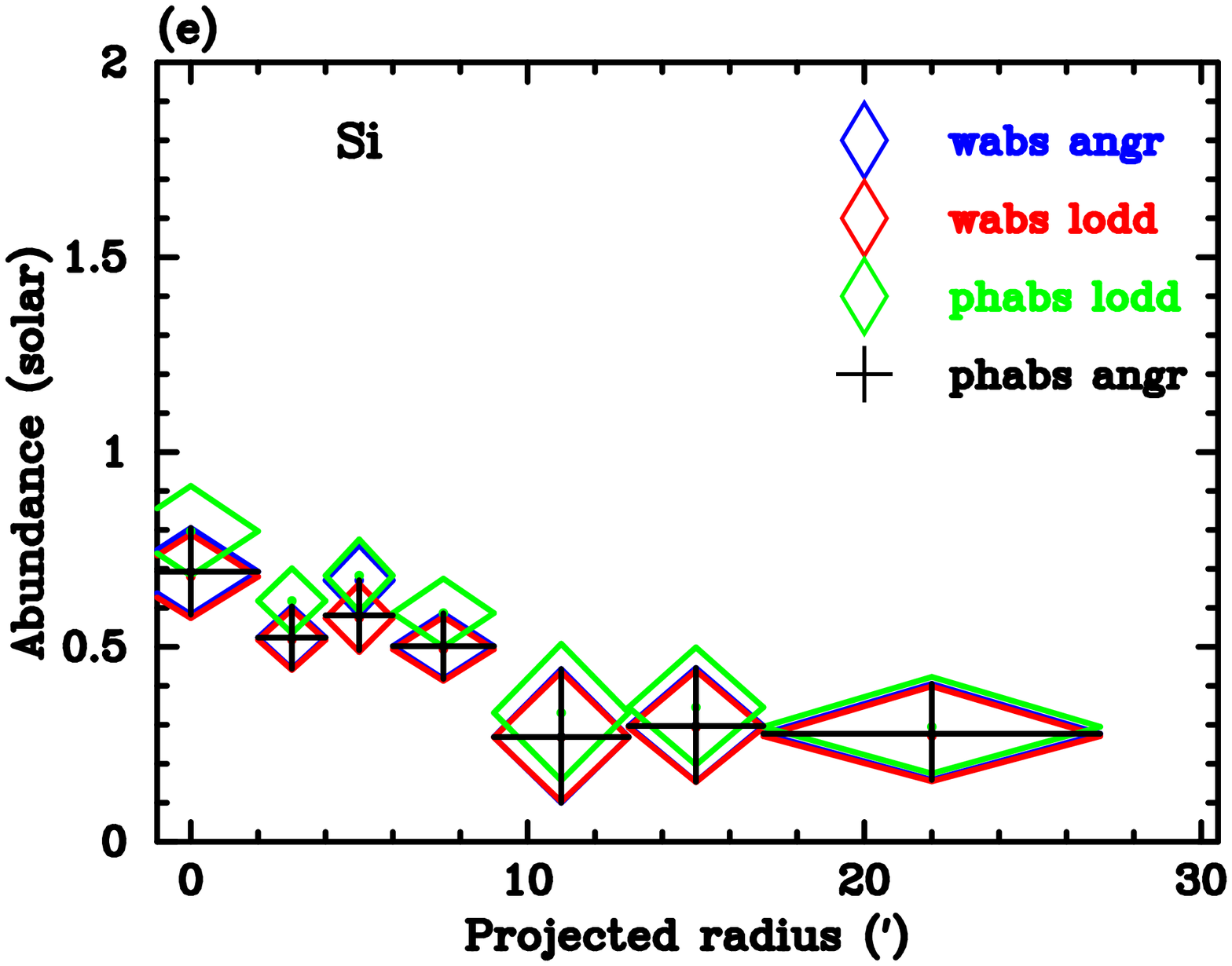}
\end{minipage}%
\begin{minipage}{0.33\textwidth}
\FigureFile(\textwidth,\textwidth){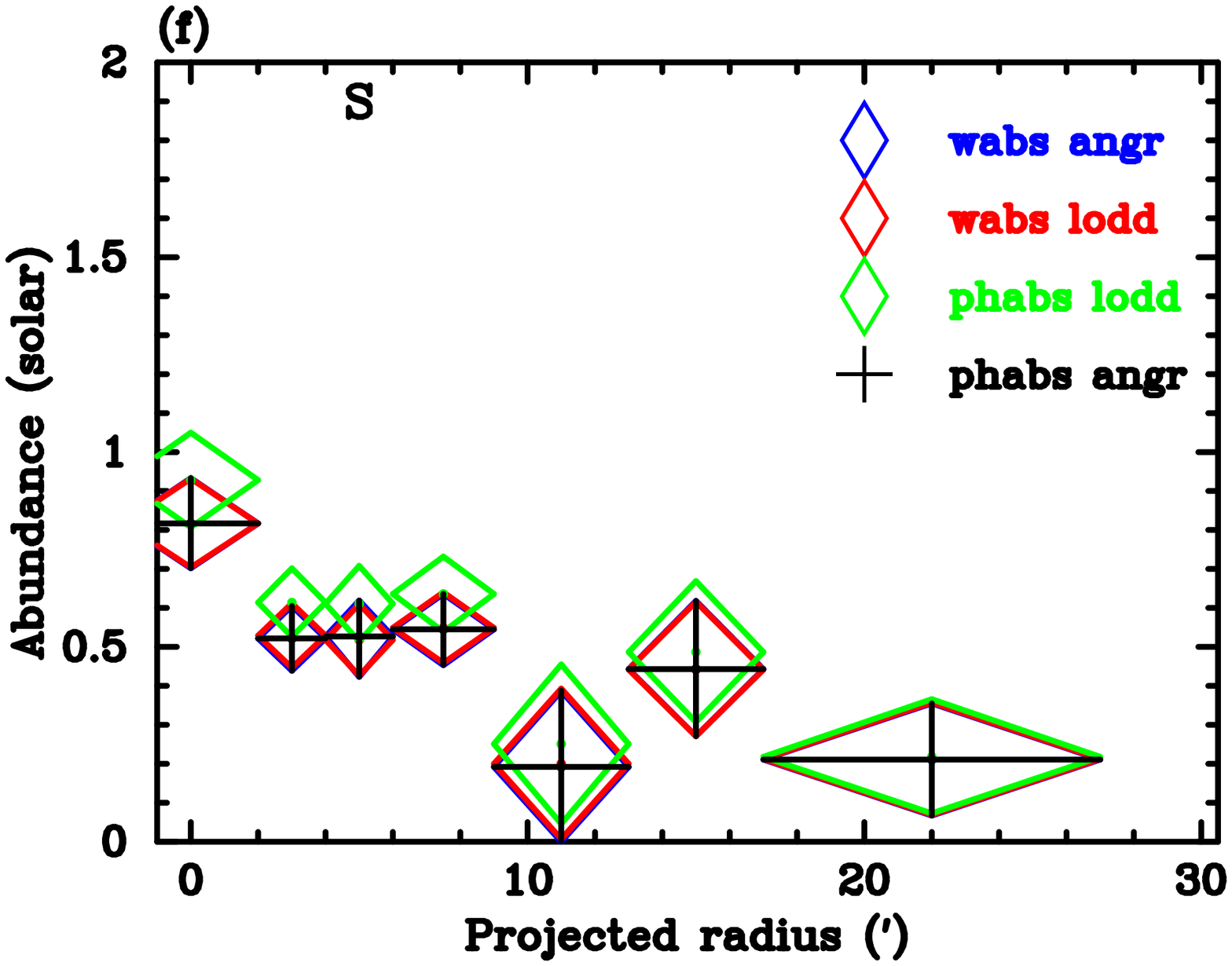}
\end{minipage}

\begin{minipage}{0.33\textwidth}
\FigureFile(\textwidth,\textwidth){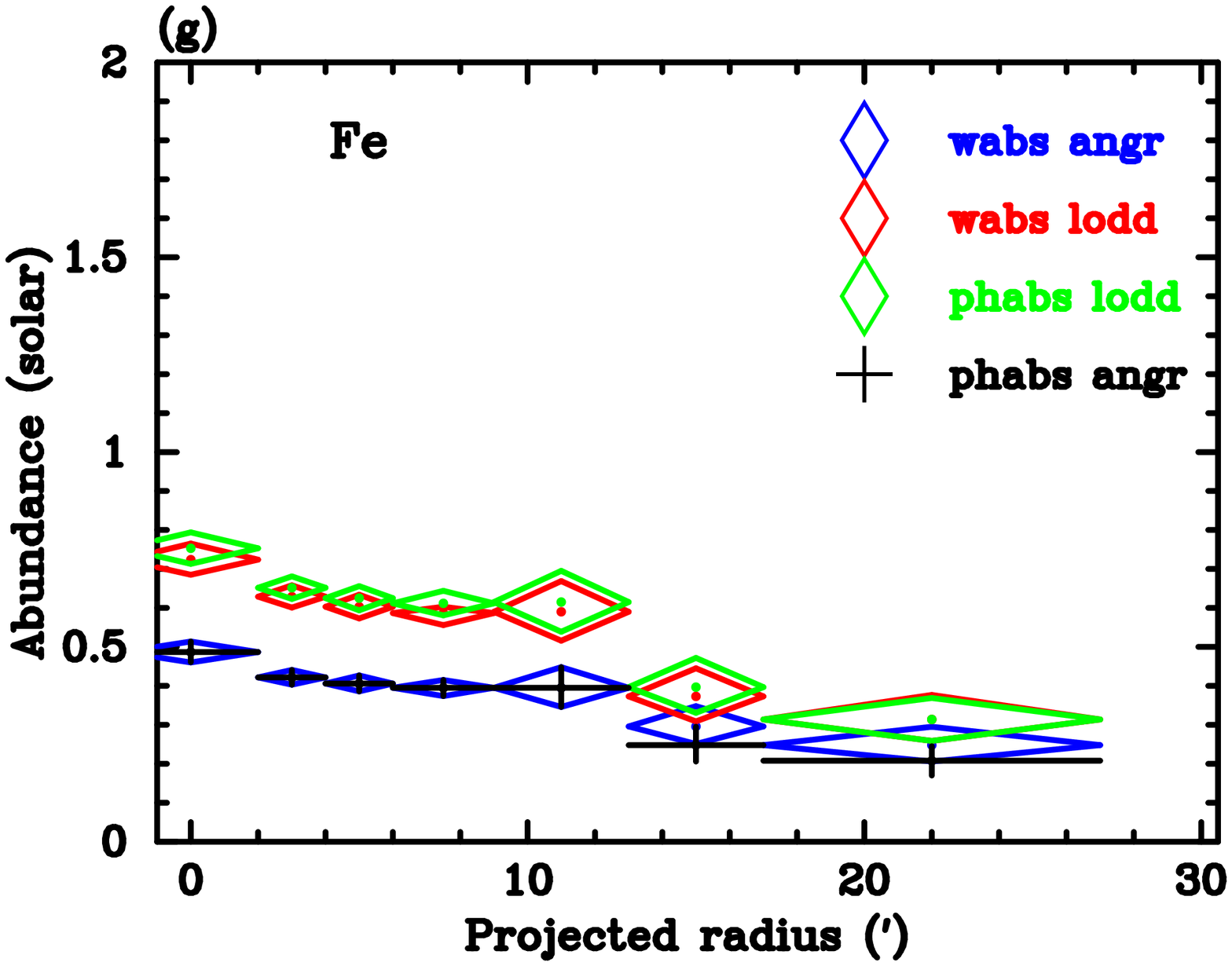}
\end{minipage}%
\begin{minipage}{0.33\textwidth}
\FigureFile(\textwidth,\textwidth){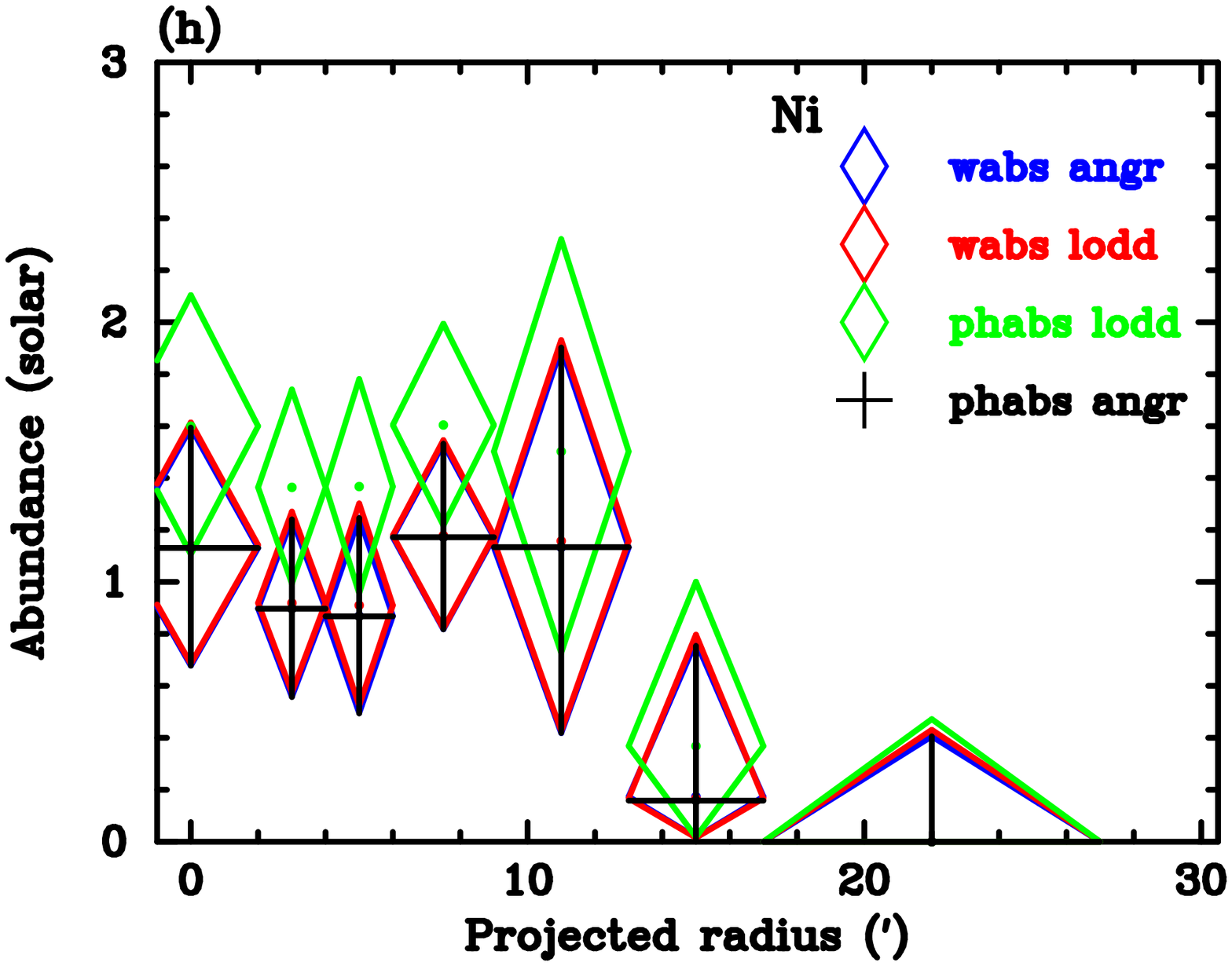}
\end{minipage}

\caption{
Same as figure~\ref{fig:result} of the Suzaku result
for the black crosses ({\it phabs} {\it angr}).
Other fit results are also plotted in blue, red, and green diamonds,
when different abundance ratio ({\it lodd}) and/or
different absorption model ({\it wabs}) are assumed.
}\label{fig:lodd-result}
\end{figure*}

\begin{figure*}[t]
\begin{minipage}{0.33\textwidth}
\FigureFile(\textwidth,\textwidth){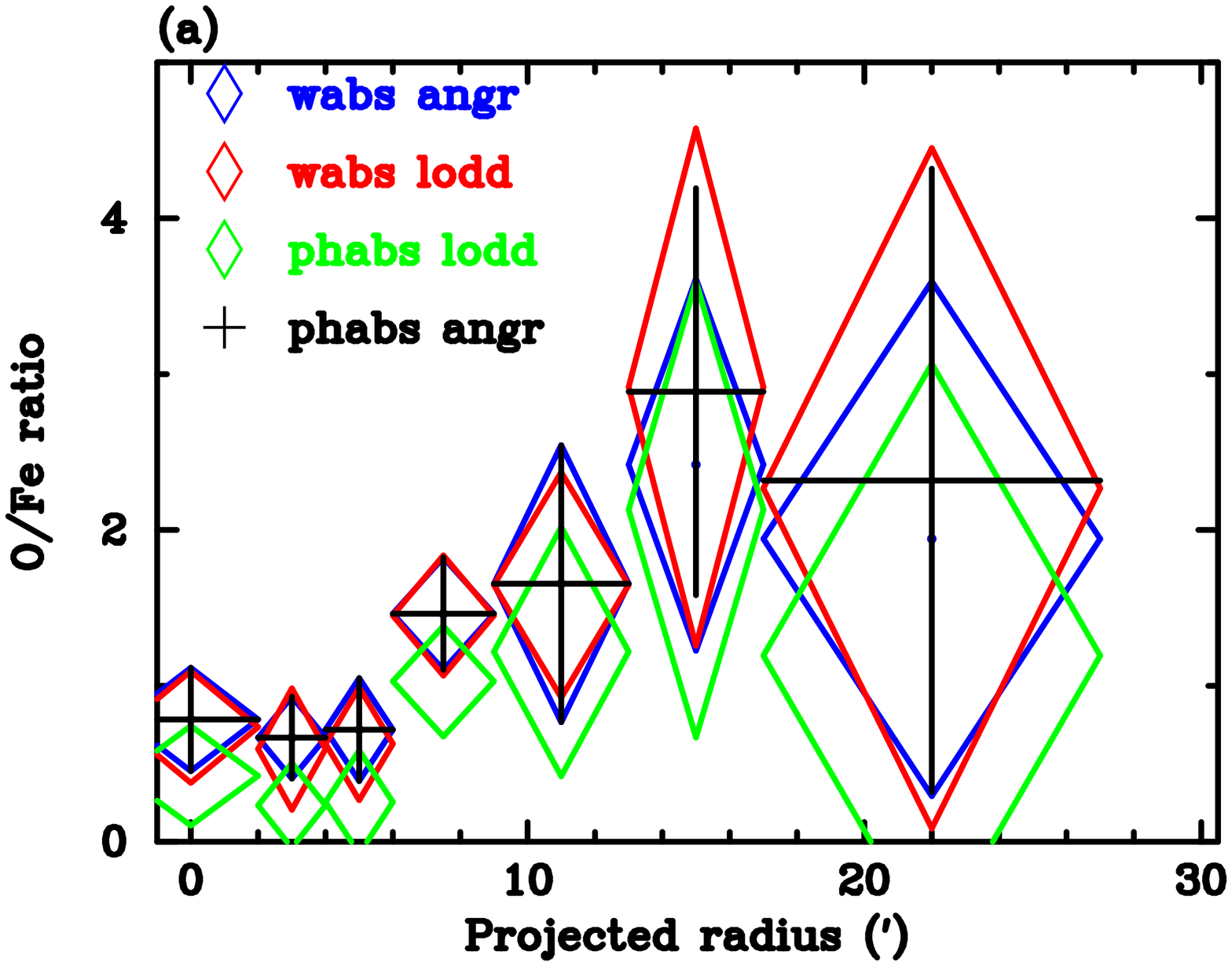}
\end{minipage}%
\begin{minipage}{0.33\textwidth}
\FigureFile(\textwidth,\textwidth){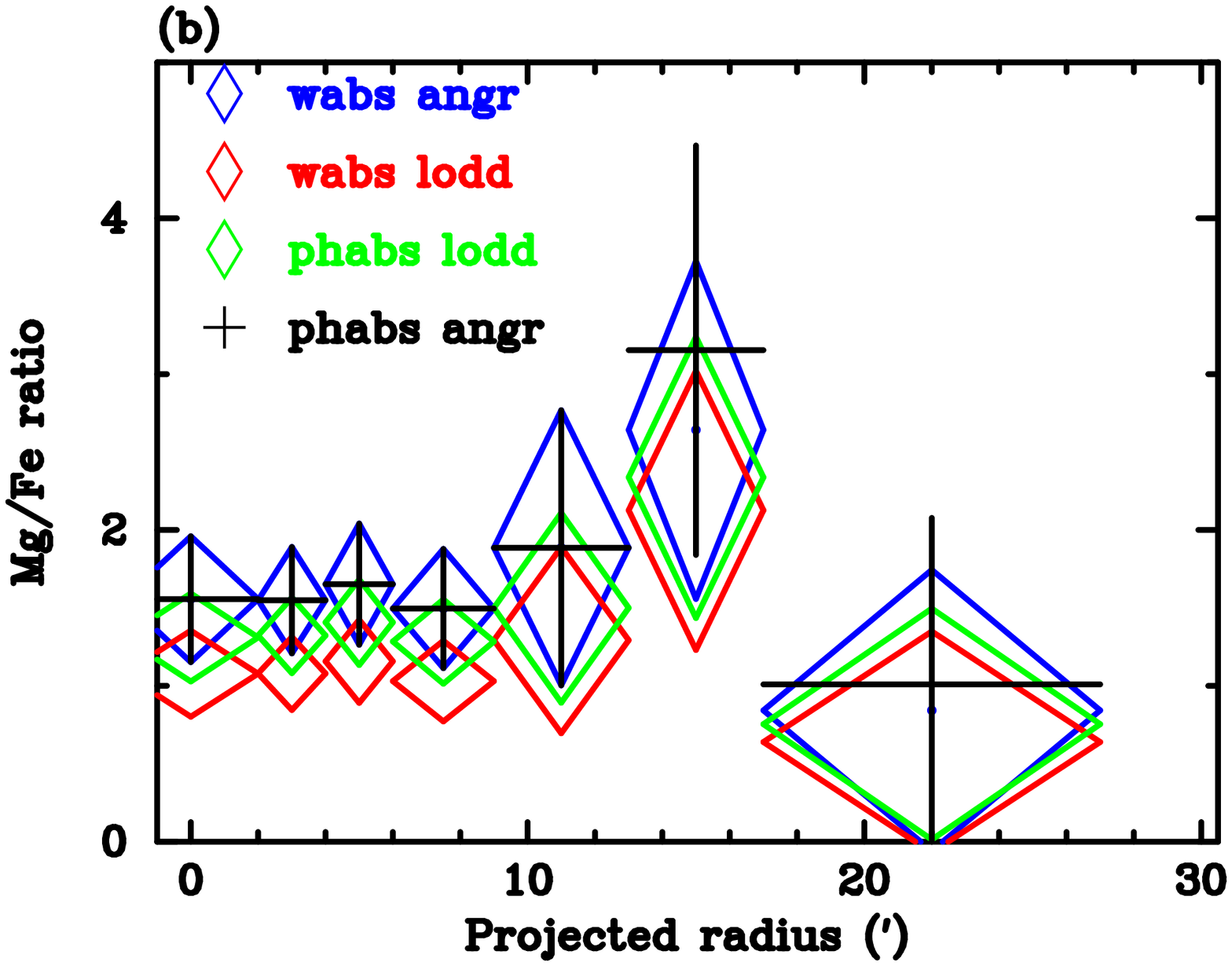}
\end{minipage}%
\begin{minipage}{0.33\textwidth}
\FigureFile(\textwidth,\textwidth){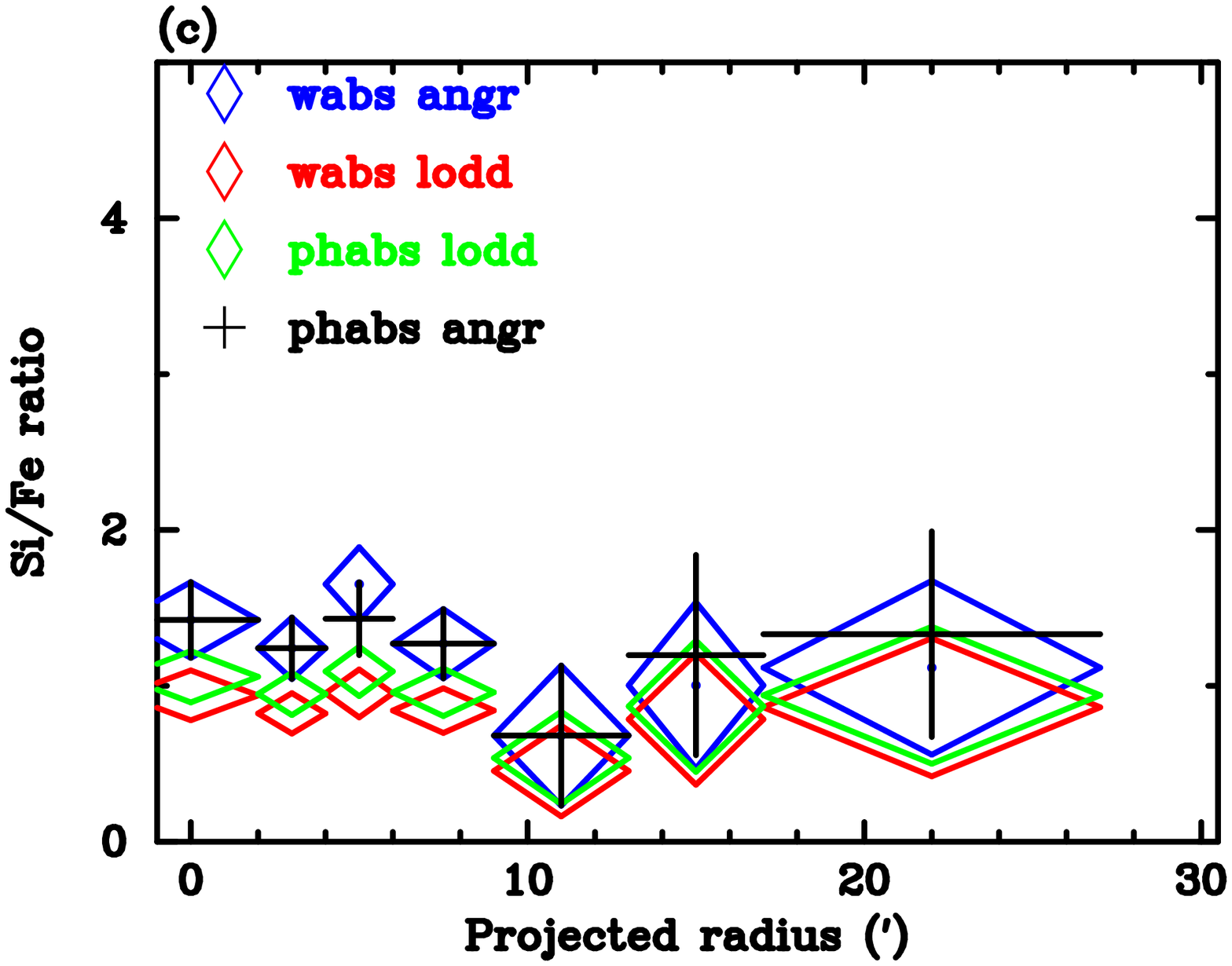}
\end{minipage}%
\caption{
(a) O/Fe ratio, (b) Mg/Fe ratio, and (c) Si/Fe ratio,
which are same as figure~\ref{fig:ratio}(a)
for the black crosses ({\it phabs} {\it angr}) in these figures (a)--(c).
Other fit results are also plotted in blue, red, and green diamonds,
when different abundance ratio ({\it lodd}) and/or
different absorption model ({\it wabs}) are assumed.
}\label{fig:lodd-ratio}
\end{figure*}

\begin{figure*}[t]
\begin{minipage}{0.33\textwidth}
\FigureFile(\textwidth,\textwidth){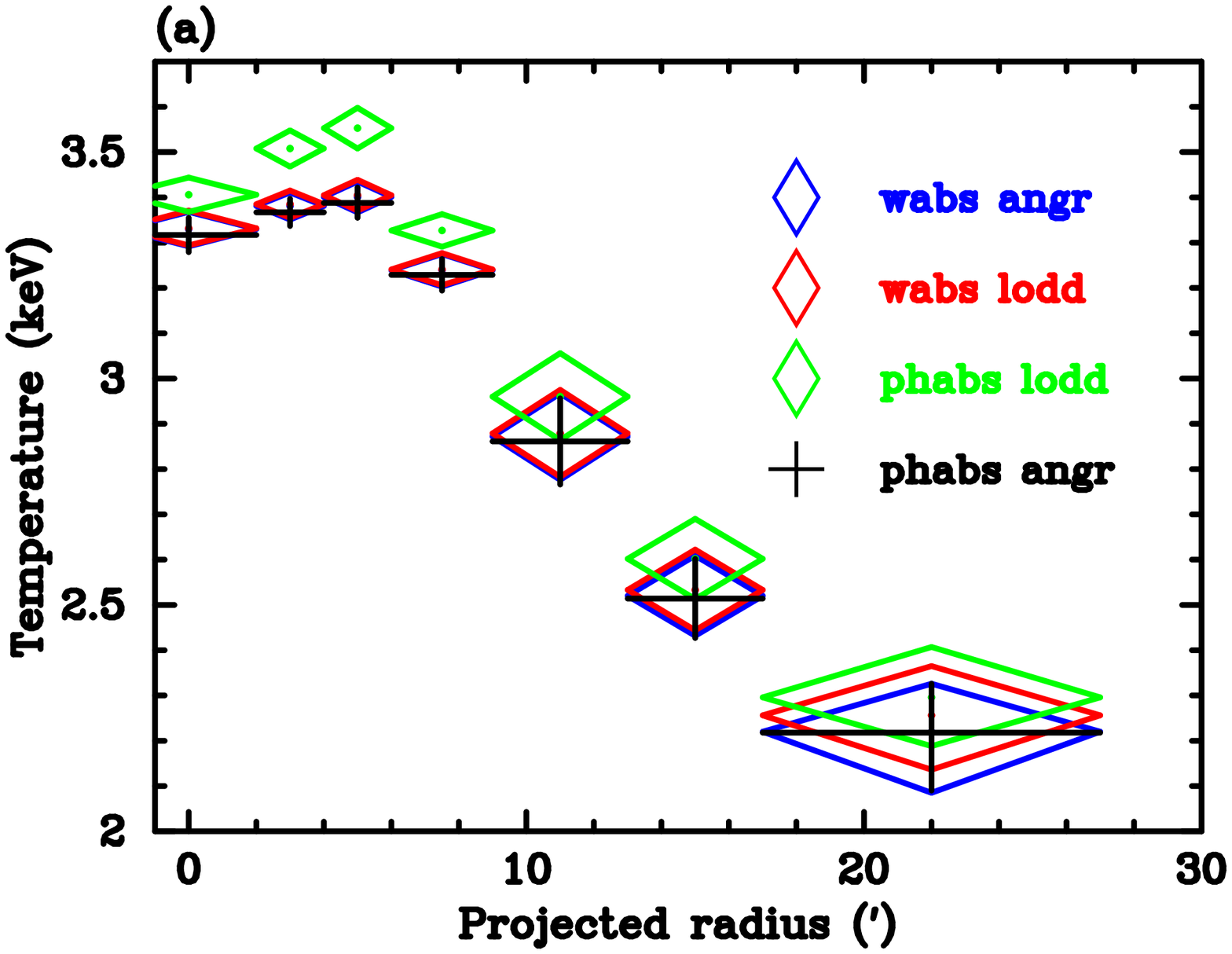}
\end{minipage}%
\begin{minipage}{0.33\textwidth}
\FigureFile(\textwidth,\textwidth){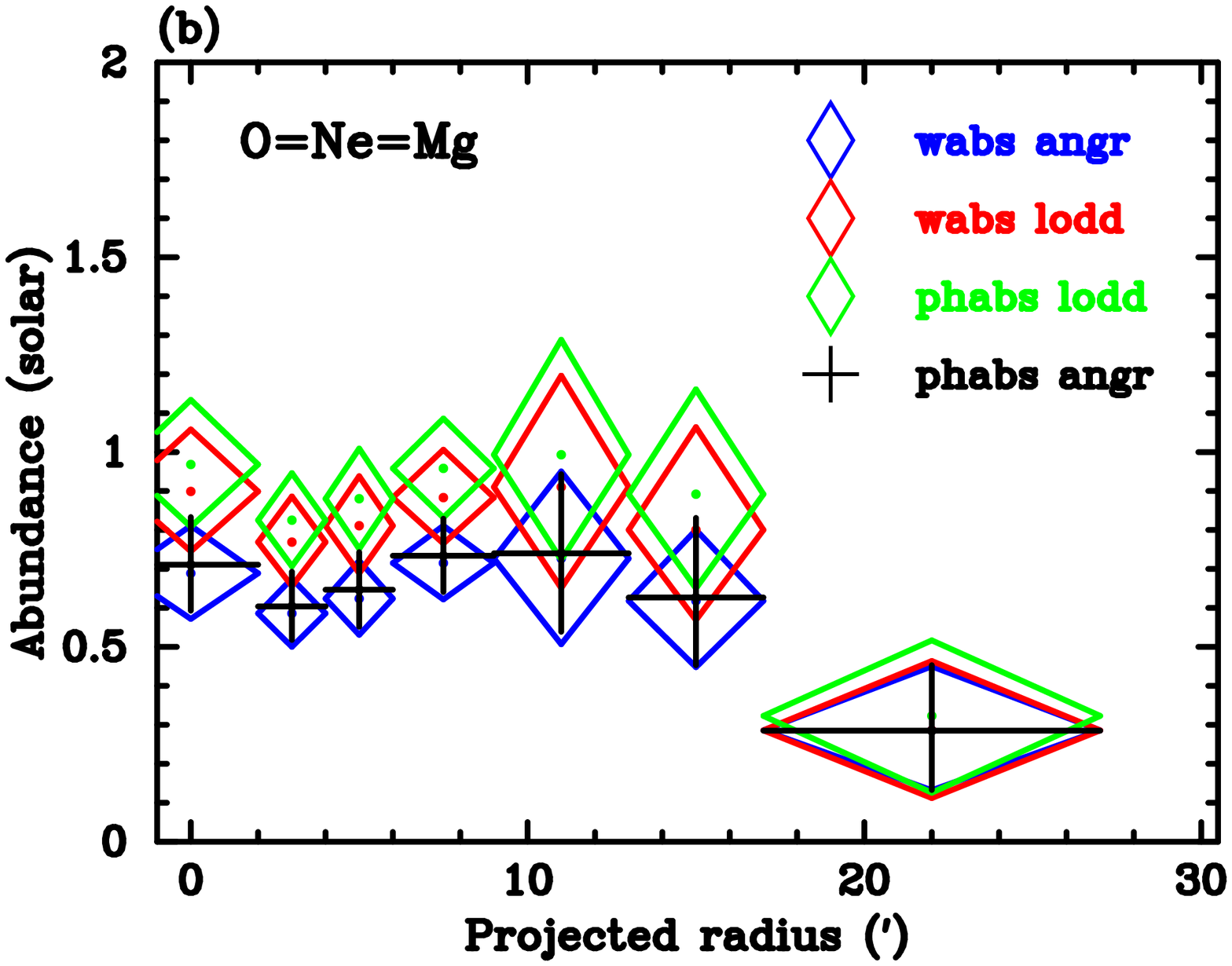}
\end{minipage}%
\begin{minipage}{0.33\textwidth}
\FigureFile(\textwidth,\textwidth){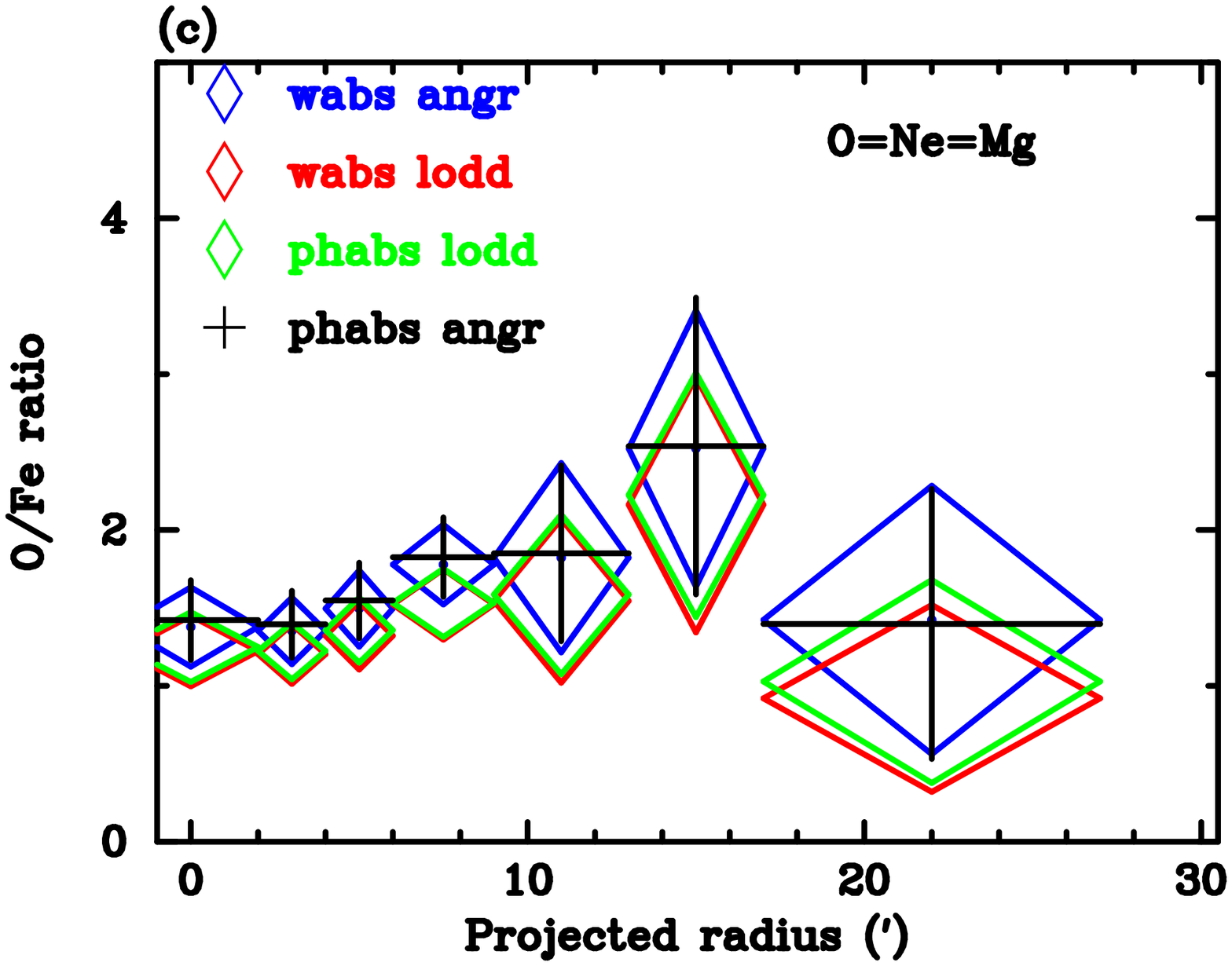}
\end{minipage}%
\caption{
Radial profiles of (a) temperature, (b) O = Ne = Mg abundance,
and (c) abundance ratio to Fe,
when the abundances of O, Ne, and Mg are constrained to
have the same value for the assumed abundance tables of
{\it angr} or {\it lodd}.
Meaning of the markers are same as figures~\ref{fig:lodd-result} and
\ref{fig:lodd-ratio}.
}\label{fig:lodd-onemg}
\end{figure*}

\begin{table*}[t]
\caption{List of $\chi^2$/dof for each fit.}
\label{tab:chi}
\begin{center}
\begin{tabular}{lccccccccc}
\hline\hline
\makebox[6em][l]{Region} & \multicolumn{4}{c}{O, Ne, Mg independent} && \multicolumn{4}{c}{Constrained to be O=Ne=Mg}\\
\hline
 & \makebox[0in][c]{\it phabs angr} & \makebox[0in][c]{\it wabs angr} & \makebox[0in][c]{\it wabs lodd} & \makebox[0in][c]{\it phabs lodd} && \makebox[0in][c]{\it phabs angr} & \makebox[0in][c]{\it wabs angr} & \makebox[0in][c]{\it wabs lodd} & \makebox[0in][c]{\it phabs lodd} \\
\hline
0--2$'$ $\dotfill$   & 1240/992 & 1240/992 & 1215/992 & 1274/992 && 1278/996 & 1260/996 & 1258/996 & 1335/996\\
2--4$'$ $\dotfill$   & 1290/992 & 1290/992 & 1267/992 & 1356/992 && 1358/996 & 1346/996 & 1343/996 & 1464/996\\
4--6$'$ $\dotfill$   & 1345/992 & 1345/992 & 1312/992 & 1400/992 && 1418/996 & 1396/996 & 1397/996 & 1513/996\\
6--9$'$ $\dotfill$   & 1329/992 & 1329/992 & 1298/992 & 1395/992 && 1365/996 & 1340/996 & 1359/996 & 1464/996\\
9--13$'$ $\dotfill$  & 1057/992 & 1057/992 & 1053/992 & 1059/992 && 1058/996 & 1057/996 & 1057/996 & 1063/996\\
13--17$'$ $\dotfill$ & 1082/992 & 1082/992 & 1077/992 & 1078/992 && 1086/996 & 1085/996 & 1080/996 & 1080/996\\
\hline
\end{tabular}
\end{center}
\end{table*}

There are several abundance tables usable in XSPEC\@,
such as {\it angr} \citep{anders89}, {\it aneb} \citep{anders82},
{\it feld} \citep{feldman92}, and {\it lodd} \citep{lodders03}.
The elemental abundances relative to hydrogen in number
defined by these tables are
plotted in figure~\ref{fig:abund-abs}(a), and ratios of each element
to {\it angr}\/ is shown in figure~\ref{fig:abund-abs}(b).
Although {\it feld}\/ is not used in our spectral analysis,
it is added to compare recent Suzaku results
\citep{matsushita06a, tawara06} which adopt this table.
As seen in figure~\ref{fig:abund-abs}(b),
{\it angr}\/ and {\it feld}\/ are essentially the same
except for Fe abundance, which is 1.44 times larger
for {\it angr}\/ than {\it feld}.

The {\it angr}\/ table is the de facto of XSPEC,
however, Fe abundance is significantly different from others.
This is because {\it angr}\/ is based on the solar photospheric
measurement of Fe\emissiontype{I} \citep{blackwell84},
which disagrees with the meteorites measurement adopted in {\it aneb}.
Discussions afterwards 
seem to prefer the lower abundance by meteorites,
because Fe abundance determined by Fe\emissiontype{II},
which is dominant in the solar photosphere, is consistent
with the meteorites value \citep{Biemont1991,Raassen1998}.

It is notable that C, N, O, and Ne abundances in {\it lodd}\/
is significantly lower ($\sim 60$\% of {\it angr}).
These values are based on the three-dimensional time-dependent
hydrodynamical model solar atmosphere, in which departures
from local thermodynamical equilibrium (LTE) and
blend of Ni\emissiontype{I} line with the forbidden
[O\emissiontype{I}] $\lambda6300$ are considered
\citep{Allende-Prieto2001,Allende-Prieto2002}.
The {\it lodd}\/ table in XSPEC is the recommended elemental
abundance of the solar photosphere (table~1 in \cite{lodders03}),
however, he states that the photo-solar abundances as
representative of the solar system should be slightly larger
(by $\sim 1.2$ times for C, N, and O) for helium and heavy-element
due to the settling effects.

The decrease of these abundant elements in
the Inter Stellar Medium (ISM) also leads to the difference
in the assumed Galactic absorption as seen in
figure~\ref{fig:abund-abs}(c).
The {\it phabs}\/ absorption model with {\it lodd} abundance table
gives by $\sim 15$\% smaller absorption at the maximum
in our noticed energy range of 0.4--7.1~keV in the spectral fitting,
while the difference between {\it wabs}\/ and
{\it phabs}\/ with {\it angr}\/ is negligible.
This is because we have assumed the neutral hydrogen column density
of $N_{\rm H} = 4.9\times 10^{20}$ cm$^{-2}$ \citep{dickey90}
with ``1 solar'' abundance for the ISM\@.
In fact, the ISM consists of monoatomic gas, molecules, and grains,
so that it is suggested to have smaller abundance than the solar abundance
by \citet{Wilms2000}.
\citet{Shaver1983} have reported a radial abundance
gradient in the Milky Way Galaxy of $\sim 16$\% per kpc.

Considering these uncertainties, we have tested the spectral fit
of A~1060 with four kinds of combinations in
the absorption model and the abundance table,
{\it phabs} {\it angr}, {\it wabs} {\it angr}, {\it wabs} {\it lodd},
and {\it phabs} {\it lodd}.
We further tested the fit when the abundances of O, Ne, and Mg are linked
to have the same value relative to the assumed ``solar'' abundance,
in two reasons: they are all supposed to be primarily the SN~II products;
O and Ne abundances are susceptible to the Galactic component in
outer annuli. Note that the ${\rm O : Ne : Mg}$ ratios are
significantly different between {\it angr} ($1:0.145:0.045$)
and {\it lodd} ($1:0.151:0.072$).

Results are summarized in figures~\ref{fig:lodd-result}--\ref{fig:lodd-onemg}
and a list of $\chi^2$/dof is presented in table~\ref{tab:chi}.
As expected from figure~\ref{fig:abund-abs}(c),
the {\it phabs} {\it angr}\/ and {\it wabs} {\it angr}\/ results
are almost identical. The {\it phabs} {\it lodd} gives slightly
larger temperature than others, which leads the O abundance profile
in figure~\ref{fig:lodd-result}(b) to be different behavior from others.
Other discrepancies in the derived abundances can be explained by
difference in the definition of ``1 solar''.

Since $\chi^2$ values for {\it phabs} {\it lodd} are generally larger than
other models, the Galactic absorption model by {\it phabs} {\it lodd}
is considered to be inappropriate with our X-ray data.
On the other hand, {\it wabs} {\it lodd} model shows the minimum $\chi^2$
in table~\ref{tab:chi}. This is interesting because the {\it wabs}
absorption model assumes {\it aneb} abundance table built-in the code.
This may indicates that our modeling of the Galactic component is
too naive, and/or the Galactic abundance gradient is responsible
so that effective elemental abundance between absorption and emission
are different.

It is notable that shape of the radial abundance profiles
are quite similar among these four models in figure~\ref{fig:lodd-onemg}
when $\rm O=Ne=Mg$ is assumed.
Difference from Fe or Si is evident as seen in
figures~\ref{fig:lodd-ratio}(c) and \ref{fig:lodd-onemg}(c):
Fe and Si abundances decreases with radius with roughly constant Si/Fe ratio,
while $\rm O=Ne=Mg$ are almost constant within $r\lesssim 17'$.
The higher $\rm (O=Ne=Mg)/Fe$ ratio than ``1 solar'' is also suggested
for both {\it angr}\/ and {\it lodd}\/ abundance tables.

\end{document}